# Risk prediction models for discrete ordinal outcomes: calibration and the impact of the proportional odds assumption

Michael EDLINGER [a,b], Maarten VAN SMEDEN [c,d], Hannes F ALBER [e,f], Maria WANITSCHEK [g], Ben VAN CALSTER [a,h,i] *

[a] Department of Development and Regeneration, KU Leuven, Belgium

[b] Department of Medical Statistics, Informatics, and Health Economics, Medical University Innsbruck, Austria

[c] Julius Centre for Health Science and Primary Care, University Medical Centre Utrecht, the Netherlands

[d] Department of Clinical Epidemiology, Leiden University Medical Centre, the Netherlands

[e] Department of Internal Medicine and Cardiology, Klinikum Klagenfurt am Wörthersee, Austria

[f] Karl Landsteiner Institute for Interdisciplinary Science, Rehabilitation Centre, Münster, Austria

[g] University Clinic of Internal Medicine III - Cardiology and Angiology, Tirol Kliniken, Innsbruck, Austria

[h] EPI-centre, KU Leuven, Leuven, Belgium

[i] Department of Biomedical Data Sciences, Leiden University Medical Centre, the Netherlands

[*] Correspondence to: Ben Van Calster, Department of Development and Regeneration, KU Leuven, Herestraat 49 Box 805, 3000 Leuven, Belgium; ben.vancalster@kuleuven.be

Short title: calibration of risk prediction models for ordinal outcomes

Word count main text: 5468 (excluding abstract, appendix and references)

Funding: Research Foundation – Flanders (FWO) (grant G0B4716N), Internal Funds KU Leuven (grants C24/15/037 and C24M/20/064).

## Abstract

Calibration is a vital aspect of the performance of risk prediction models, but research in the context of ordinal outcomes is scarce. This study compared calibration measures for risk models predicting a discrete ordinal outcome, and investigated the impact of the proportional odds assumption on calibration and overfitting. We studied the multinomial, cumulative, adjacent category, continuation ratio, and stereotype logit/logistic models. To assess calibration, we investigated calibration intercepts and slopes, calibration plots, and the estimated calibration index. Using large sample simulations, we studied the performance of models for risk estimation under various conditions, assuming that the true model has either a multinomial logistic form or a cumulative logit proportional odds form. Small sample simulations were used to compare the tendency for overfitting between models. As a case study, we developed models to diagnose the degree of coronary artery disease (five categories) in symptomatic patients. When the true model was multinomial logistic, proportional odds models often yielded poor risk estimates, with calibration slopes deviating considerably from unity even on large model development datasets. The stereotype logistic model improved the calibration slope, but still provided biased risk estimates for individual patients. When the true model had a cumulative logit proportional odds form, multinomial logistic regression provided biased risk estimates, although these biases were modest. Non-proportional odds models require more parameters to be estimated from the data, and hence suffered more from overfitting. Despite larger sample size requirements, we generally recommend multinomial logistic regression for risk prediction modeling of discrete ordinal outcomes.

## 1. Introduction

Risk prediction modeling is ubiquitous in the medical literature. Most of these prediction models are developed for dichotomous outcomes, estimating the risk that a condition is present (diagnostic) or will develop within a certain time horizon (prognostic). However, several clinically important outcomes are ordinal in nature, with a finite and often limited number of ordered categories. One example is the extent of coronary artery disease in symptomatic patients, for which Edlinger et al recently developed a risk prediction model.[1] The diagnosis can be any of five increasingly severe conditions: no coronary artery disease, non-obstructive stenosis, one-vessel disease, two-vessel disease, or three-vessel disease. Another example is the modified Rankin scale to assess function recovery after stroke, as in a model by Risselada et al.[2] This scale has seven ordered categories: death, severe disability, moderately severe disability, moderate disability, slight disability, no significant disability despite symptoms, or no symptoms at all. Such outcomes are often dichotomized, although we would generally not recommend that for the following reasons: (1) it leads to a loss of information, (2) the merged categories may require different clinical management, and (3) merging categories may result in an extremely heterogeneous ‘supercategory’.

The default statistical model for ordinal outcomes is the cumulative logit model with proportional odds (CL-PO), which is commonly referred to as ‘ordinal logistic regression’. Several alternative logistic models for ordered categories exist, such as the adjacent category, continuation ratio, and stereotype models, which make different assumptions about the structure of ordinality.[3-10]

Alternatively, the multinomial logistic regression (MLR) can be used for modeling ordinal and other multi-category outcomes, ignoring the ordinality of the outcome.

For dichotomous outcomes, there is a large body of methodological literature and guidance on how prediction models should be constructed and how their performance should be evaluated in terms of discrimination and calibration.[11-17] Methods to assess discrimination and calibration have been extended to models for nominal outcomes.[18-20] For ordinal outcomes, discrimination measures have been proposed but calibration has been barely addressed.[21-24] Harrell and colleagues discussed the development of a risk prediction model for an ordinal outcome using CL-PO.[21] Calibration was assessed for a dichotomized version of the outcome, such that the standard methods for binary outcomes could be applied. More research on calibration is required, in particular because calibration is the Achilles heel of prediction modelling.[25]

In the present work, we study the performance of a variety of regression algorithms to develop prediction models for discrete ordinal outcomes. We (1) evaluate different approaches to investigate calibration, (2) study the impact of the proportional odds assumption on risk estimates and calibration statistics, and (3) explore the impact on overfitting when using simpler models that assume proportional odds versus more complex models without assuming proportional odds.

This paper is structured as follows. Regression models for discrete ordinal outcomes are described in section 2, and measures for predictive performance in section 3. Section 4 presents a simulation study to assess the impact of model choice on estimated risks, model calibration, and overfitting, and to compare approaches to quantify model calibration. Section 5 presents a case study, and in section 6 we discuss our findings.

## 2. Regression models for discrete ordinal outcomes

### 2.1. Regression models

We predict an outcome $Y$ with $K$ categories ($k = 1, \dots, K$) using $Q$ predictors $X_q$ ($q = 1, \dots, Q$), $\mathbf{X} = \left[X_1, \dots, X_Q\right]^T$. For simplicity, we will assume in notation that the models are modeling the predictors as linear and additive effects, but our work can easily be generalized to allow for alternative functional forms and interaction terms.

#### 2.1.1. *Multinomial logistic regression*

A generic model for categorical outcomes is multinomial logistic regression (MLR), which ignores the ordinality of the outcome. MLR models the outcome as follows:[10]

$$\log\left(\frac{P(Y=k)}{P(Y=1)}\right) = \alpha_{MLR,k} + \boldsymbol{\beta}_{MLR,k}^T \mathbf{X} = L_{MLR,k} \tag{1}$$

for $k = 2, \dots, K$ and where $\boldsymbol{\beta}_{MLR,k}^T = \left[\beta_{MLR,k,1}, \dots, \beta_{MLR,k,Q}\right]$ and where $L$ is called a linear predictor. One outcome category is used as the reference, and all other categories are contrasted with this reference category. We use $Y = 1$ as the reference, but the choice does not affect the estimated risks.

#### 2.1.2. *Cumulative logit models*

The likely most commonly used regression model for ordinal outcomes is the cumulative logit with proportional odds (CL-PO):[10]

$$\log\left(\frac{P(Y \geq k)}{P(Y < k)}\right) = \alpha_{CLPO,k} + \boldsymbol{\beta}_{CLPO}^{T}\mathbf{X} = L_{CLPO,k} \qquad (2)$$

for $k = 2, \ldots, K$ and where $\boldsymbol{\beta}_{CLPO}^{T} = \left[\beta_{CLPO,1}, \ldots, \beta_{CLPO,Q}\right]$. Due to the proportional odds assumption, every predictor effect is modeled using only one parameter, irrespective of $k$. This means that predictor effects are assumed constant over $k$ on the log-odds scale. The model has $K - 1$ intercepts.

The cumulative logit model can also be formulated without the proportional odds assumption, leading to the CL-NP model:[10]

$$\log\left(\frac{P(Y \geq k)}{P(Y < k)}\right) = \alpha_{CLNP,k} + \boldsymbol{\beta}_{CLNP,k}^{T}\mathbf{X} = L_{CLNP,k} \qquad (3)$$

for $k = 2, \ldots, K$ and where $\boldsymbol{\beta}_{CLNP,k}^{T} = \left[\beta_{CLNP,k,1}, \ldots, \beta_{CLNP,k,Q}\right]$. Here, the predictor effects depend on $k$, such that $K - 1$ parameters are estimated for each predictor. Note that CL-NP may lead to invalid models where the estimated risk that $Y \geq k$ is higher than the estimated risk that $Y \geq k - 1$.[8,10]

2.1.3. *Adjacent category models*

An alternative method to model ordinality is to target pairwise probabilities of adjacent categories, rather than cumulative probabilities. Assuming proportional odds, the adjacent category with proportional odds model (AC-PO) model is[10]

$$\log\left(\frac{P(Y = k+1)}{P(Y = k)}\right) = \alpha_{ACPO,k} + \boldsymbol{\beta}_{ACPO}^{T}\mathbf{X} = L_{ACPO,k} \qquad (4)$$

for $k = 1, \dots, K-1$ and where $\boldsymbol{\beta}_{ACPO}^{T} = \left[\beta_{ACPO,1}, \dots, \beta_{ACPO,Q}\right]$. Proportional odds in this setup refers to identical effects for moving up one category, instead of identical effects for every dichotomization of $Y$.

The adjacent model setup can also be applied without the proportional odds assumption, leading to the adjacent category without proportional odds model (AC-NP):

$$\log\left(\frac{P(Y=k+1)}{P(Y=k)}\right) = \alpha_{ACNP,k} + \boldsymbol{\beta}_{ACNP,k}^{T}\mathbf{X} = L_{ACNP,k} \tag{5}$$

for $k = 1, \dots, K-1$ and where $\boldsymbol{\beta}_{ACNP,k}^{T} = [\beta_{ACNP,k,1}, \cdots, \beta_{ACNP,k,Q}]$. This model is equivalent to MLR.

2.1.4. *Continuation ratio models*

Instead of cumulative or pairwise probabilities, conditional probabilities can be targeted. Continuation ratio models estimate the probability of a given outcome category conditional on the outcome being at least that category. The continuation ratio model with proportional odds assumptions (CR-PO) is[10]

$$\log\left(\frac{P(Y>k)}{P(Y\geq k)}\right) = \alpha_{CRPO,k} + \boldsymbol{\beta}_{CRPO}^{T}\mathbf{X} = L_{CRPO,k} \tag{6}$$

for $k = 1, \dots, K-1$ and where $\boldsymbol{\beta}_{CRPO}^{T} = \left[\beta_{CRPO,1}, \dots, \beta_{CRPO,Q}\right]$. Without proportional odds, the continuation ratio model is (CR-NP):

$$\log\left(\frac{P(Y>k)}{P(Y\geq k)}\right) = \alpha_{CRNP,k} + \boldsymbol{\beta}_{CRNP,k}^{T}\mathbf{X} = L_{CRNP,k} \tag{7}$$

for $k = 1, \dots, K-1$ and where $\boldsymbol{\beta}_{CRNP,k}^{T} = \left[\beta_{CRNP,k,1}, \dots, \beta_{CRNP,k,Q}\right]$.

2.1.5. *Stereotype logistic model*

Anderson introduced a model that finds a compromise between MLR and AC-PO, by relaxing the proportional odds assumption on the level of the $K - 1$ equations rather than on the level of each predictor separately.[7] The stereotype logistic model (SLM) is written as:

$$\log\left(\frac{P(Y=k)}{P(Y=1)}\right) = \alpha_{SLM,k} + \phi_k \boldsymbol{\beta}_{SLM}^T \mathbf{X} = L_{SLM,k} \qquad (8)$$

for $k = 2, \dots, K$ and where $\boldsymbol{\beta}_{SLM}^T = \left[\beta_{SLM,1}, \dots, \beta_{SLM,Q}\right]$ and $Y = 1$ is used as the reference. The model estimates one coefficient per predictor, but estimates $K - 1$ scaling factors $\phi$. Every predictor coefficient is multiplied by $\phi_k$. To avoid identifiability problems, a constraint has to be imposed on the scaling factors, which typically is that $\phi_2 = 1$. In principle, the model is an ordered model if the scaling factors are monotonically increasing or decreasing. While this could be imposed as an additional constraint during model fitting, it is not necessary and may cause computational problems.[3,7]

2.2. A comparison of the number of parameters

For any particular application, the number of parameters (regression model coefficients including intercepts) of the above defined models varies. The models without a proportional odds assumption (MLR, CL-NP, AC-NP, CR-NP) require $(Q + 1)(K - 1)$ parameters, models with proportional odds (CL-PO, AC-PO, CR-PO) require $Q + K - 1$ parameters. SLM falls in between with $Q + 2K - 3$ parameters. Table 1 presents the number of parameters for illustrative values of $Q$ and $K$.

## 3. Predictive performance measures for discrete ordinal outcomes models

The estimated risk of category $k$ is denoted by $\hat{P}_k$, with the estimated risk for individual $i$ in a data set of size $N$ $(i = 1, \ldots, N)$ denoted as $\hat{p}_{i,k}$. These risks are model-specific, conditional on $\mathbf{X}$ and the estimated model parameters. Hence $\hat{P}_k = P(Y = k|\mathbf{X}, \hat{\boldsymbol{\theta}}.)$, where $\hat{\boldsymbol{\theta}}.$ includes all parameters estimated from the model of choice (Equations 1-8). E.g. $\hat{\boldsymbol{\theta}}_{SLM}$ includes all $Q + 2K - 3$ estimated intercepts, model coefficients and scaling factors. The Appendix provides more details on how to calculate the risks for the different types of models. Analogously, the estimated risk that the outcome category has at least value $k$ is denoted as $\hat{V}_k = \sum_{j=k}^{K} \hat{P}_j$, with the estimated risk for individual $i$ denoted as $\hat{v}_{i,k}$.

### 3.1. Calibration of risk models for ordinal outcomes

#### *3.1.1. Calibration intercepts and slopes per outcome category or dichotomy*

A simple approach that capitalizes on the well-known calibration tools for binary outcomes, is to evaluate risk model calibration for every outcome category separately by defining a binary outcome $Y_k$ that equals 1 if $Y = k$ and 0 otherwise.[26] The calibration intercept and calibration slope can be computed by the following binary logistic calibration model:

$$log\left(\frac{P(Y_k=1)}{P(Y_k=0)}\right) = a_c + b_c \times logit(\hat{P}_k). \quad (9)$$

The calibration slope equals $b_c$, the calibration intercept equals $a_c$ when $b_c$ is fixed to 1. Alternatively, the outcome can be dichotomized as $Y_{\geq k}$ (1 if $Y \geq k$ and 0 otherwise), and a calibration model for the dichotomized outcome can be defined as:

$$log\left(\frac{P(Y_{\geq k}=1)}{P(Y_{\geq k}=0)}\right) = a_d + b_d \times logit(\hat{V}_k). \quad (10)$$

Calibration intercepts and slopes can be obtained as for $Y_k$.

Due to the ordinal nature of the outcome, $Y_{\geq k}$ may appear more sensible than $Y_k$, although this may depend on the actual clinical decisions that the model is intended to support.

*3.1.2. Model-specific calibration intercepts and calibration slopes*

When making a prediction model for a binary outcome using standard maximum likelihood logistic regression, the calibration intercept and calibration slope are by definition 0 and 1 when evaluated on the development dataset (i.e. the exact same dataset that was used to develop the prediction model).[26] A model with intercept of 0 and slope of 1 has been defined as 'weak calibration'.[26] Thus, maximum likelihood binary logistic regression for a binary outcome is by definition weakly calibrated on the development dataset. When making a prediction model for an ordinal outcome, and assessing calibration per outcome category ($Y_k$) or per outcome dichotomy ($Y_{\geq k}$) (Equations 9-10), calibration intercepts and slopes are no longer 0 and 1 on the development dataset. Procedures with intercept 0 and slope 1 on the development dataset are possible, but depend on the regression model used to develop the prediction model for the ordinal outcome. Such procedures are therefore not generic; therefore, we describe them for each ordinal regression model separately.

For MLR, the model-specific calibration model is of the following form:[18]

$$\log\left(\frac{P(Y=k)}{P(Y=1)}\right) = a_{MLR,k} + \sum_{j=2}^{K} b_{MLR,k,j}\hat{L}_{MLR,j} \quad (11)$$

for $k = 2, \ldots, K$ and $\hat{L}_{MLR,j}$ are the linear predictors from the fitted MLR prediction model (Equation 1). The calibration intercepts equal $a_{MLR,k}$, when fixing the corresponding calibration slope $b_{MLR,k,j=k}$ to 1 and the remaining slopes $b_{MLR,k,j\neq k}$ to 0. The calibration slopes equal $b_{MLR,k,j=k}$, when fixing the remaining slopes $b_{MLR,k,j\neq k}$ to 0. When this model is used to evaluate calibration of the MLR model on the development dataset, weak calibration holds: the calibration slopes are $b_{MLR,k,j=k}$ equal 1 and the calibration intercepts $a_{MLR,k}$ equal 0. See Van Hoorde and colleagues for further elaboration in the context of prediction models for nominal outcomes.[18]

For CL-PO, the $K - 1$ linear predictors are identical except for the intercepts (Equation 2). Hence for each linear predictor $\hat{L}_{CLPO,j}$, $j = 2, \ldots, K$, separate CL-PO calibration models are fit as follows:

$$\log\left(\frac{P(Y \geq k)}{P(Y < k)}\right) = a_{CLPO,k} + b_{CLPO,j}\hat{L}_{CLPO,j}, \text{ with } k = 2, \ldots, K. \quad (12)$$

The calibration slopes equal $b_{CLPO,j}$, and the calibration intercepts equal $a_{CLPO,k=j}$ when $b_{CLPO,j}$ is fixed to 1. Similarly, for fitted AC-PO and CR-PO prediction models (Equations 4 and 6), $K - 1$ separate AC-PO or CR-PO calibration models are fit for each linear predictor $\hat{L}_{\cdot,j}$, $j = 1, \ldots, K - 1$:

$$\text{AC-PO: } \log\left(\frac{P(Y = k+1)}{P(Y = k)}\right) = a_{ACPO,k} + b_{ACPO,j}\hat{L}_{ACPO,j}, \text{ with } k = 1, \ldots, K - 1 \quad (13)$$

$$\text{CR-PO: } \log\left(\frac{P(Y > k)}{P(Y \geq k)}\right) = a_{CRPO,k} + b_{CRPO,j}\hat{L}_{CRPO,j}, \text{ with } k = 1, \ldots, K - 1. \quad (14)$$

Calibration intercepts and slopes are calculated as for the CL-PO model (Equation 12). For fitted prediction models based on AC-NP, CR-NP, and SLM (Equations 5, 7, and 8), the setup is analogous to that for prediction models based on MLR. Calibration models are as follows:

$$\text{AC-NP: } \log\left(\frac{P(Y=k+1)}{P(Y=k)}\right) = a_{ACNP,k} + \sum_{j=1}^{K-1} b_{ACNP,k,j}\hat{L}_{ACNP,j}, \text{ with } k = 1,\ldots,K-1 \quad (15)$$

$$\text{CR-NP: } \log\left(\frac{P(Y>k)}{P(Y\geq k)}\right) = a_{CRNP,k} + \sum_{j=1}^{K-1} b_{CRNP,k,j}\hat{L}_{CRNP,j}, \text{ with } k = 1,\ldots,K-1 \quad (16)$$

$$\text{SLM: } \log\left(\frac{P(Y=k)}{P(Y=1)}\right) = a_{SLM,k} + \sum_{j=2}^{K} b_{SLM,k,j}\hat{L}_{SLM,j}, \text{ with } k = 2,\ldots,K. \quad (17)$$

Calibration intercepts and slopes are calculated as for the MLR calibration model (Equation 11). For every model, weak calibration holds on the development dataset: calibration intercepts are 0 and calibration slopes are 1.

3.1.3. *Flexible multinomial calibration plots*

To generate flexible calibration curves for risk models with multi-category outcomes based on any model, Van Hoorde and colleagues suggested a flexible recalibration model that is extended from the MLR recalibration model.[18] The model is as follows:

$$\log\left(\frac{P(Y=k)}{P(Y=1)}\right) = a_{flex,k} + \sum_{j=2}^{K} s_{k,j}\left(\hat{Z}_j\right) \quad (18)$$

where $k = 2,\ldots,K$, $\hat{Z}_j = \log\left(\hat{P}_j/\hat{P}_1\right)$ obtained from the fitted model, and $\mathbf{s_j} = \left[s_{1,j}\left(\hat{Z}_j\right) \;\; \ldots \;\; s_{K-1,j}\left(\hat{Z}_j\right)\right]^T$ a vector spline smoother.[27,28] The probabilities resulting from this flexible recalibration model are labeled the observed proportions $\hat{O}_k = P(Y = k|\hat{P}_1,\ldots,\hat{P}_K,\hat{\mathbf{a}}_{\boldsymbol{flex}},\hat{\mathbf{s}})$, with $k = 1,\ldots,K$, and where $\hat{\mathbf{a}}_{\boldsymbol{flex}}$ are the estimated values for $a_{flex,k}$ and $\hat{\mathbf{s}}$ are the fitted spline smoothers $\mathbf{s_j}$. For individual *i*, the observed proportions are denoted as $\hat{o}_{i,k}$. See Discussion and Supplementary Material for information about alternative flexible recalibration models.

For each outcome category *k*, a calibration plot can be constructed that relates the estimated model-based risks $\hat{P}_k$ (horizonal axis) to the observed proportion $\hat{O}_k$ (vertical axis). Contrary to binary outcomes, there is no one-to-one relationship between $\widehat{P}_k$ and $\widehat{O}_k$ for ordered or unordered multicategory outcomes.[18] Either the result can be plotted as a calibration scatter plot, or the scatter plot can be smoothed to present the results as calibration plots. Using $\widehat{O}_k$ and $\widehat{P}_k$, it is also possible to make calibration plots per outcome dichotomy, should that be of interest.

Flexible calibration curves for an outcome category may be obtained more simply by replacing $b_c$ in Equation 9 with a splines or loess fit, as described elsewhere for binary outcomes.[26] Because this approach ignores the multicategory nature of the outcome, it cannot be used to generate calibration scatter plots but may approximate smoothed calibration scatter plots based on Equation 18.

3.1.4. *Estimated calibration index*

Single-number summaries of calibration plots exist for binary outcomes, such as Harrell's E statistics.[11] The estimated calibration index (ECI) was introduced as a single-number summary of calibration for nominal outcomes, but can also be used for ordinal or binary outcomes.[29] The ECI is the average squared difference between $\widehat{p}_{i,k}$ and $\widehat{o}_{i,k}$, where the latter are based on a flexible recalibration model (Equation 18). Originally, ECI was defined as follows:

$$\text{ECI} = \frac{\sum_{i=1}^{N}\sum_{k=1}^{K}\left(\widehat{p}_{i,k}-\widehat{o}_{i,k}\right)^2}{NK} * \frac{100K}{2}. \tag{19}$$

The second part of the formula ensures that ECI is scaled between 0 and 100. Here, 0 indicates that $\widehat{p}_{i,k} = \widehat{o}_{i,k}$ for all *i* and *k* and 100 the theoretical worst-case scenario where for each case the estimated risk of one outcome category is 1 and the observed proportion of another outcome category is 1. This is an extreme scaling; in the current work we use a different one, where the maximal value of ECI refers to a model that has no predictive ability. In that case, all $\hat{o}_{i,k}$ equal the event rate of outcome category $k$ ($\bar{Y}_k$). If we set the maximal value to 1 instead of 100, this rescaled ECI is defined as follows:

$$\frac{\sum_{i=1}^{N}\sum_{k=1}^{K}\left(\hat{p}_{i,k}-\hat{o}_{i,k}\right)^2}{\sum_{i=1}^{N}\sum_{k=1}^{K}\left(\hat{p}_{i,k}-\bar{Y}_k\right)^2}. \tag{20}$$

3.2. *Discrimination*

To evaluate model discrimination, we used the ordinal C statistic (ORC).[24] Despite being designed for ordinal outcomes, the ORC equals the average C statistic for all pairs of outcome categories, and is interpreted as the probability to separate two cases from two randomly chosen outcome categories. As with the binary C statistic, ORC=0.5 implies no and ORC=1 perfect discriminative performance. To calculate pairwise C statistics, we have to express the prediction of the outcome through a single number. For proportional odds models, this can be based on $\hat{\boldsymbol{\beta}}_{.}^{\boldsymbol{T}}\mathbf{X}$. For any model, we can also use the expected value of the outcome prediction, $\sum_{k=1}^{K} k\hat{P}_k$. For all pairs of outcome categories, a pairwise C statistic is calculated as the standard binary C statistic for cases belonging to one of the two outcome categories using the single number prediction.

## 4. Monte Carlo simulation study

4.1. Methods

We use the Aims-Data-Estimands-Methods-Performance (ADEMP) structure to provide a structured overview of the simulation study.[30]

*Aims*. The aims are to (1) study the impact of the choice of model on estimated risks and model calibration, (2) study the impact of the model choice on model overfitting, and (3) evaluate different approaches to calculate calibration slopes for regression models predicting an ordinal outcome.

*Data generating mechanism*. We simulate data assuming a true model that has either MLR or CL-PO form. Under CL-PO proportional odds holds for cumulative logits only. For data under MLR form, we specified four main scenarios involving a model with $Q = 4$ continuous predictors $X_q$ and an outcome $Y$ with $K = 3$ categories. For simplicity, every predictor is independently normally distributed conditional on the outcome category, i.e. $X_{q,k} \sim \mathcal{N}\left(\mu_{q,k}, 1\right)$. The four scenarios vary by outcome prevalence (balanced or imbalanced) and whether the means of each predictor are equidistant between outcome categories or not, in a full factorial approach (Table 2, Table S1). Equidistant means imply that proportional odds hold for adjacent category logits, but not for cumulative logits. The true ORC for these scenarios is 0.74. For data under CL-PO form, we specified three main scenarios with 4 continuous predictors, 3 outcome categories ($Q = 4$, $K = 3$), and an ORC of 0.74 under the data generating model (Table 2, Table S2). The scenarios vary by outcome prevalence (balanced, imbalanced, highly imbalanced). We identified additional scenarios by varying factors non-factorially in an effort to maximize the effect on miscalibration and on differences between models. We investigated the effect of having $K = 4$ (and $Q = 3$), highly non-equidistant means (only for data under MLR form), highly imbalanced outcome distribution, and low discrimination (ORC=0.66). Finally, we added scenarios with only binary predictors and scenarios in which noise predictors are included.

*Estimands/targets of analysis*. The focus in this simulation is on large sample and out-of-sample calibration performance, but we also assess discrimination and prediction error.

*Methods*. We focus on the MLR, CL-PO, AC-PO, and SLM models to limit the amount of results (Equations 1, 2, 4, and 8). To approach true model coefficients and performance, a large dataset with 200,000 observations was simulated for each scenario. Models were fitted (developed) and performance evaluated (validated) on this single large dataset. Next, to assess the impact of overfitting, we simulated 200 new datasets of size 100 and 500 for all main scenarios, developed the models on each dataset, and evaluated performance on the large dataset with size 200,000. We report the mean value for each performance measure. The chosen sample sizes are partly arbitrary, but see Supplementary Material for further explanation.

*Performance measures*. We report the calibration intercepts and slopes by outcome category, by outcome dichotomy, and by linear predictor (i.e. algorithm-specific). Further, we report the root mean squared prediction error (rMSPE) and ORC. For rMSPE we use the true risks $p_{i,k}$ in each scenario (which are known under the data generating model):

$$\frac{1}{NK}\sum_{i=1}^{N}\sum_{k=1}^{K}\left(\hat{p}_{i,k}-p_{i,k}\right)^2. \tag{21}$$

The ECI was only reported for the large sample evaluation, not for evaluating overfitting. The statistical analyses were performed using the R statistical software, version 4.0.1. The package for fitting the logistic regression models was VGAM, using functions vglm and rrvglm for SLM.[31] The complete R code is available on GitHub (https://github.com/benvancalster/OrdinalCalibration).

## 4.2. Results

### *4.1.1. MLR truth – main scenarios*

In the large sample simulations, when true predictor means were equidistant (scenarios 1-2), risk estimates corresponded almost perfectly with true risk for the MLR, AC-PO, and SLM models (Figures 1-2). When the true predictor means were non-equidistant (scenarios 3-4), only MLR obtained risk estimates that corresponded closely to the true risks (Figures 3-4). Regarding calibration intercepts and slopes (Table 3), we observed that these were near perfect for MLR. For the SLM model, the scaling factors also resulted in near perfect calibration intercepts and slopes even though estimated risks deviated from the true risks for scenarios 3 and 4. For AC-PO, calibration slopes per outcome category and per outcome dichotomy were off for scenarios 3-4. Scenarios 1-2 did not pose problems for AC-PO, because the equidistant means imply that proportional odds hold in terms of adjacent categories. For CL-PO, calibration intercepts were fine but calibration slopes per outcome category or per outcome dichotomy were off for all scenarios. Interestingly, the model-specific calibration intercepts and slopes were 0 and 1, respectively, for all models and scenarios. Hence, the miscalibration problems for CL-PO and AC-PO were not reflected in these measures. The reason is that these calculations are model-specific, and thus that they quantify calibration under the assumption that proportional odds hold (for cumulative logits in case of CL-PO, or for adjacent category logits in case of AC-PO). Flexible calibration curves are presented in Figures S1-4.

The ECI and rMSPE results were lowest (i.e. best) for the MLR model throughout, and substantially higher for CL-PO and AC-PO under non-equidistant means, and slightly increased for CL-PO even under equidistant means (Table 3). For SLM, ECI was low throughout, but rMSPE was increased under non-equidistant means. The discrimination differed only slightly, with ORC providing slightly higher values for MLR in scenarios 3 and 4. For completeness, the large-sample estimated model coefficients are given in Table S3.

The results of the small sample simulations were in line with expectations (Tables 4-5). When comparing the large sample performance to the average validation performance of models developed on small samples (N=100), the MLR models had the strongest decrease in performance, CL-PO and AC-PO the least. MLR models, which have the highest number of parameters, even had worse validation performance than the three other types of models. The effects of overfitting were smaller when development datasets had a sample size of 500.

*4.1.2. MLR truth – additional scenarios*

In the additional scenarios, the above findings show a similar pattern (Table S4, Figures S5-S18). MLR continued to provide near perfect risk estimates, but risk estimates for other models were clearly distorted. For CL-PO and AC-PO, it was not difficult to find scenarios where calibration slopes for intermediate outcome categories ($Y = 2$ if $K = 3$, or $Y \in \{2,3\}$ if $K = 4$) are highly problematic. In scenario 8, the calibration slope for $Y = 2$ was even negative for CL-PO and AC-PO. In scenarios 6-8, with $Q = 3$ and $Y = 4$, one can clearly see how SLM's scaling factors helped to ascertain good calibration intercepts, slopes, and plots, despite distorted individual risk estimates. Having binary predictors or a number of noise predictors did not change the findings.

*4.1.2. CL-PO truth – main and additional scenarios*

We present results for the main scenarios in the main text (Figures 5-7), and for all other scenarios in the Supplementary material (Figures S19-S33). In the large sample situations, risk estimates corresponded almost perfectly with true risks for CL-PO (Figures 5-7 and S19-S24). Other models had distorted risk estimates, but the distortion was generally modest. Calibration intercepts and slopes were near perfect for CL-PO, but not for the other models (Tables 3 and S5). For MLR, calibration slopes of around 1.3-1.4 were observed for category 2 in several simulation settings (scenarios 1, 4, 5, 8, 9), but the scatter plots of estimated versus true risk as well as the calibration plots (Figures S25-S33) show that estimated risks were less strongly biased than when fitting CL-PO models under MLR truth. ECI and rMSPE were best for CL-PO and worst for AC-PO (Tables 3 and S6). ECI and rMSPE results for MLR, AC-PO and SLM were better than what was obtained when CL-PO models were fitted under MLR truth. Results for small sample simulations were similar to those under MLR truth: again, MLR had the strongest decrease in performance and CL-PO and AC-PO the least (Tables 4-5). Model coefficients for all large sample models are given in Table S6.

## 5. Case study: prediction of coronary artery disease

### 5.1. Methods

The Coronary Artery disease Risk Determination In Innsbruck by diaGnostic ANgiography (CARDIIGAN) cohort includes patients with suspected coronary artery disease that were recruited between 2004 and 2008 at the University Clinic of Cardiology in Innsbruck (Austria).[32] A prediction model based on the CL-PO model was developed with the CARDIIGAN data, concerning the diagnosis of non-obstructive coronary artery and multi-vessel disease in five ordinal disease categories: no coronary artery disease, non-obstructive stenosis, one-vessel disease, two-vessel disease, and three-vessel disease.[1] This outcome has clinical relevance because different categories require different treatment decisions.[32] The patient group involved 4,888 individuals, presenting with symptoms at the hospital, who had not had a known previous coronary artery or other heart disease and without coronary revascularization in the past. For earlier studies, the missing values had already been multiply imputed;[33] in the current illustration we used one of the imputed data sets for convenience.

We applied the following algorithms: MLR (identical to AC-NP), CL-PO, AC-PO, CR-PO, CR-NP, and SLM. The proportional odds assumption in the CL-PO framework was tested per variable using a likelihood ratio test. We used the enhanced bootstrap with 200 bootstrap samples to internally validate the models.[11] We used eleven predictors covering demographic information, symptoms, comorbidities and biomarkers (Table S7). This means that 15 coefficients (including intercepts) have to be estimated for the proportional odds models, 18 for SLM, and 48 for non-proportional odds models. If we focus on the smallest outcome category (three-vessel disease, n=429), this implies an EPP (events per parameter excluding intercepts) of 39 for proportional odds models, 31 for SLM, and 10 for non-proportional odds models. See Supplementary Material for example R code to fit models and evaluate performance. The complete R code is available on GitHub (https://github.com/benvancalster/OrdinalCalibration).

## 5.2. Results

The likelihood ratio tests suggested violations of the proportional odds assumption for the CL-PO model, mainly for age and hypertension (Table S8). All models had an optimism-corrected ORC of 0.693-0.694 (Table 6, see Table S9 for model coefficients). The risk estimates varied strongly between methods, and this was most obvious for the outcome category 'non-obstructive stenosis' (Figures 8 and S34-S37). For proportional odds models, risk estimates for intermediate outcome categories were capped at some point (see also Table S10). Apparent calibration curves (per outcome category as well as per outcome dichotomy) deviated most from the ideal diagonal line for AC-PO and CR-PO, to a lesser extent for CL-PO, and least for MLR, CR-NP, and SLM (Figures 9-10 show calibration scatter plots per outcome category and outcome dichotomy, Figures S38-43 also provide flexible calibration plots). The bootstrap-corrected slopes per outcome category or per outcome dichotomy deviated from the target value of 1 most strongly for CR-PO and AC-PO, and least strongly for MLR, CR-NP, and SLM (Table 6). The model-specific calibration slopes largely reflect overfitting, because for proportional odds models this assessment assumes the model's proportional odds assumption holds. Hence, model-specific calibration slopes were closer to 1 for the proportional odds models and SLM, which required fewer parameters than MLR and CR-NP.

# 6. Discussion

In this study we focused on calibration of risk prediction models for discrete ordinal outcomes, and on the impact of assuming proportional odds on risk estimation and calibration performance. The results show that assuming proportional odds leads to (sometimes strongly) distorted risk estimates, calibration slopes, and calibration plots when the true model had MLR form and hence the proportional odds assumption was violated. Naturally, MLR models yielded appropriate risk estimates in these settings. In contrast, when the true model had the CL-PO form, the MLR model had distorted risk estimates and calibration. The deviations for MLR under CL-PO truth were less dramatic than the deviations of the CL-PO model when the true model had the MLR form.

Perhaps surprisingly, when the true model had the CL-PO form, other proportional odds models such as AC-PO also had deviating risk estimates. This highlights the importance of the specific form of proportional odds that is assumed, which varies between the cumulative, adjacent category and continuation ratio logit models. The SLM model, which can be seen as a compromise between MLR and AC-PO models, also showed distorted risk estimates when the true model had the MLR form. Due to its scaling factors, SLM did yield appropriate calibration intercepts and calibration slopes. When the true model had the CL-PO form, however, SLM did not improve upon MLR. Our small sample size simulations showed that in smaller samples, the models that do not assume proportional odds suffer from more overfitting, due to the higher number of parameters that need to be estimated when proportional odds are not assumed.

For binary outcomes modeled with maximum likelihood logistic regression, the calibration intercept and slope are by definition 0 and 1 on the data on which the model is developed. This is a well-known property of calibration intercepts and slopes, which was previously extended to models for nominal outcomes based on multinomial logistic regression.[18] In this paper we further generalized in this work to models for ordinal outcomes under the label 'model-specific' calibration assessment. For proportional odds models, this approach assesses calibration under the assumption that proportional odds hold. Violations of the assumption are therefore not considered, which makes this approach inappropriate for quantifying calibration of ordinal prediction models. For other models, this approach performs satisfactorily, but a general drawback is that it is less intuitive than simple calibration assessment for each outcome category or dichotomy.

Based on our findings, we generally recommend non-proportional odds models such as MLR for developing risk prediction models for an ordinal outcome. We are inclined to believe that proportional odds assumptions will often not hold in the practice of medical risk prediction. But even when it does, we argue that the loss in efficiency and increased risk of overfitting associated with using MLR is less problematic than the opposite problem, i.e. the risk of severe miscalibration when using proportional odds models (even under moderate deviation from the proportional odds assumption). However, MLR has more parameters and hence needs a larger sample size in order to obtain a reliable risk prediction model.[34] Sample size determination methods for prediction models based on MLR are currently underway. This will help to plan model development studies for ordinal outcomes by calculating the minimum sample size needed to use MLR. If this minimum sample size is too high given the resources for a given project, it can be discussed whether a proportional odds model would be defendable or whether no model should be developed until more resources become available. A compromise to assuming strict proportional odds may be the SLM model, which uses less parameters than MLR. This model can help to improve calibration slopes and flexible calibration curves, although risks on the individual level may still be distorted.

To assess calibration, we recommend to calculate the calibration intercepts and slopes per outcome category or per outcome dichotomy. Whether to focus on outcome categories or dichotomies depends on the specific (clinical) context, i.e. on how risk estimates are used in clinical practice to decide upon patient management. If each outcome category is associated with a different management option, calibration per outcome category is preferred. When the management decision is binary, and depends on whether $P(Y \geq k)$ exceeds a given threshold, calibration per dichotomy may be preferred. For internal validation, these estimates can be based on bootstrapping.[11] When externally validating a model, flexible calibration plots (scatter plots as well as flexible calibration curves) are recommended because they provide a more general overview of calibration:[29] whereas calibration intercepts and slopes assess weak calibration, calibration curves assess moderate calibration.[26] Again, calibration plots can be constructed per outcome category or dichotomy. We based the flexible calibration plots on a flexible recalibration model with an MLR-like setup (Equation 18). In Supplementary Material, we compared this approach to other approaches to assess whether non-MLR prediction models are disadvantaged by this setup. Differences between the approaches were small. One may prefer to replace the $\widehat{Z}_j$ in Equation 18 with $\text{logit}(\hat{V}_j)$ to acknowledge the ordinal nature of the outcome (i.e. the third approach in the Supplementary Material). Finally, when different models are compared at external validation, the ECI is an attractive single summary measure. Of course, summarizing performance into a single number always has limitations.

We did not address partial proportional odds models, in which the proportional odds assumption can be relaxed for some but not all predictors.[4,35] This usually requires the use of a test for proportional odds per variable, e.g. likelihood ratio tests or the Brant test.[36] However, by evaluating the proportional odds assumption one considers the same number of parameters as in non-proportional odds models. Future studies could look into the power of these tests to detect deviations from proportional odds assumptions that would result in important miscalibration and distorted predictions. Further, the use of CL-PO has been advocated in settings outside of prediction models. For instance, they can be used to model continuous outcomes, in particular when these outcomes have skewed or semi-continuous distributions and in randomized controlled trials to improve statistical efficiency.[37,38] While our focus is in risk prediction modeling and hence our results do not directly generalize to these settings, our finding that the type of proportional odds assumption matters (e.g. on the level of cumulative logits versus adjacent category logits) seems to warrant further investigation.

To conclude, when the proportional odds assumptions do not strictly hold, as we believe is often the case in practical application of risk prediction models, the use of proportional odds models to develop prediction models for discrete ordinal outcomes can result in poor risk estimates and poor calibration. For the development of risk prediction models, we therefore warn readers against using proportional odds models without careful argumentation, and to consider multinomial logistic regression to model ordered categorical outcomes.

**Funding**

ME and BVC were supported by Research Foundation – Flanders (FWO) grant G0B4716N. BVC was supported by Internal Funds KU Leuven grant C24M/20/064. The funding bodies had no role in the design of the study, data collection, statistical analysis, interpretation of data, or in writing of the manuscript.

**Data availability**

For the CAD data, collaboration is welcomed and data sharing can be agreed upon by contacting Michael Edlinger (michael.edlinger@i-med.ac.at).

**ORCID**

Michael Edlinger: https://orcid.org/0000-0001-8801-3268

Maarten van Smeden: https://orcid.org/0000-0002-5529-1541

Hannes F Alber: https://orcid.org/0000-0002-5842-1591

Maria Wanitschek: Not available

Ben Van Calster: https://orcid.org/0000-0003-1613-7450

**Appendix: calculating estimated probabilities for each type of model**

MLR: $P(Y = k) = \frac{exp(L_{MLR,k})}{1+\sum_{j=2}^{K} exp(L_{MLR,j})}$, with $L_{MLR,1}$ set to 0.

CL-PO: $P(Y \geq k) = \frac{exp(L_{CLPO,k})}{1+exp(L_{CLPO,k})}$, and $P(Y = k) = P(Y \geq k) - P(Y \geq k + 1)$. Note that $P(Y \geq K + 1) = 0$.

CL-NP: analogous as for CL-PO.

AC-PO: $P(Y = k) = \frac{exp(\sum_{j=1}^{k} L_{ACPO,j})}{1+\sum_{r=1}^{K-1} exp(\sum_{s=1}^{r} L_{ACPO,s})}$, with $L_{ACPO,K}$ set to 0.

AC-NP: analogous as for AC-PO.

CR-PO: for $k = 1,\ldots,K - 1$, $P(Y = k) = \frac{exp(L_{CRPO,k})}{1+exp(L_{CRPO,k})} \times (1 - P(Y < k))$. Note that $P(Y < 1) = 0$. Finally, $P(Y = K) = 1 - \sum_{k=1}^{K-1} P(Y = k)$.

CR-NP: analogous as for CR-PO.

SLM: $P(Y = k) = \frac{exp(L_{SLM,k})}{1+\sum_{j=2}^{j=K} exp(L_{SLM,j})}$, with $L_{SLM,1} = 0$.

Table 1. Number of parameters to be estimated for some values of $K$ and $Q$.

| $K$ | $Q$ | Number of parameters | | |
|---|---|---|---|---|
| | | **Proportional odds models** | **Stereotype logistic model** | **Non-proportional odds models** |
| 3 | 3 | 5 | 6 | 8 |
| 3 | 5 | 7 | 8 | 12 |
| 3 | 10 | 12 | 13 | 22 |
| 5 | 3 | 7 | 10 | 16 |
| 5 | 5 | 9 | 12 | 24 |
| 5 | 10 | 14 | 17 | 44 |
| 10 | 3 | 12 | 20 | 36 |
| 10 | 5 | 14 | 22 | 54 |
| 10 | 10 | 19 | 27 | 99 |

$K$, number of outcome categories; $Q$, number of predictors.

Table 2. Overview of simulation scenarios.

| **Scenario** | ***Q*** | ***K*** | **ORC** | **Outcome distribution** | **Means of $X_{p,k}$*** |
|---|---|---|---|---|---|
| | | | | | |
| TRUE MODEL HAS MLR FORM | | | | | |
| Basic | | | | | |
| 1 | 4 continuous | 3 | 0.74 | Balanced | Equidistant |
| 2 | 4 continuous | 3 | 0.74 | Imbalanced | Equidistant |
| 3 | 4 continuous | 3 | 0.74 | Balanced | Non-equidistant |
| 4 | 4 continuous | 3 | 0.74 | Imbalanced | Non-equidistant |
| Additional | | | | | |
| 5 | 4 continuous | 3 | 0.74 | Imbalanced | Highly non-equidistant |
| 6 | 3 continuous | 4 | 0.74 | Imbalanced | Highly non-equidistant |
| 7 | 3 continuous | 4 | 0.66 | Imbalanced | Highly non-equidistant |
| 8 | 3 continuous | 4 | 0.66 | Highly imbalanced | Highly non-equidistant |
| 9 | 4 binary | 3 | 0.74 | Imbalanced | Non-equidistant |
| 10 | 3 binary | 4 | 0.74 | Imbalanced | Highly non-equidistant |
| 11 | 8 continuous (4 true + 4 noise) | 3 | 0.74 | Imbalanced | Non-equidistant |
| | | | | | |
| TRUE MODEL HAS CL-PO FORM | | | | | |
| Basic | | | | | |
| 1 | 4 continuous | 3 | 0.74 | Balanced | NA |
| 2 | 4 continuous | 3 | 0.74 | Imbalanced | NA |
| 3 | 4 continuous | 3 | 0.74 | Highly imbalanced | NA |
| Additional | | | | | |
| 4 | 3 continuous | 4 | 0.74 | Imbalanced | NA |
| 5 | 3 continuous | 4 | 0.66 | Imbalanced | NA |
| 6 | 3 continuous | 4 | 0.66 | Highly imbalanced | NA |
| 7 | 4 binary | 3 | 0.74 | Imbalanced | NA |
| 8 | 3 binary | 4 | 0.74 | Imbalanced | NA |
| 9 | 8 continuous (4 true + 4 noise) | 3 | 0.74 | Balanced | NA |

ORC, ordinal C statistic; MLR, multinomial logistic regression; CL-PO, cumulative logit model with proportional odds.

* For binary predictors, equidistance does not refer to means per outcome category, but to logit(prevalence) per outcome category.

Table 3. Apparent performance based on a large dataset of n=200,000 for the main simulation scenarios.

| | **CALIBRATION INTERCEPTS AND SLOPES** | | | | | | | **SINGLE NUMBER METRICS** | | |
|---|---|---|---|---|---|---|---|---|---|---|
| | **Per outcome category** | | | **Per outcome dichotomy** | | **Model-specific** | | | | |
| **Model** | $Y = 1$ | $Y = 2$ | $Y = 3$ | $Y > 1$ | $Y > 2$ | **LP1** | **LP2** | **ECI** | **rMSPE** | **ORC** |
| ***MLR truth scenario 1: balanced outcome, equidistant means*** | | | | | | | | | | |
| MLR | 0.00 / 1.00 | 0.00 / 0.99 | 0.00 / 1.00 | 0.00 / 1.00 | 0.00 / 1.00 | 0.00 / 1.00 | 0.00 / 1.00 | 0.000 | 0.002 | 0.741 |
| CL-PO | 0.00 / 1.02 | -0.01 / 0.75 | 0.00 / 1.02 | 0.00 / 1.02 | 0.00 / 1.02 | 0.00 / 1.00 | 0.00 / 1.00 | 0.006 | 0.012 | 0.741 |
| AC-PO | 0.00 / 1.00 | 0.00 / 0.99 | 0.00 / 1.00 | 0.00 / 1.00 | 0.00 / 1.00 | 0.00 / 1.00 | 0.00 / 1.00 | 0.000 | 0.001 | 0.741 |
| SLM | 0.00 / 1.00 | 0.00 / 0.99 | 0.00 / 1.00 | 0.00 / 1.00 | 0.00 / 1.00 | 0.00 / 1.00 | 0.00 / 1.00 | 0.000 | 0.001 | 0.741 |
| ***MLR truth scenario 2: imbalanced outcome, equidistant means*** | | | | | | | | | | |
| MLR | 0.00 / 1.00 | 0.00 / 1.00 | 0.00 / 1.00 | 0.00 / 1.00 | 0.00 / 1.00 | 0.00 / 1.00 | 0.00 / 1.00 | 0.000 | 0.002 | 0.740 |
| CL-PO | 0.01 / 0.96 | -0.01 / 0.79 | -0.01 / 1.14 | -0.01 / 0.96 | -0.01 / 1.14 | 0.00 / 1.00 | 0.00 / 1.00 | 0.010 | 0.016 | 0.740 |
| AC-PO | 0.00 / 1.00 | 0.00 / 1.01 | 0.00 / 0.99 | 0.00 / 1.00 | 0.00 / 0.99 | 0.00 / 1.00 | 0.00 / 1.00 | 0.000 | 0.002 | 0.740 |
| SLM | 0.00 / 1.00 | 0.00 / 1.00 | 0.00 / 1.00 | 0.00 / 1.00 | 0.00 / 1.00 | 0.00 / 1.00 | 0.00 / 1.00 | 0.000 | 0.002 | 0.740 |
| ***MLR truth scenario 3: balanced outcome, non-equidistant means*** | | | | | | | | | | |
| MLR | 0.00 / 1.00 | 0.00 / 1.00 | 0.00 / 1.00 | 0.00 / 1.00 | 0.00 / 1.00 | 0.00 / 1.00 | 0.00 / 1.00 | 0.000 | 0.002 | 0.741 |
| CL-PO | -0.03 / 1.21 | -0.01 / 0.75 | 0.03 / 0.86 | 0.03 / 1.21 | 0.03 / 0.86 | 0.00 / 1.00 | 0.00 / 1.00 | 0.049 | 0.075 | 0.738 |
| AC-PO | 0.00 / 1.19 | 0.00 / 0.95 | 0.00 / 0.84 | 0.00 / 1.19 | 0.00 / 0.84 | 0.00 / 1.00 | 0.00 / 1.00 | 0.046 | 0.074 | 0.738 |
| SLM | 0.00 / 1.00 | 0.00 / 1.00 | 0.00 / 1.00 | 0.00 / 1.00 | 0.00 / 1.00 | 0.00 / 1.00 | 0.00 / 1.00 | 0.000 | 0.063 | 0.738 |
| ***MLR truth scenario 4: imbalanced outcome, non-equidistant means*** | | | | | | | | | | |
| MLR | 0.00 / 1.00 | 0.00 / 1.00 | 0.00 / 1.00 | 0.00 / 1.00 | 0.00 / 1.00 | 0.00 / 1.00 | 0.00 / 1.00 | 0.000 | 0.002 | 0.737 |
| CL-PO | -0.02 / 1.11 | 0.01 / 1.13 | 0.03 / 0.85 | 0.02 / 1.11 | 0.03 / 0.85 | 0.00 / 1.00 | 0.00 / 1.00 | 0.032 | 0.058 | 0.735 |
| AC-PO | 0.00 / 1.17 | 0.00 / 1.47 | 0.00 / 0.76 | 0.00 / 1.17 | 0.00 / 0.76 | 0.00 / 1.00 | 0.00 / 1.00 | 0.059 | 0.064 | 0.736 |
| SLM | 0.00 / 1.00 | 0.00 / 1.00 | 0.00 / 1.00 | 0.00 / 1.00 | 0.00 / 1.00 | 0.00 / 1.00 | 0.00 / 1.00 | 0.000 | 0.047 | 0.733 |
| ***CL-PO truth scenario 1: balanced outcome*** | | | | | | | | | | |
| MLR | 0.00 / 0.99 | 0.00 / 1.38 | 0.00 / 0.99 | 0.00 / 0.99 | 0.00 / 0.99 | 0.00 / 1.00 | 0.00 / 1.00 | 0.006 | 0.014 | 0.740 |
| CL-PO | 0.00 / 1.00 | 0.00 / 1.01 | 0.00 / 0.99 | 0.00 / 1.00 | 0.00 / 0.99 | 0.00 / 1.00 | 0.00 / 1.00 | 0.000 | 0.003 | 0.740 |
| AC-PO | 0.00 / 1.00 | 0.00 / 1.38 | 0.00 / 0.99 | 0.00 / 1.00 | 0.00 / 0.99 | 0.00 / 1.00 | 0.00 / 1.00 | 0.006 | 0.014 | 0.740 |
| SLM | 0.00 / 0.99 | 0.00 / 1.38 | 0.00 / 0.99 | 0.00 / 0.99 | 0.00 / 0.99 | 0.00 / 1.00 | 0.00 / 1.00 | 0.006 | 0.014 | 0.740 |
| ***CL-PO truth scenario 2: imbalanced outcome*** | | | | | | | | | | |
| MLR | 0.00 / 1.00 | 0.00 / 1.09 | 0.00 / 0.99 | 0.00 / 1.00 | 0.00 / 0.99 | 0.00 / 1.00 | 0.00 / 1.00 | 0.005 | 0.013 | 0.740 |
| CL-PO | 0.00 / 1.00 | 0.00 / 1.01 | 0.00 / 0.99 | 0.00 / 1.00 | 0.00 / 0.99 | 0.00 / 1.00 | 0.00 / 1.00 | 0.000 | 0.003 | 0.740 |
| AC-PO | 0.00 / 1.07 | 0.00 / 1.34 | 0.00 / 0.88 | 0.00 / 1.07 | 0.00 / 0.88 | 0.00 / 1.00 | 0.00 / 1.00 | 0.012 | 0.018 | 0.740 |
| SLM | 0.00 / 1.00 | 0.00 / 1.09 | 0.00 / 0.99 | 0.00 / 1.00 | 0.00 / 0.99 | 0.00 / 1.00 | 0.00 / 1.00 | 0.006 | 0.013 | 0.740 |
| ***CL-PO truth scenario 3: highly imbalanced outcome*** | | | | | | | | | | |
| MLR | 0.00 / 1.00 | 0.00 / 1.02 | 0.00 / 0.98 | 0.00 / 1.00 | 0.00 / 0.98 | 0.00 / 1.00 | 0.00 / 1.00 | 0.004 | 0.009 | 0.742 |
| CL-PO | 0.00 / 1.00 | 0.00 / 1.00 | 0.00 / 1.00 | 0.00 / 1.00 | 0.00 / 1.00 | 0.00 / 1.00 | 0.00 / 1.00 | 0.000 | 0.002 | 0.742 |
| AC-PO | 0.00 / 1.08 | 0.00 / 1.22 | 0.00 / 0.77 | 0.00 / 1.08 | 0.00 / 0.77 | 0.00 / 1.00 | 0.00 / 1.00 | 0.015 | 0.017 | 0.742 |
| SLM | 0.00 / 1.00 | 0.00 / 1.02 | 0.00 / 0.98 | 0.00 / 1.00 | 0.00 / 0.98 | 0.00 / 1.00 | 0.00 / 1.00 | 0.004 | 0.009 | 0.742 |

MLR, multinomial logistic regression; CL-PO, cumulative logit model with proportional odds; AC-PO, adjacent category logit model with proportional odds; SLM, stereotype logit model; LP, linear predictor; ECI, estimated calibration index; rMSPE, root mean squared prediction error; ORC, ordinal C statistic; CAD, coronary artery disease.

Table 4. Validation performance based on small development datasets of n=100 for the main simulation scenarios (reported as the average performance on a large validation dataset for 200 simulated development datasets).

| | CALIBRATION INTERCEPTS AND SLOPES | | | | | | | SINGLE NUMBER METRICS | |
|---|---|---|---|---|---|---|---|---|---|
| | Per outcome category | | | Per outcome dichotomy | | Model-specific | | | |
| Model | $Y = 1$ | $Y = 2$ | $Y = 3$ | $Y > 1$ | $Y > 2$ | LP1 | LP2 | rMSPE | ORC |
| ***MLR truth scenario 1: balanced outcome, equidistant means*** | | | | | | | | | |
| MLR | 0.00 / 0.78 | 0.03 / 0.31 | -0.02 / 0.80 | 0.00 / 0.78 | -0.02 / 0.80 | 0.02 / 0.73 | -0.03 / 0.76 | 0.104 | 0.727 |
| CL-PO | 0.00 / 0.86 | 0.03 / 0.55 | -0.02 / 0.86 | 0.00 / 0.86 | -0.02 / 0.86 | 0.02 / 0.84 | 0.00 / 0.84 | 0.080 | 0.728 |
| AC-PO | 0.00 / 0.84 | 0.03 / 0.73 | -0.02 / 0.85 | 0.00 / 0.84 | -0.02 / 0.85 | 0.02 / 0.83 | -0.04 / 0.83 | 0.077 | 0.728 |
| SLM | 0.00 / 0.85 | 0.03 / 0.56 | -0.02 / 0.85 | 0.00 / 0.85 | -0.02 / 0.85 | 0.02 / 1.04 | -0.02 / 0.82 | 0.085 | 0.728 |
| ***MLR truth scenario 2: imbalanced outcome, equidistant means*** | | | | | | | | | |
| MLR | -0.01 / 0.80 | 0.03 / 0.45 | 0.01 / 0.73 | 0.01 / 0.80 | 0.01 / 0.73 | 0.02 / 0.82 | -0.01 / 0.58 | 0.104 | 0.724 |
| CL-PO | 0.00 / 0.80 | 0.02 / 0.64 | 0.00 / 0.96 | 0.00 / 0.80 | 0.00 / 0.96 | 0.00 / 0.84 | -0.04 / 0.84 | 0.079 | 0.726 |
| AC-PO | -0.01 / 0.84 | 0.03 / 0.82 | 0.01 / 0.83 | 0.01 / 0.84 | 0.01 / 0.83 | 0.02 / 0.83 | -0.01 / 0.83 | 0.077 | 0.725 |
| SLM | -0.01 / 0.84 | 0.02 / 0.70 | 0.01 / 0.88 | 0.01 / 0.84 | 0.01 / 0.88 | 0.02 / 0.92 | 0.02 / 0.82 | 0.085 | 0.725 |
| ***MLR truth scenario 3: balanced outcome, non-equidistant means*** | | | | | | | | | |
| MLR | 0.03 / 0.85 | -0.03 / 0.56 | 0.02 / 0.79 | -0.03 / 0.85 | 0.02 / 0.79 | -0.04 / 0.83 | 0.03 / 0.67 | 0.105 | 0.725 |
| CL-PO | -0.01 / 1.07 | -0.03 / 0.60 | 0.06 / 0.76 | 0.01 / 1.07 | 0.06 / 0.76 | 0.04 / 0.89 | -0.04 / 0.89 | 0.110 | 0.725 |
| AC-PO | 0.02 / 1.05 | -0.02 / 0.76 | 0.03 / 0.75 | -0.02 / 1.05 | 0.03 / 0.75 | -0.03 / 0.87 | 0.03 / 0.87 | 0.108 | 0.725 |
| SLM | 0.03 / 0.86 | -0.03 / 0.65 | 0.02 / 0.90 | -0.03 / 0.86 | 0.02 / 0.90 | -0.03 / 0.89 | 0.00 / 0.86 | 0.109 | 0.722 |
| ***MLR truth scenario 4: imbalanced outcome, non-equidistant means*** | | | | | | | | | |
| MLR | 0.02 / 0.83 | 0.01 / 0.69 | 0.00 / 0.70 | -0.02 / 0.83 | 0.00 / 0.70 | 0.00 / 0.84 | 0.00 / 0.54 | 0.101 | 0.724 |
| CL-PO | -0.02 / 0.95 | 0.03 / 0.94 | 0.02 / 0.73 | 0.02 / 0.95 | 0.02 / 0.73 | 0.02 / 0.86 | -0.02 / 0.86 | 0.095 | 0.724 |
| AC-PO | 0.00 / 0.99 | 0.02 / 1.20 | -0.01 / 0.65 | 0.00 / 0.99 | -0.01 / 0.65 | 0.01 / 0.84 | -0.02 / 0.84 | 0.097 | 0.724 |
| SLM | 0.01 / 0.85 | 0.02 / 0.83 | -0.01 / 0.86 | -0.01 / 0.85 | -0.01 / 0.86 | 0.01 / 0.89 | -0.01 / 0.90 | 0.096 | 0.721 |
| ***CL-PO truth scenario 1: balanced outcome*** | | | | | | | | | |
| MLR | 0.01 / 0.79 | 0.00 / 0.38 | 0.02 / 0.80 | -0.01 / 0.79 | 0.02 / 0.80 | 0.00 / 0.75 | 0.01 / 0.73 | 0.108 | 0.726 |
| CL-PO | 0.00 / 0.87 | 0.01 / 0.75 | 0.01 / 0.86 | 0.00 / 0.87 | 0.01 / 0.86 | 0.03 / 0.86 | -0.03 / 0.86 | 0.080 | 0.728 |
| AC-PO | 0.01 / 0.85 | 0.01 / 1.01 | 0.01 / 0.85 | -0.01 / 0.85 | 0.01 / 0.85 | 0.00 / 0.84 | 0.00 / 0.84 | 0.079 | 0.728 |
| SLM | 0.01 / 0.85 | 0.01 / 0.75 | 0.01 / 0.88 | -0.01 / 0.85 | 0.01 / 0.88 | 0.00 / 0.68 | 0.00 / 0.83 | 0.089 | 0.727 |
| ***CL-PO truth scenario 2: imbalanced outcome*** | | | | | | | | | |
| MLR | 0.02 / 0.84 | -0.01 / 0.63 | 0.03 / 0.73 | -0.02 / 0.84 | 0.03 / 0.73 | -0.02 / 0.86 | 0.03 / 0.54 | 0.103 | 0.724 |
| CL-PO | 0.02 / 0.87 | 0.00 / 0.85 | 0.00 / 0.87 | -0.02 / 0.87 | 0.00 / 0.87 | 0.03 / 0.87 | -0.03 / 0.87 | 0.080 | 0.726 |
| AC-PO | 0.02 / 0.92 | 0.00 / 1.12 | 0.01 / 0.76 | -0.02 / 0.92 | 0.01 / 0.76 | -0.01 / 0.85 | 0.01 / 0.85 | 0.081 | 0.725 |
| SLM | 0.02 / 0.87 | -0.01 / 0.92 | 0.03 / 0.88 | -0.02 / 0.87 | 0.03 / 0.88 | -0.02 / 0.91 | 0.01 / 0.84 | 0.087 | 0.725 |
| ***CL-PO truth scenario 3: highly imbalanced outcome*** | | | | | | | | | |
| MLR | -0.02 / 0.77 | 0.07 / 0.69 | -0.04 / 0.46 | 0.02 / 0.77 | -0.04 / 0.46 | 0.05 / 0.80 | 0.01 / 0.23 | 0.100 | 0.723 |
| CL-PO | -0.03 / 0.82 | 0.06 / 0.83 | -0.04 / 0.83 | 0.03 / 0.82 | -0.04 / 0.83 | -0.02 / 0.82 | -0.01 / 0.82 | 0.075 | 0.726 |
| AC-PO | -0.03 / 0.87 | 0.06 / 1.00 | -0.05 / 0.64 | 0.03 / 0.87 | -0.05 / 0.64 | 0.05 / 0.81 | -0.08 / 0.81 | 0.077 | 0.726 |
| SLM | -0.03 / 0.81 | 0.06 / 0.83 | -0.01 / 0.72 | 0.03 / 0.81 | -0.01 / 0.72 | 0.05 / 0.82 | 0.01 / 0.75 | 0.085 | 0.725 |

MLR, multinomial logistic regression; CL-PO, cumulative logit model with proportional odds; AC-PO, adjacent category logit model with proportional odds; SLM, stereotype logit model; LP, linear predictor; ECI, estimated calibration index; rMSPE, root mean squared prediction error; ORC, ordinal C statistic; CAD, coronary artery disease.

Table 5. Validation performance based on small development datasets of n=500 for the main simulation scenarios (reported as the average performance on a large validation dataset for 200 simulated development datasets).

| | CALIBRATION INTERCEPTS AND SLOPES | | | | | | | SINGLE NUMBER METRICS | |
|---|---|---|---|---|---|---|---|---|---|
| | Per outcome category | | | Per outcome dichotomy | | Model-specific | | | |
| **Model** | $Y=1$ | $Y=2$ | $Y=3$ | $Y>1$ | $Y>2$ | **LP1** | **LP2** | **rMSPE** | **ORC** |
| | ***MLR truth scenario 1: balanced outcome, equidistant means*** | | | | | | | | |
| MLR | 0.01 / 0.97 | -0.01 / 0.67 | 0.00 / 0.97 | -0.01 / 0.97 | 0.00 / 0.97 | -0.01 / 0.95 | 0.01 / 0.97 | 0.047 | 0.738 |
| CL-PO | 0.01 / 1.00 | -0.01 / 0.72 | 0.01 / 1.00 | -0.01 / 1.00 | 0.01 / 1.00 | 0.01 / 0.98 | -0.01 / 0.98 | 0.038 | 0.738 |
| AC-PO | 0.01 / 0.98 | -0.01 / 0.95 | 0.00 / 0.98 | -0.01 / 0.98 | 0.00 / 0.98 | -0.01 / 0.98 | 0.01 / 0.98 | 0.034 | 0.738 |
| SLM | 0.01 / 0.98 | -0.01 / 0.87 | 0.00 / 0.99 | -0.01 / 0.98 | 0.00 / 0.99 | -0.01 / 0.99 | 0.00 / 0.98 | 0.037 | 0.738 |
| | ***MLR truth scenario 2: imbalanced outcome, equidistant means*** | | | | | | | | |
| MLR | 0.00 / 0.95 | 0.00 / 0.83 | 0.00 / 0.95 | 0.00 / 0.95 | 0.00 / 0.95 | 0.00 / 0.95 | 0.00 / 0.93 | 0.044 | 0.736 |
| CL-PO | 0.02 / 0.92 | -0.01 / 0.75 | -0.01 / 1.10 | -0.02 / 0.92 | -0.01 / 1.10 | 0.01 / 0.96 | -0.01 / 0.96 | 0.038 | 0.737 |
| AC-PO | 0.00 / 0.96 | 0.00 / 0.97 | 0.00 / 0.96 | 0.00 / 0.96 | 0.00 / 0.96 | 0.00 / 0.96 | 0.00 / 0.96 | 0.033 | 0.737 |
| SLM | 0.00 / 0.96 | 0.00 / 0.95 | 0.00 / 0.97 | 0.00 / 0.96 | 0.00 / 0.97 | 0.00 / 0.96 | 0.00 / 0.96 | 0.036 | 0.737 |
| | ***MLR truth scenario 3: balanced outcome, non-equidistant means*** | | | | | | | | |
| MLR | 0.00 / 0.97 | -0.01 / 0.88 | 0.01 / 0.97 | 0.00 / 0.97 | 0.01 / 0.97 | -0.01 / 0.97 | 0.01 / 0.94 | 0.045 | 0.738 |
| CL-PO | -0.03 / 1.19 | -0.01 / 0.72 | 0.05 / 0.84 | 0.03 / 1.19 | 0.05 / 0.84 | 0.01 / 0.98 | -0.01 / 0.98 | 0.082 | 0.736 |
| AC-PO | 0.00 / 1.17 | -0.01 / 0.92 | 0.01 / 0.83 | 0.00 / 1.17 | 0.01 / 0.83 | -0.01 / 0.98 | 0.01 / 0.98 | 0.081 | 0.736 |
| SLM | 0.00 / 0.98 | -0.01 / 0.92 | 0.01 / 1.00 | 0.00 / 0.98 | 0.01 / 1.00 | -0.01 / 0.98 | 0.00 / 0.98 | 0.073 | 0.735 |
| | ***MLR truth scenario 4: imbalanced outcome, non-equidistant means*** | | | | | | | | |
| MLR | 0.01 / 0.97 | -0.01 / 0.94 | 0.00 / 0.94 | -0.01 / 0.97 | 0.00 / 0.94 | -0.01 / 0.98 | 0.00 / 0.87 | 0.043 | 0.735 |
| CL-PO | -0.01 / 1.08 | 0.00 / 1.10 | 0.02 / 0.83 | 0.01 / 1.08 | 0.02 / 0.83 | 0.01 / 0.98 | 0.00 / 0.98 | 0.067 | 0.733 |
| AC-PO | 0.01 / 1.14 | 0.00 / 1.42 | 0.00 / 0.75 | -0.01 / 1.14 | 0.00 / 0.75 | -0.01 / 0.97 | 0.00 / 0.97 | 0.072 | 0.733 |
| SLM | 0.01 / 0.98 | 0.00 / 0.99 | 0.00 / 0.99 | -0.01 / 0.98 | 0.00 / 0.99 | -0.01 / 0.98 | -0.01 / 0.97 | 0.060 | 0.730 |
| | ***CL-PO truth scenario 1: balanced outcome*** | | | | | | | | |
| MLR | 0.02 / 0.94 | 0.00 / 0.90 | -0.02 / 0.96 | -0.02 / 0.94 | -0.02 / 0.96 | -0.01 / 0.94 | -0.02 / 0.96 | 0.048 | 0.738 |
| CL-PO | 0.02 / 0.97 | 0.00 / 0.96 | -0.02 / 0.96 | -0.02 / 0.97 | -0.02 / 0.96 | 0.02 / 0.97 | 0.02 / 0.97 | 0.034 | 0.738 |
| AC-PO | 0.02 / 0.96 | 0.00 / 1.30 | -0.02 / 0.96 | -0.02 / 0.96 | -0.02 / 0.96 | -0.01 / 0.96 | -0.02 / 0.96 | 0.036 | 0.738 |
| SLM | 0.02 / 0.95 | 0.01 / 1.19 | -0.02 / 0.97 | -0.02 / 0.95 | -0.02 / 0.97 | -0.01 / 0.97 | -0.03 / 0.96 | 0.039 | 0.738 |
| | ***CL-PO truth scenario 2: imbalanced outcome*** | | | | | | | | |
| MLR | 0.00 / 0.97 | -0.01 / 0.97 | 0.03 / 0.91 | 0.00 / 0.97 | 0.03 / 0.91 | 0.00 / 0.98 | 0.03 / 0.84 | 0.046 | 0.737 |
| CL-PO | 0.00 / 0.97 | -0.01 / 0.97 | 0.03 / 0.96 | 0.00 / 0.97 | 0.03 / 0.96 | 0.00 / 0.97 | -0.03 / 0.97 | 0.034 | 0.738 |
| AC-PO | 0.00 / 1.03 | -0.01 / 1.27 | 0.03 / 0.85 | 0.00 / 1.03 | 0.03 / 0.85 | 0.00 / 0.96 | 0.03 / 0.96 | 0.038 | 0.737 |
| SLM | 0.00 / 0.97 | -0.01 / 1.07 | 0.03 / 0.94 | 0.00 / 0.97 | 0.03 / 0.94 | 0.00 / 0.98 | 0.03 / 0.95 | 0.038 | 0.738 |
| | ***CL-PO truth scenario 3: highly imbalanced outcome*** | | | | | | | | |
| MLR | -0.02 / 0.96 | 0.02 / 0.96 | 0.01 / 0.87 | 0.02 / 0.96 | 0.01 / 0.87 | 0.02 / 0.98 | 0.00 / 0.64 | 0.041 | 0.739 |
| CL-PO | -0.02 / 0.96 | 0.02 / 0.97 | 0.01 / 0.97 | 0.02 / 0.96 | 0.01 / 0.97 | -0.01 / 0.96 | -0.02 / 0.96 | 0.033 | 0.739 |
| AC-PO | -0.02 / 1.04 | 0.02 / 1.17 | 0.01 / 0.74 | 0.02 / 1.04 | 0.01 / 0.74 | 0.02 / 0.96 | -0.01 / 0.96 | 0.037 | 0.739 |
| SLM | -0.02 / 0.96 | 0.02 / 0.99 | 0.01 / 0.97 | 0.02 / 0.96 | 0.01 / 0.97 | 0.02 / 0.97 | 0.02 / 0.97 | 0.035 | 0.739 |

MLR, multinomial logistic regression; CL-PO, cumulative logit model with proportional odds; AC-PO, adjacent category logit model with proportional odds; SLM, stereotype logit model; LP, linear predictor; ECI, estimated calibration index; rMSPE, root mean squared prediction error; ORC, ordinal C statistic; CAD, coronary artery disease.

Table 6. Results for the case study on coronary artery disease (CAD).

| Performance statistic | MLR | CL-PO | AC-PO | CR-PO | CR-NP | SLM |
|---|---|---|---|---|---|---|
| | | | **Apparent performance** | | | |
| Calibration intercepts and slopes per outcome category | | | | | | |
| 1 (No CAD) | 0.00 / 1.00 | -0.02 / 1.13 | 0.00 / 1.38 | 0.00 / 1.48 | 0.00 / 1.00 | 0.00 / 1.00 |
| 2 (Non-obstructive stenosis) | 0.00 / 1.06 | 0.00 / 0.82 | 0.00 / 0.63 | -0.04 / 0.65 | 0.00 / 1.09 | 0.00 / 1.16 |
| 3 (One-vessel disease) | 0.00 / 0.97 | 0.02 / 0.82 | 0.00 / 1.11 | 0.03 / 1.08 | 0.00 / 0.93 | 0.00 / 0.96 |
| 4 (Two-vessel disease) | 0.00 / 1.02 | 0.01 / 1.06 | 0.00 / 1.04 | 0.06 / 1.20 | 0.00 / 1.04 | 0.00 / 1.03 |
| 5 (Three-vessel disease) | 0.00 / 0.98 | 0.00 / 0.89 | 0.00 / 0.70 | -0.02 / 0.61 | 0.00 / 0.98 | 0.00 / 0.98 |
| Calibration intercepts and slopes per outcome dichotomy | | | | | | |
| 2-5 vs 1 | 0.00 / 1.00 | 0.02 / 1.13 | 0.00 / 1.38 | 0.00 / 1.48 | 0.00 / 1.00 | 0.00 / 1.00 |
| 3-5 vs 1-2 | 0.00 / 0.99 | 0.02 / 0.91 | 0.00 / 0.94 | 0.04 / 0.99 | 0.00 / 1.00 | 0.00 / 1.00 |
| 4-5 vs 1-3 | 0.00 / 1.00 | 0.00 / 0.95 | 0.00 / 0.83 | 0.03 / 0.83 | 0.00 / 1.01 | 0.00 / 1.00 |
| 5 vs 1-4 | 0.00 / 0.98 | 0.00 / 0.89 | 0.00 / 0.70 | -0.02 / 0.61 | 0.00 / 0.98 | 0.00 / 0.98 |
| Calibration intercepts and slopes, model-specific | | | | | | |
| Linear predictor 1 | 0.00 / 1.00 | 0.00 / 1.00 | 0.00 / 1.00 | 0.00 / 1.00 | 0.00 / 1.00 | 0.00 / 1.00 |
| Linear predictor 2 | 0.00 / 1.00 | 0.00 / 1.00 | 0.00 / 1.00 | 0.00 / 1.00 | 0.00 / 1.00 | 0.00 / 1.00 |
| Linear predictor 3 | 0.00 / 1.00 | 0.00 / 1.00 | 0.00 / 1.00 | 0.00 / 1.00 | 0.00 / 1.00 | 0.00 / 1.00 |
| Linear predictor 4 | 0.00 / 1.00 | 0.00 / 1.00 | 0.00 / 1.00 | 0.00 / 1.00 | 0.00 / 1.00 | 0.00 / 1.00 |
| ECI | 0.005 | 0.030 | 0.141 | 0.194 | 0.005 | 0.004 |
| ORC | 0.696 | 0.695 | 0.695 | 0.695 | 0.695 | 0.694 |
| | | | **Bootstrap-corrected performance** | | | |
| Calibration intercepts and slopes per outcome category | | | | | | |
| 1 (No CAD) | 0.00 / 0.99 | -0.02 / 1.12 | 0.00 / 1.37 | 0.00 / 1.47 | 0.00 / 0.98 | 0.00 / 0.99 |
| 2 (Non-obstructive stenosis) | 0.00 / 0.99 | -0.01 / 0.81 | 0.00 / 0.62 | -0.04 / 0.64 | 0.00 / 1.02 | 0.00 / 1.13 |
| 3 (One-vessel disease) | 0.00 / 0.89 | 0.02 / 0.80 | 0.00 / 1.09 | 0.03 / 1.06 | 0.00 / 0.86 | 0.00 / 0.95 |
| 4 (Two-vessel disease) | 0.00 / 0.97 | 0.01 / 1.05 | 0.00 / 1.02 | 0.06 / 1.17 | 0.00 / 0.97 | 0.00 / 1.03 |
| 5 (Three-vessel disease) | 0.00 / 0.94 | 0.01 / 0.87 | 0.01 / 0.68 | 0.00 / 0.60 | 0.01 / 0.93 | 0.00 / 0.98 |
| Calibration intercepts and slopes per outcome dichotomy | | | | | | |
| 2-5 vs 1 | 0.00 / 0.99 | 0.02 / 1.12 | 0.00 / 1.37 | 0.00 / 1.47 | 0.00 / 0.98 | 0.00 / 0.99 |
| 3-5 vs 1-2 | 0.00 / 0.98 | 0.02 / 0.90 | 0.00 / 0.93 | 0.04 / 0.98 | 0.00 / 0.98 | 0.00 / 0.99 |
| 4-5 vs 1-3 | 0.00 / 0.97 | 0.01 / 0.93 | 0.01 / 0.81 | 0.03 / 0.81 | 0.01 / 0.98 | 0.00 / 1.00 |
| 5 vs 1-4 | 0.00 / 0.94 | 0.01 / 0.87 | 0.01 / 0.68 | 0.00 / 0.60 | 0.01 / 0.93 | 0.00 / 0.98 |
| Calibration intercepts and slopes, model-specific | | | | | | |
| Linear predictor 1 | 0.00 / 0.95 | 0.00 / 0.99 | 0.00 / 0.98 | 0.00 / 0.98 | 0.00 / 0.98 | -0.01 / 0.99 |
| Linear predictor 2 | 0.00 / 0.96 | 0.00 / 0.99 | 0.00 / 0.98 | 0.00 / 0.98 | 0.00 / 0.96 | 0.00 / 0.98 |
| Linear predictor 3 | 0.01 / 0.96 | -0.01 / 0.99 | 0.00 / 0.98 | 0.01 / 0.98 | 0.01 / 0.89 | 0.00 / 0.99 |
| Linear predictor 4 | 0.00 / 0.96 | -0.01 / 0.99 | 0.01 / 0.98 | 0.01 / 0.98 | 0.01 / 0.65 | 0.00 / 0.99 |
| ORC | 0.694 | 0.693 | 0.693 | 0.693 | 0.693 | 0.693 |

MLR, multinomial logistic regression; CL-PO, cumulative logit model with proportional odds; AC-PO, adjacent category logit model with proportional odds; CR-PO, continuation ratio logit model with proportional odds; CR-NP, continuation ration logit model without proportional odds; SLM, stereotype logit model; ECI, estimated calibration index; ORC, ordinal C statistic; CAD, coronary artery disease.

Figure 1. Scatter plots of true risks versus estimated risks for simulation scenario 1 when the true model has the form of a multinomial logistic regression. The plots are based on a random subset of 1,000 cases from all 200,000 cases.

MLR, multinomial logistic regression; CL-PO, cumulative logit model with proportional odds; AC-PO, adjacent category logit model with proportional odds; SLM, stereotype logit model.

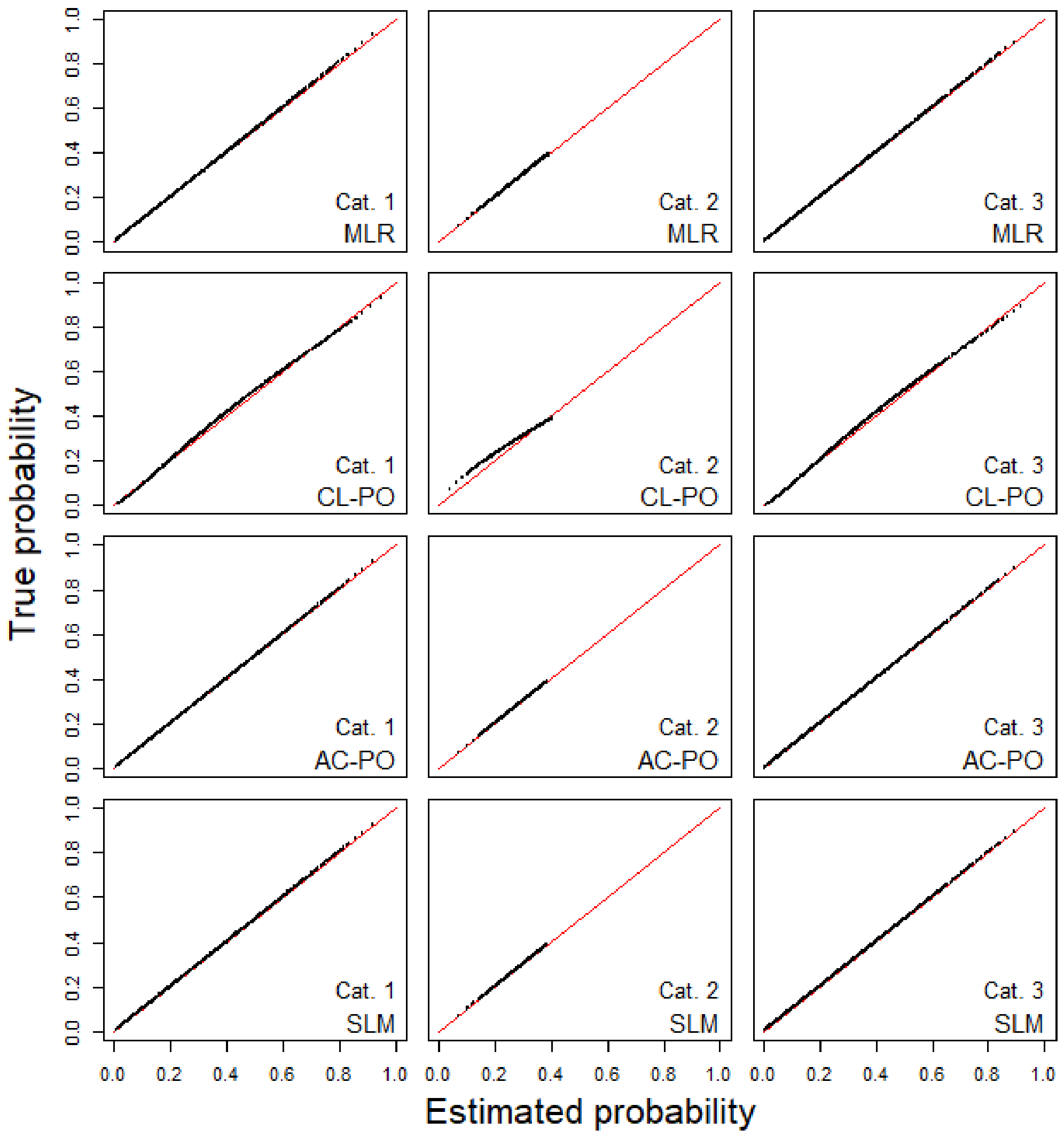

Figure 2. Scatter plots of true risks versus estimated risks for simulation scenario 2 when the true model has the form of a multinomial logistic regression. The plots are based on a random subset of 1,000 cases from all 200,000 cases.

MLR, multinomial logistic regression; CL-PO, cumulative logit model with proportional odds; AC-PO, adjacent category logit model with proportional odds; SLM, stereotype logit model.

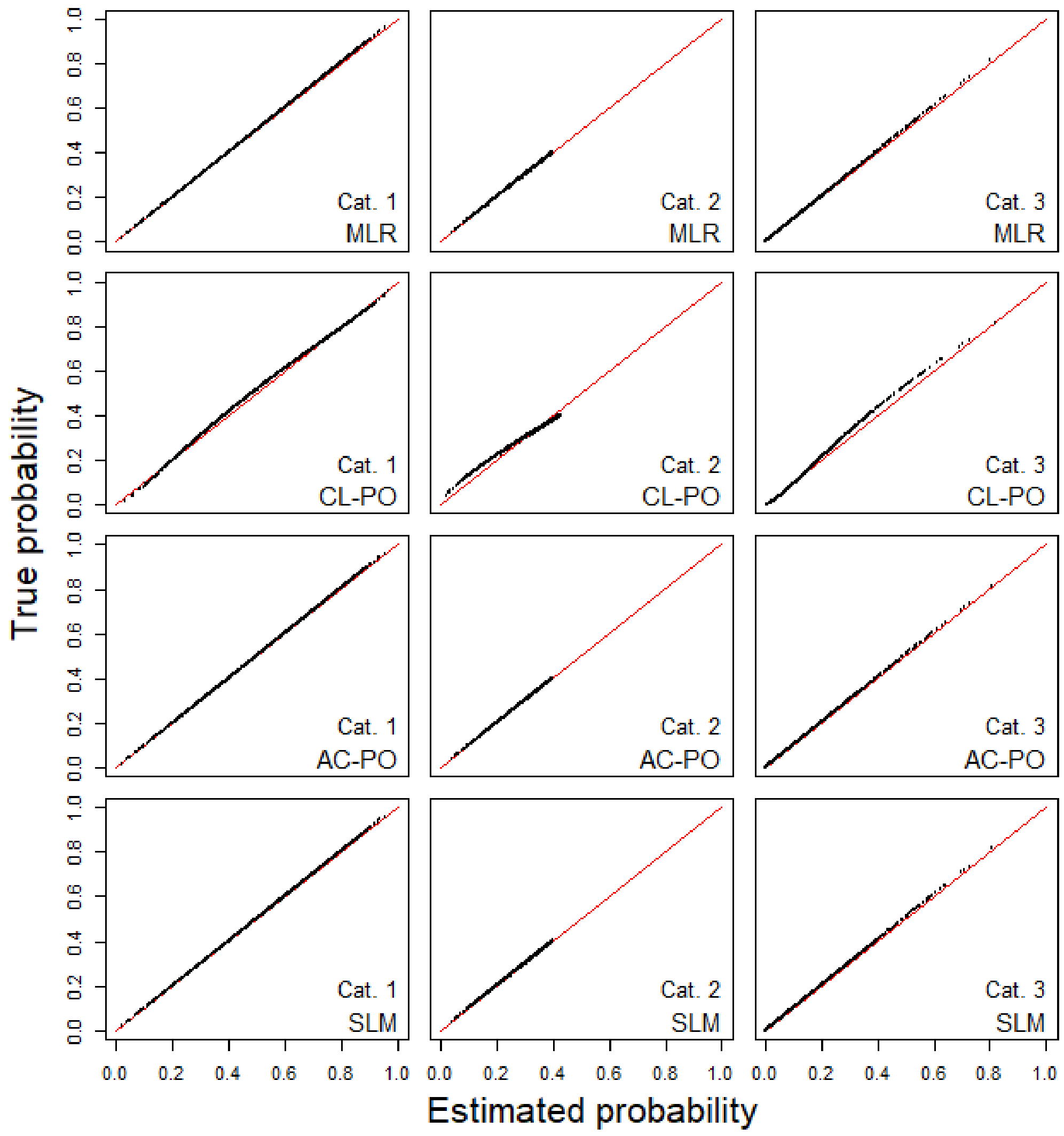

Figure 3. Scatter plots of true risks versus estimated risks for simulation scenario 3 when the true model has the form of a multinomial logistic regression. The plots are based on a random subset of 1,000 cases from all 200,000 cases.

MLR, multinomial logistic regression; CL-PO, cumulative logit model with proportional odds; AC-PO, adjacent category logit model with proportional odds; SLM, stereotype logit model.

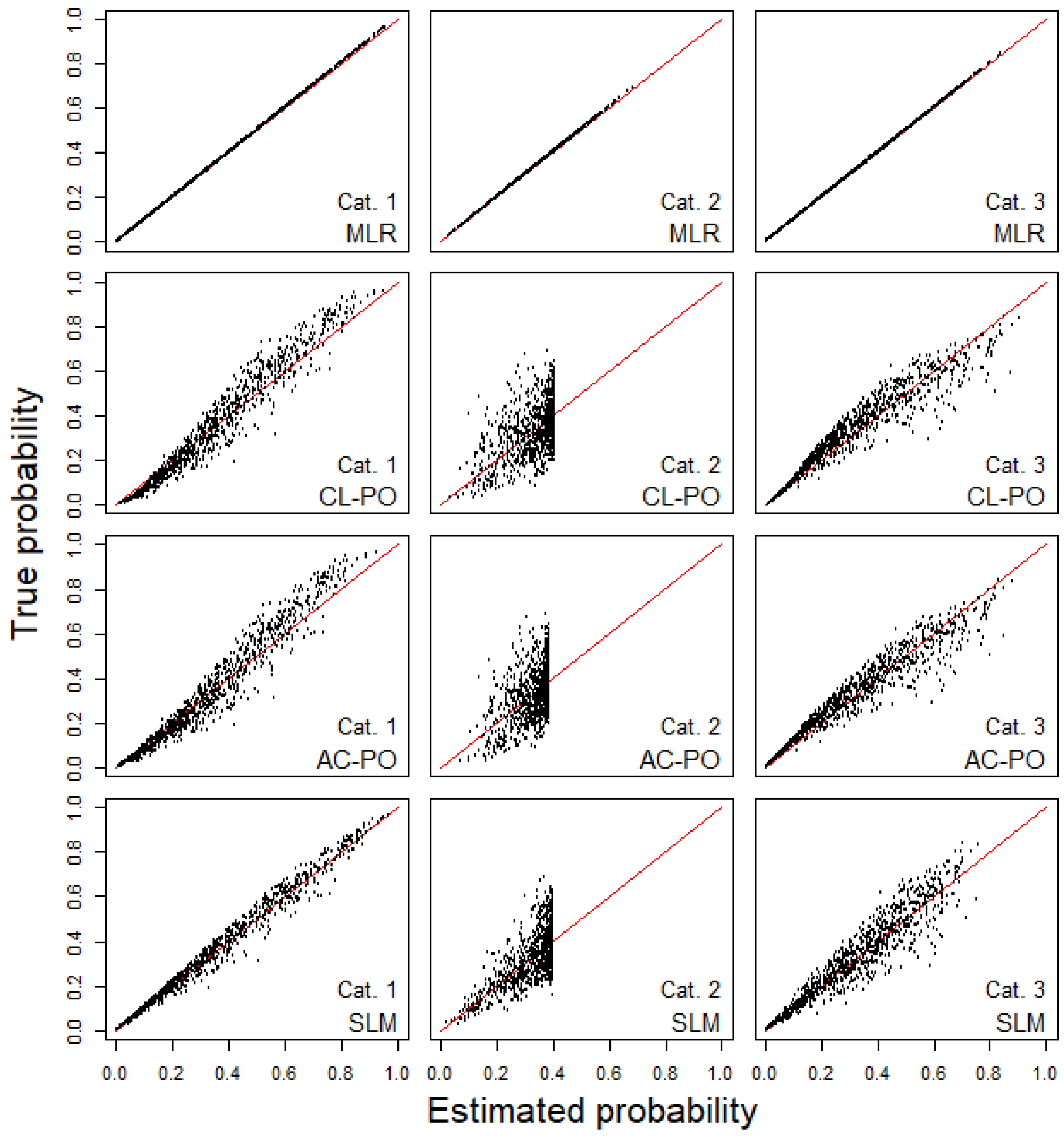

Figure 4. Scatter plots of true risks versus estimated risks for simulation scenario 4 when the true model has the form of a multinomial logistic regression. The plots are based on a random subset of 1,000 cases from all 200,000 cases.

MLR, multinomial logistic regression; CL-PO, cumulative logit model with proportional odds; AC-PO, adjacent category logit model with proportional odds; SLM, stereotype logit model.

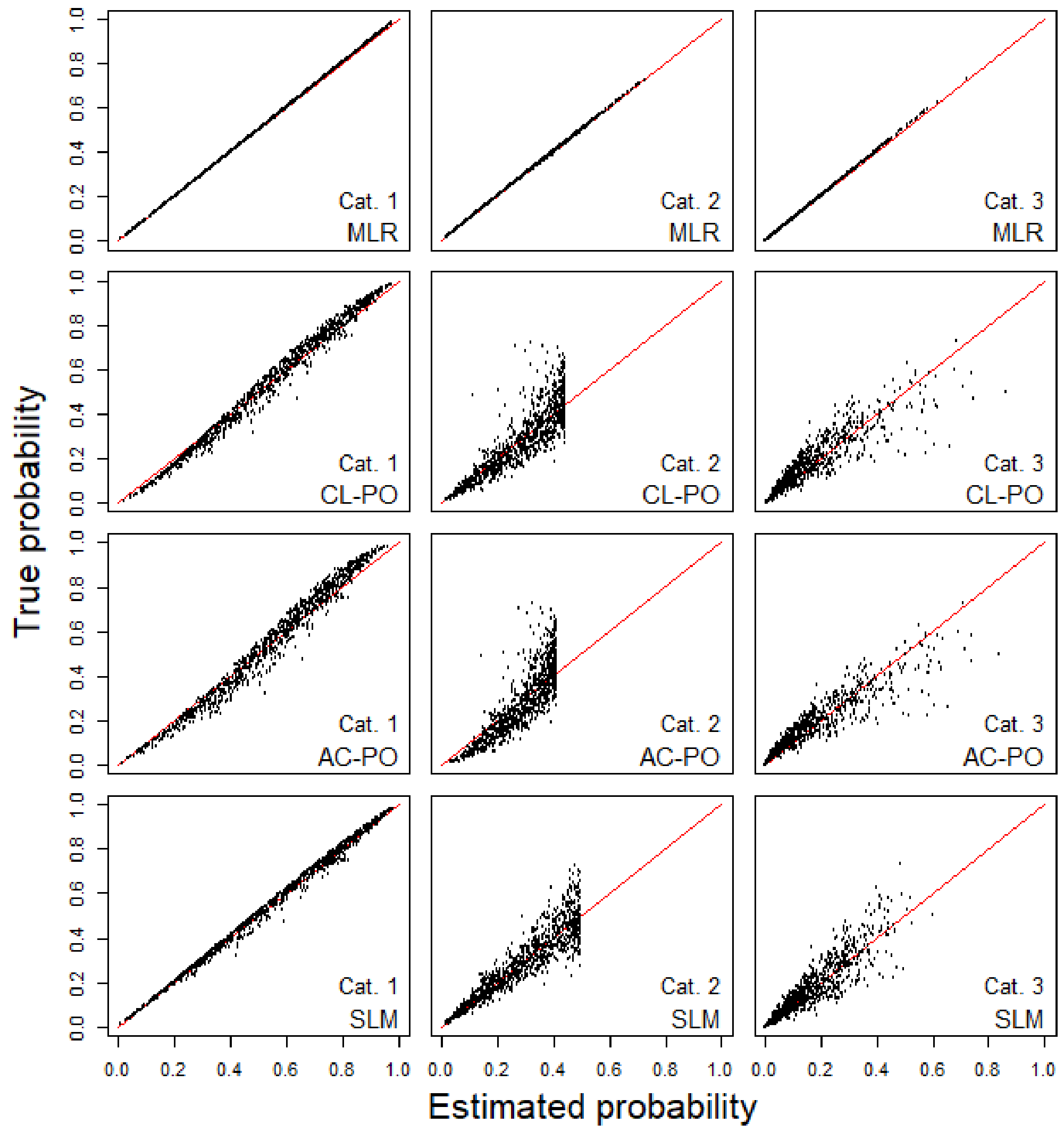

Figure 5. Scatter plots of true risks versus estimated risks for simulation scenario 1 when the true model has the form of a cumulative logit model with proportional odds. The plots are based on a random subset of 1,000 cases from all 200,000 cases.

MLR, multinomial logistic regression; CL-PO, cumulative logit model with proportional odds; AC-PO, adjacent category logit model with proportional odds; SLM, stereotype logit model.

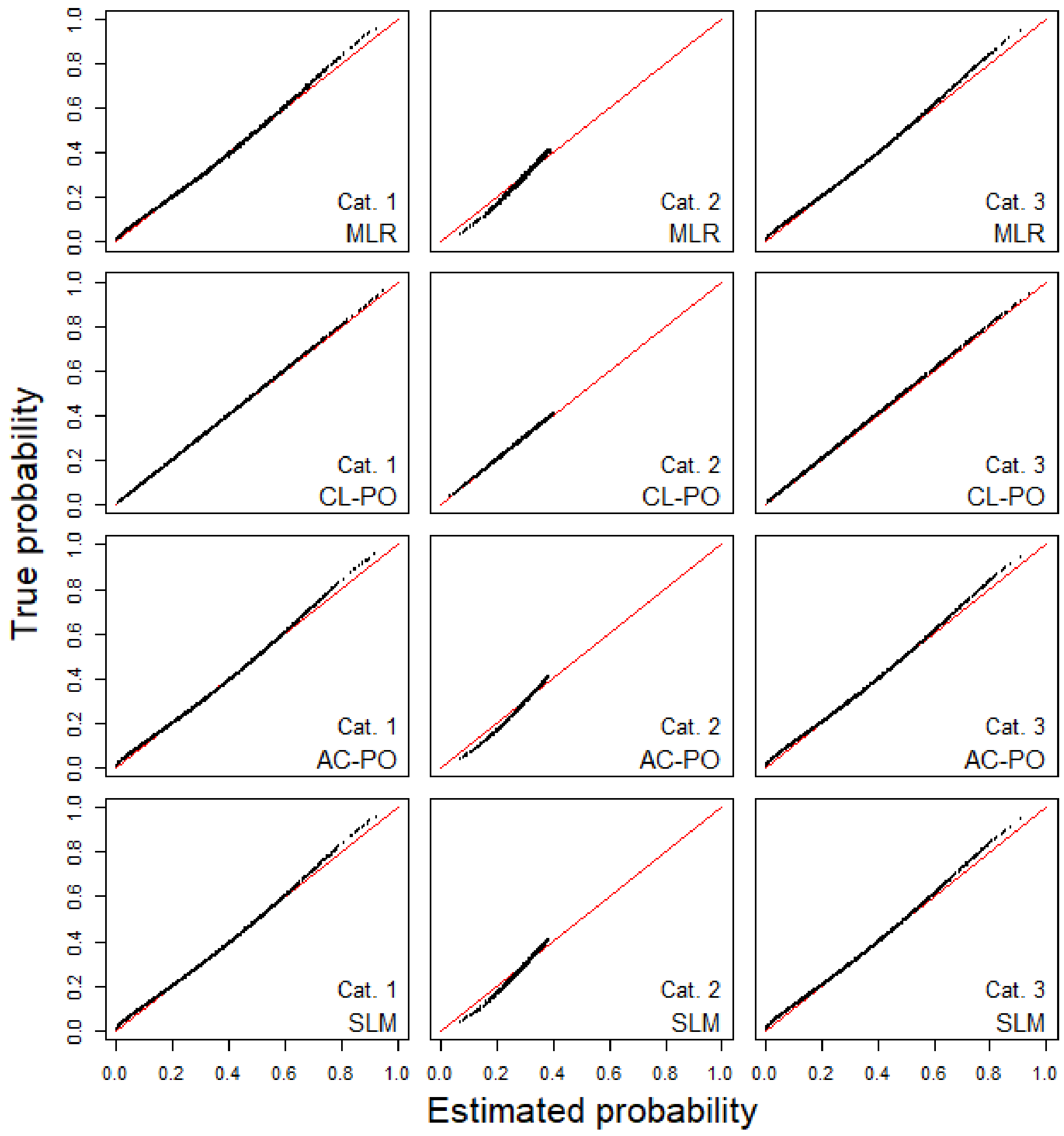

Figure 6. Scatter plots of true risks versus estimated risks for simulation scenario 2 when the true model has the form of a cumulative logit model with proportional odds. The plots are based on a random subset of 1,000 cases from all 200,000 cases.

MLR, multinomial logistic regression; CL-PO, cumulative logit model with proportional odds; AC-PO, adjacent category logit model with proportional odds; SLM, stereotype logit model.

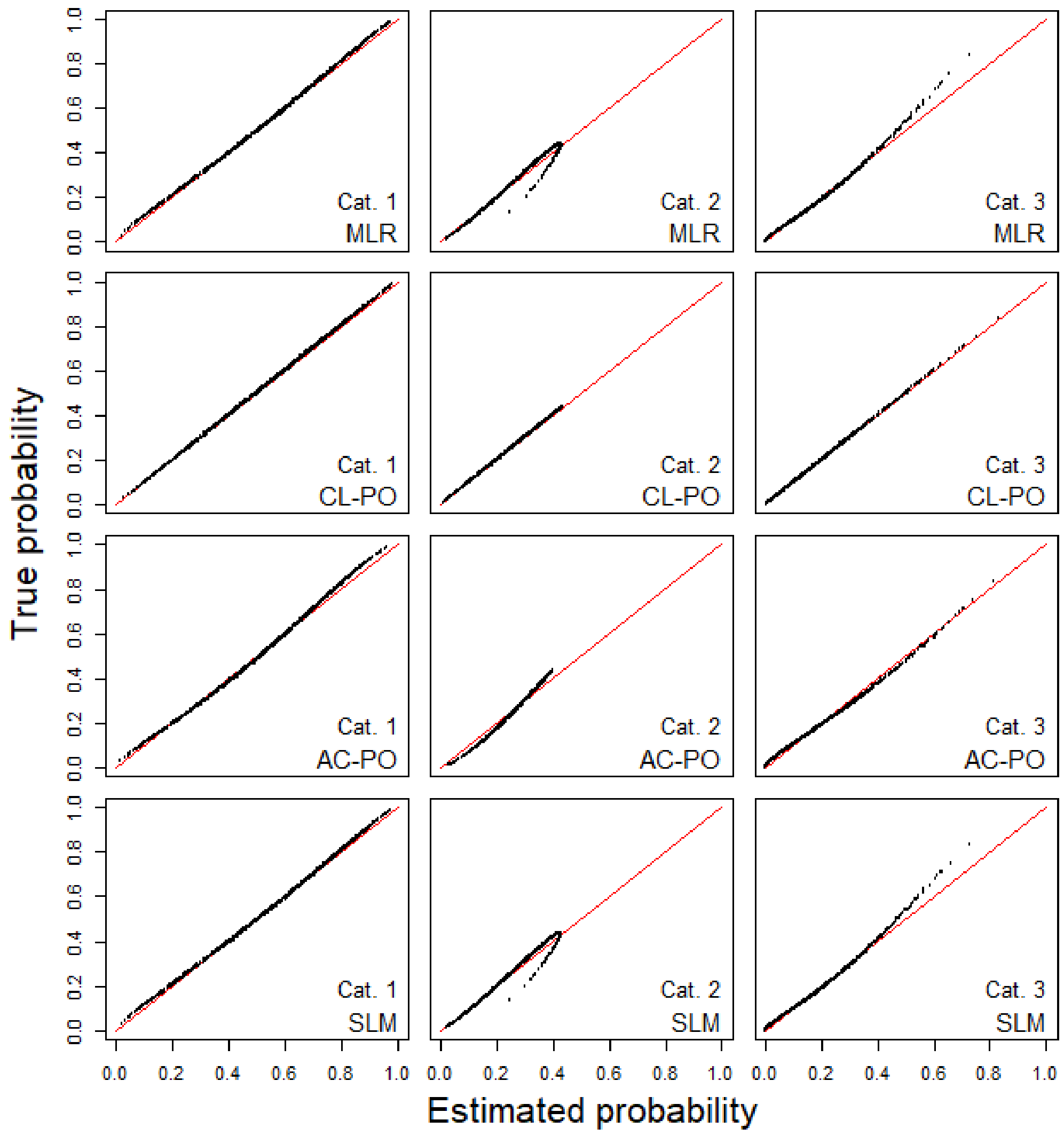

Figure 7. Scatter plots of true risks versus estimated risks for simulation scenario 3 when the true model has the form of a cumulative logit model with proportional odds. The plots are based on a random subset of 1,000 cases from all 200,000 cases.

MLR, multinomial logistic regression; CL-PO, cumulative logit model with proportional odds; AC-PO, adjacent category logit model with proportional odds; SLM, stereotype logit model.

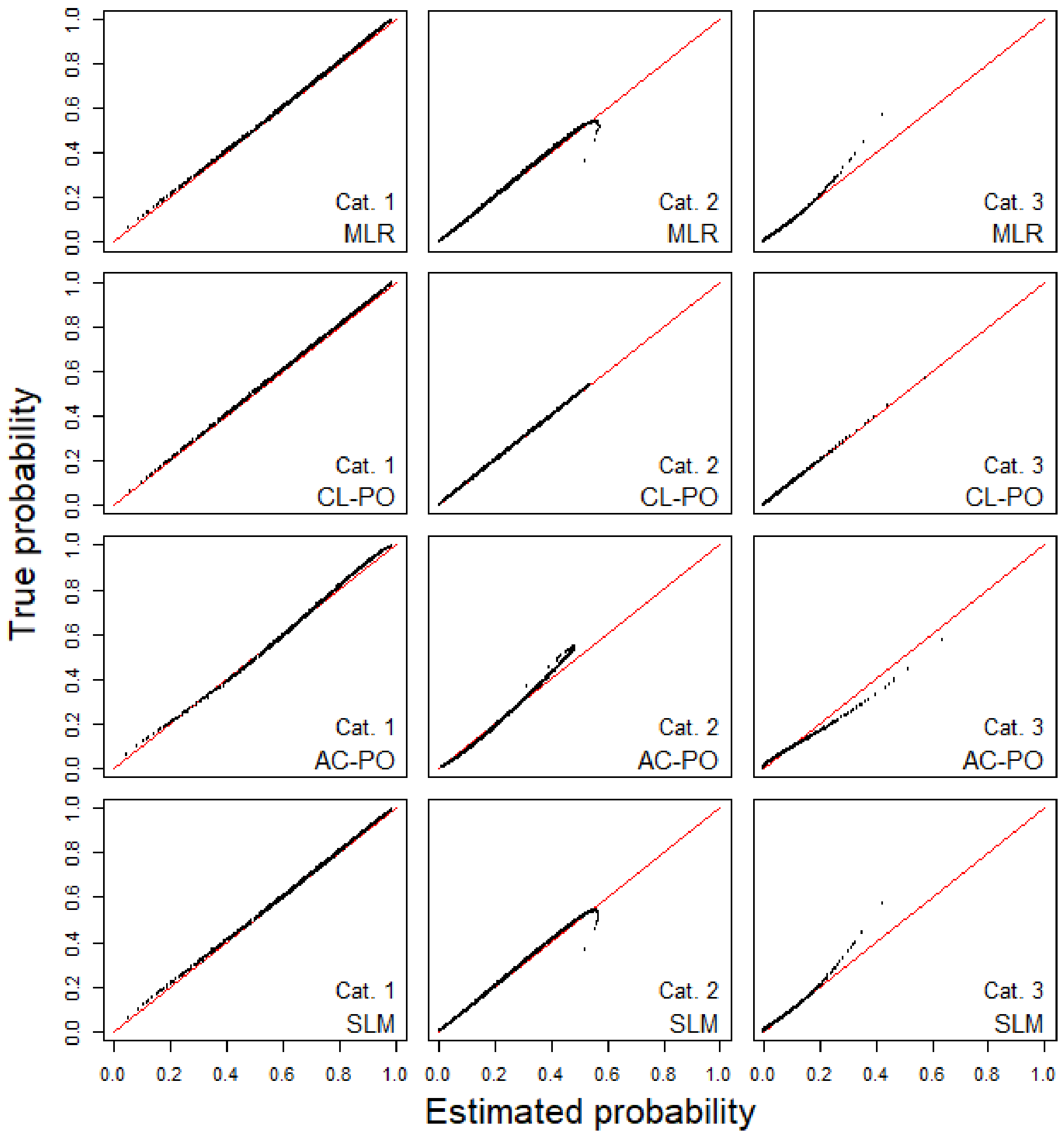

Figure 8. Scatter plot of estimated probabilities for having non-obstructive stenosis in the case study (n=4,888).

MLR, multinomial logistic regression; CL-PO, cumulative logit model with proportional odds; AC-PO, adjacent category logit model with proportional odds; CR-PO, continuation ratio logit model with proportional odds; CR-NP, continuation ratio logit model without proportional odds; SLM, stereotype logit model

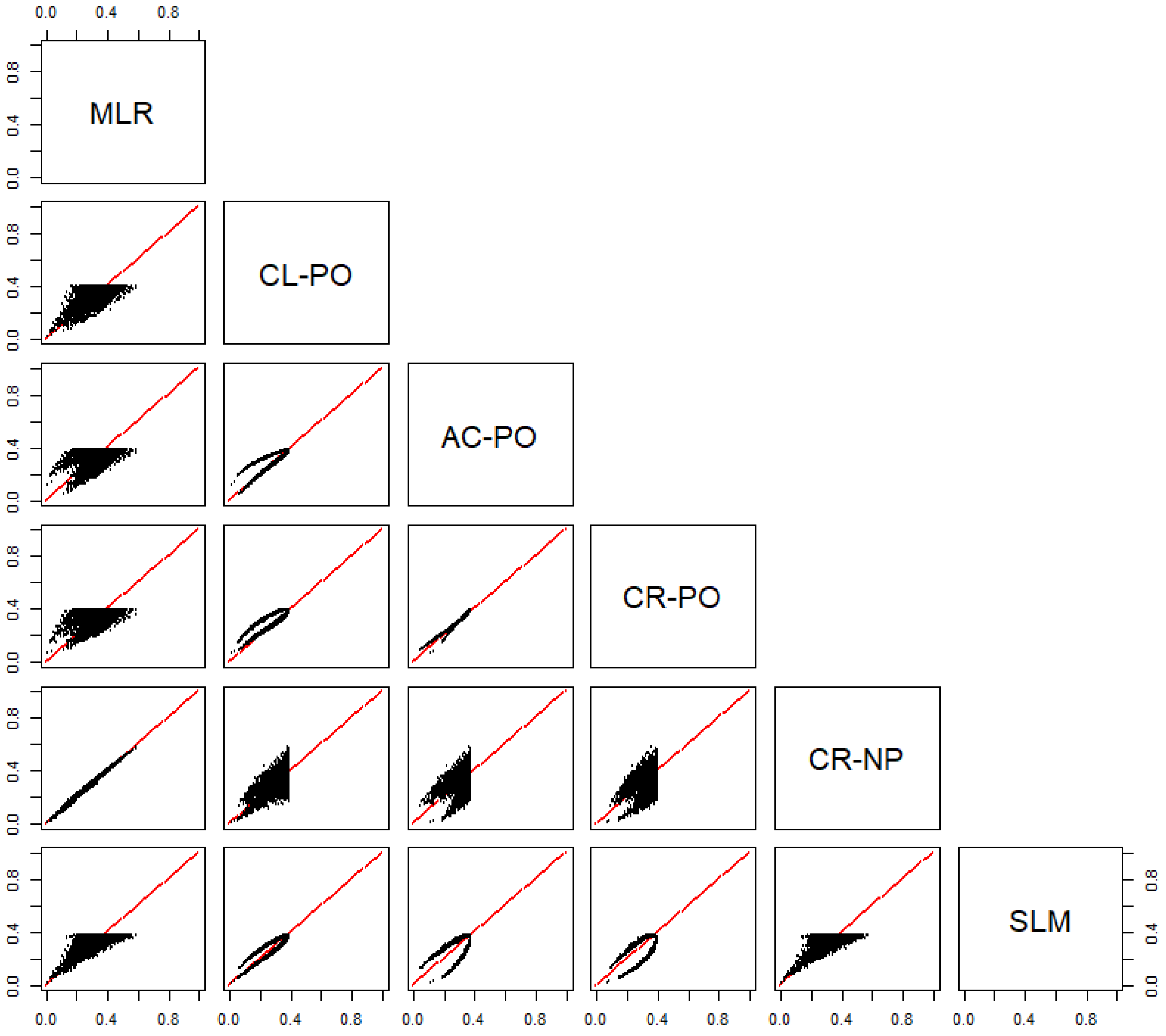

Figure 9. Calibration scatter plots per outcome category for the models in the case study (green for no coronary artery disease, orange for non-obstructive stenosis, red for one-vessel disease, brown for two-vessel disease, black for three-vessel disease). These plots are generated for the model development data (i.e. apparent validation, n=4,888).

MLR, multinomial logistic regression; CL-PO, cumulative logit model with proportional odds; AC-PO, adjacent category logit model with proportional odds; CR-PO, continuation ratio logit model with proportional odds; CR-NP, continuation ratio logit model without proportional odds; SLM, stereotype logit model

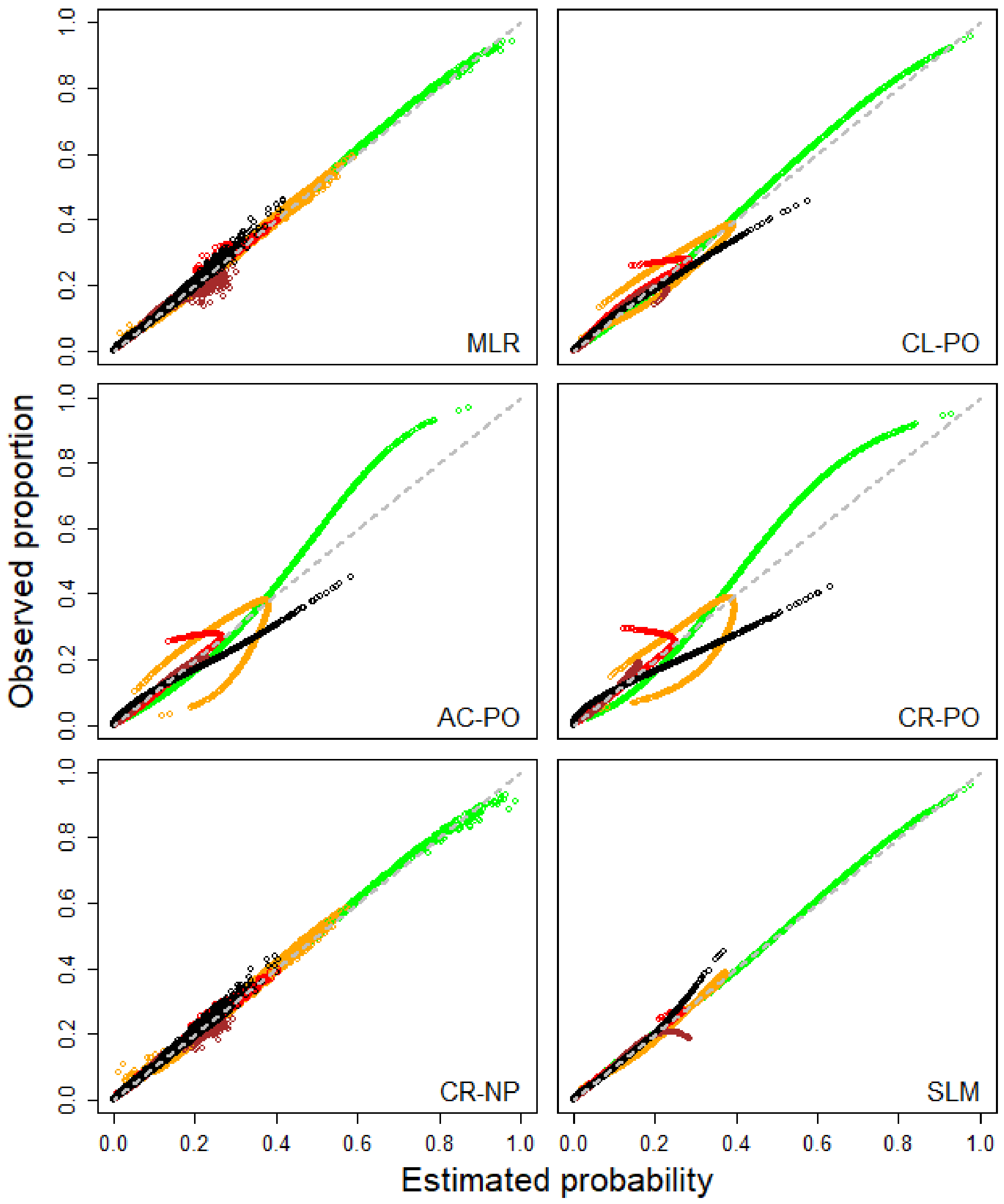

Figure 10. Calibration scatter plots per outcome dichotomy for the models in the case study (orange for non-obstructive stenosis or worse, red for one-vessel disease or worse, brown for two-vessel disease or worse, black for three-vessel disease). These plots are generated for the model development data (i.e. apparent validation, n=4,888).

MLR, multinomial logistic regression; CL-PO, cumulative logit model with proportional odds; AC-PO, adjacent category logit model with proportional odds; CR-PO, continuation ratio logit model with proportional odds; CR-NP, continuation ratio logit model without proportional odds; SLM, stereotype logit model

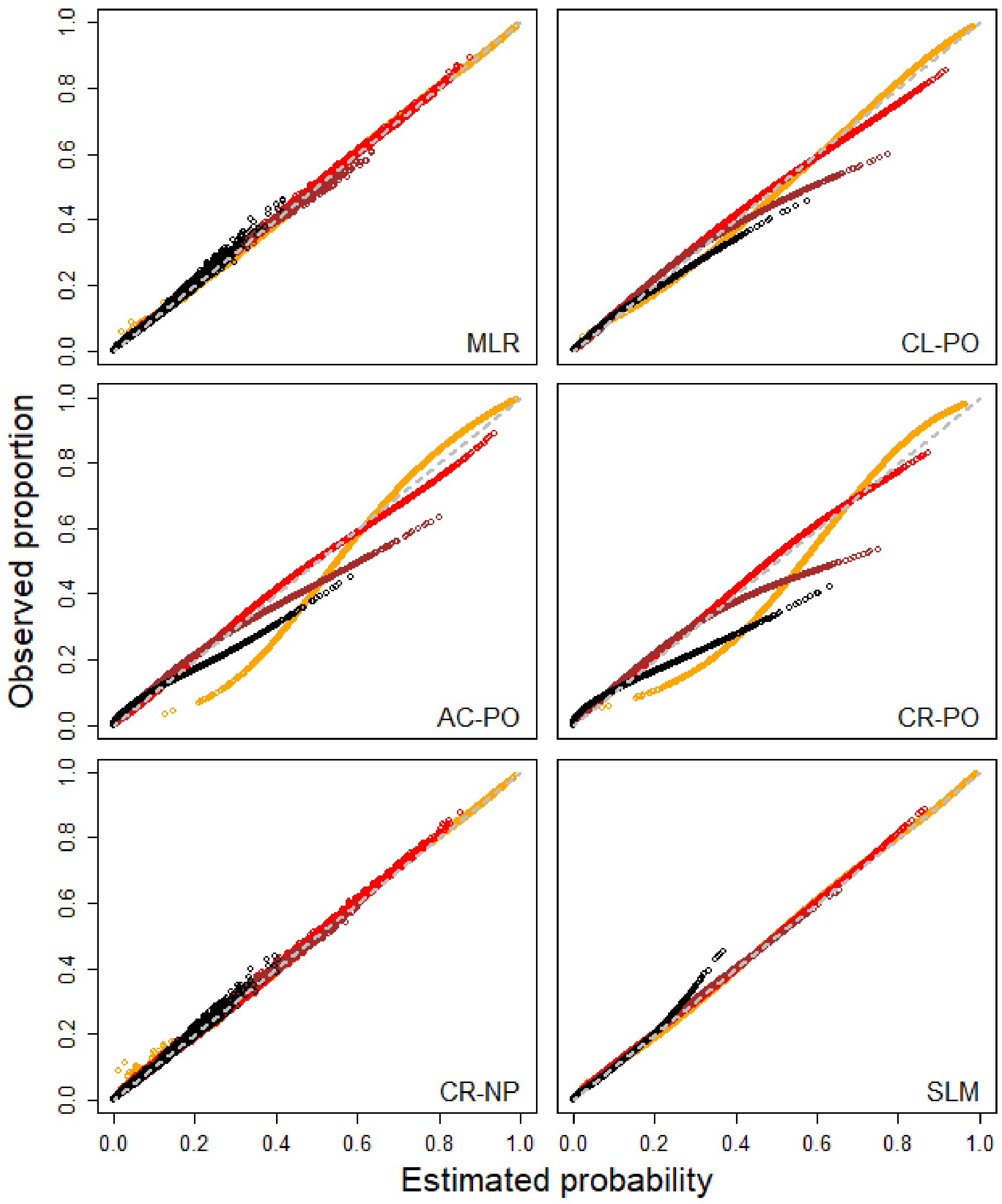

# Risk prediction models for discrete ordinal outcomes: calibration and the impact of the proportional odds assumption

Michael EDLINGER, Maarten VAN SMEDEN, Hannes F ALBER,

Maria WANITSCHEK, Ben VAN CALSTER * †

## SUPPLEMENTARY MATERIAL

* Correspondence to: Ben Van Calster, Department of Development and Regeneration, KU Leuven, Herestraat 49 Box 805, 3000 Leuven, Belgium; ben.vancalster@kuleuven.be

**TABLE OF CONTENTS**

## 1. Explanation for the sample sizes used in the simulation study

We used large datasets of 200,000 simulated cases to approach true model coefficients and model performance. When evaluating at all 20 simulation scenarios, this sample size ensures a minimum of 10,000 cases in each outcome category and scenario. Knowing that, across all scenarios, the maximum number of parameters in a model is 18 (including intercepts). This implies that there events per parameter (EPP) in the smallest outcome category was at least 1,111 (Table A2). Note that EPP is calculated without taking intercepts into account.[34] This is very high, even for models with modest true discriminatory ability.

Table A2. Events per parameter (EPP) in the smallest outcome category by N, scenario, and model.

| | EPP in smallest outcome category | | | | | | | | |
|---|---|---|---|---|---|---|---|---|---|
| | N=200,000 | | | N=500 | | | N=100 | | |
| **Scenario** | **MLR** | **CL-PO/ AC-PO** | **SLM** | **MLR** | **CL-PO/ AC-PO** | **SLM** | **MLR** | **CL-PO/ AC-PO** | **SLM** |
| MLR 1 | 8,333 | 16,667 | 13,333 | 20.8 | 41.7 | 33.3 | 4.2 | 8.3 | 6.7 |
| MLR 2 | 3,750 | 7,500 | 6,000 | 9.4 | 18.8 | 15.0 | 1.9 | 3.8 | 3.0 |
| MLR 3 | 8,333 | 16,667 | 13,333 | 20.8 | 41.7 | 33.3 | 4.2 | 8.3 | 6.7 |
| MLR 4 | 3,750 | 7,500 | 6,000 | 9.4 | 18.8 | 15.0 | 1.9 | 3.8 | 3.0 |
| MLR 5 | 3,750 | 7,500 | 6,000 | nd | nd | nd | nd | nd | nd |
| MLR 6 | 3,333 | 10,000 | 6,000 | nd | nd | nd | nd | nd | nd |
| MLR 7 | 3,333 | 10,000 | 6,000 | nd | nd | nd | nd | nd | nd |
| MLR 8 | 1,111 | 3,333 | 2,000 | nd | nd | nd | nd | nd | nd |
| MLR 9 | 3,750 | 7,500 | 6,000 | nd | nd | nd | nd | nd | nd |
| MLR 10 | 3,333 | 10,000 | 6,000 | nd | nd | nd | nd | nd | nd |
| MLR 11 | 1,875 | 3,750 | 3,333 | nd | nd | nd | nd | nd | nd |
| | | | | | | | | | |
| CL-PO 1 | 8,333 | 16,667 | 13,333 | 20.8 | 41.7 | 33.3 | 4.2 | 8.3 | 6.7 |
| CL-PO 2 | 3,750 | 7,500 | 6,000 | 9.4 | 18.8 | 15.0 | 1.9 | 3.8 | 3.0 |
| CL-PO 3 | 1,250 | 2,500 | 2,000 | 3.1 | 6.3 | 5.0 | 0.6 | 1.3 | 1.0 |
| CL-PO 4 | 3,333 | 10,000 | 6,000 | nd | nd | nd | nd | nd | nd |
| CL-PO 5 | 3,333 | 10,000 | 6,000 | nd | nd | nd | nd | nd | nd |
| CL-PO 6 | 1,111 | 3,333 | 2,000 | nd | nd | nd | nd | nd | nd |
| CL-PO 7 | 3,750 | 7,500 | 6,000 | nd | nd | nd | nd | nd | nd |
| CL-PO 8 | 3,333 | 10,000 | 6,000 | nd | nd | nd | nd | nd | nd |
| CL-PO 9 | 4,167 | 8,333 | 7,407 | nd | nd | nd | nd | nd | nd |

MLR, multinomial logistic regression; CL-PO, cumulative logit model with proportional odds; AC-PO, adjacent category logit model with proportional odds; SLM, stereotype logit model; nd, not done.

To investigate overfitting in the 7 main scenarios, we need datasets where the events per parameter is too low. We sampled datasets of size 100 and 500. When sample size is 100, the EPP in the smallest outcome category varied between 0.6 and 8.3 depending on outcome category and scenario. These values are low or very low by all standards. When the sample size is 500, the EPP in the smallest outcome category varied between 3.1 and 41.7. These values are low to acceptable.

## 2. Example R code

```
library(VGAM)

# CODE TO FIT DIFFERENT MODELS

mlr <- vglm(y ~ x1 + x2, family=multinomial(refLevel = "1"), data=dataset)
 # Multinomial logistic regression; alternative: multinom (nnet package)

clpo <- vglm(y ~ x1 + x2, family=cumulative(parallel=T, reverse=T), data=dataset)
 # Cumulative logit model with proportional odds; alternatives: polr (MASS package), clm (ordinal
   package), orm (rms package)
 # If you do not specify reverse = T, it focuses on Y ≤ k instead of Y ≥ k

acp <- vglm(y ~ x1 + x2, family=acat(parallel=T), data=dataset)
 # Adjacent category logit model with proportional odds

crp <- vglm(y ~ x1 + x2, family=cratio(parallel=T), data=dataset)
 # Continuation ratio logit model with proportional odds

crnp <- vglm(y ~ x1 + x2, family=cratio(parallel=F), data=dataset)
 # Continuation ratio logit model without proportional odds

slm=rrvglm(y ~ x1 + x2, multinomial(refLevel = "1"), data = dataset)
 # Stereotype logistic model


# CALCULATE ESTIMATED PROBABILITIES AND LINEAR PREDICTORS (MLR ONLY)

mlrpred <- predictvglm(mlr,newdata=cad,type="response")
mlrlpred <- predictvglm(mlr,newdata=cad,type="link")


# CALIBRATION INTERCEPT AND SLOPE FOR EACH OUTCOME CATEGORY

calout <- function(out,preds,k){
  cores = matrix(0,k,2)
  for (i in (1:k)){
    cores[i,1] = glm(out==i ~ 1, offset=logit(preds[,i]), family=binomial)$coefficients
    cores[i,2] = glm(out==i ~ logit(preds[,i]), family=binomial)$coefficients[2]
  }
  return(cores)
}
mlrcalout = calout(out=dataset$y,preds=mlrpred,5) # MLR, 5 CATEGORIES


# CALIBRATION INTERCEPT AND SLOPE FOR EACH OUTCOME DICHOTOMY

caldout <- function(out,preds,k){
  cores = matrix(0,k-1,2)
  for (i in (2:(k-1))){
    cores[i-1,1] = glm(out>=i ~ 1, offset=logit(rowSums(preds[,i:k])),
                       family=binomial)$coefficients
    cores[i-1,2] = glm(out>=i ~ logit(rowSums(preds[,i:k])), family=binomial)$coefficients[2]
  }
  cores[k-1,1] = glm(out>=k ~ 1, offset=logit(preds[,k]), family=binomial)$coefficients
  cores[k-1,2] = glm(out>=k ~ logit(preds[,k]),
                     family=binomial)$coefficients[2]
  return(cores)
}
mlrcaldout = caldout(out=dataset$y,preds=mlrpred,k=5) # MLR, 5 CATEGORIES


# MODEL SPECIFIC CALIBRATION INTERCEPTS AND SLOPES (MLR ONLY, 5 OUTCOME CATEGORIES)

mlrrecali <- coefficients(vglm(dataset$y ~ 1, offset = mlrlpred[,1:4],
                               family=multinomial(refLevel = "1")))[c(1:4)]
mlrrecals <- coefficients(vglm(dataset$y ~ mlrlpred[,1] + mlrlpred[,2] +
                                           mlrlpred[,3] + mlrlpred[,4],
```

```
                                constraints=list("(Intercept)"=diag(4),
                                                 "mlrlpred[, 1]"=rbind(1,0,0,0),
                                                 "mlrlpred[, 2]"=rbind(0,1,0,0),
                                                 "mlrlpred[, 3]"=rbind(0,0,1,0),
                                                 "mlrlpred[, 4]"=rbind(0,0,0,1)),
                                family=multinomial(refLevel = "1")))[c(5:8)]


# FLEXIBLE RECALIBRATION MODEL (MLR ONLY, 5 OUTCOME CATEGORIES)

mlrlp1=log(mlrpred[,2]/mlrpred[,1])
mlrlp2=log(mlrpred[,3]/mlrpred[,1])
mlrlp3=log(mlrpred[,4]/mlrpred[,1])
mlrlp4=log(mlrpred[,5]/mlrpred[,1])
mlrvgamsmps4 = vgam(dataset$y ~ sm.ps(mlrlp1,df=4) + sm.ps(c(mlrlp2),df=4) +
                                 sm.ps(c(mlrlp3),df=4) + sm.ps(c(mlrlp4),df=4),
                    family=multinomial(refLevel = "1"))


# ECI, ORIGINAL FORMULA FROM VAN HOORDE ET AL (J BIOMED INFORM 2015)

eci_bvc <- function(calout,preds,k){
  (mean((preds-fitted(calout))*(preds-fitted(calout))))*(100*k/2)
}
mlrECI = eci_bvc(calout=mlrvgamsmps4,preds=mlrpred,k=5) # MLR, 5 CATEGORIES


# ECI, ADAPTED FORMULA TO COMPARE WITH RANDOM MODEL – VERSION USED IN THIS PAPER

eci_rel <- function(calout,preds,k,outc){
  prevm=matrix((table(outc)/length(outc))[1:k],nrow=dim(preds)[1],ncol=k,byrow=T)
  ecir=mean((preds-prevm)*(preds-prevm))
  ecim=mean((preds-fitted(calout))*(preds-fitted(calout)))
  return(ecim/ecir)
}
mlrECIr = eci_rel(calout=mlrvgamsmps4,preds=mlrpred,k=5,outc=cad$o5) # MLR, 5 CATEGORIES


# CALIBRATION SCATTER PLOT PER OUTCOME CATEGORY BASED ON FLEXIBLE RECALIBRATION MODEL
# (MLR, 5 CATEGORIES)

plot(preds[,1],fitted(obs)[,1],type="p",pch=1,col="green",lwd=1,
     ylab="Observed proportion",xlab="Estimated probability", xlim=0:1 ,ylim=0:1)
points(preds[,2],fitted(obs)[,2],type="p",pch=1,col="orange")
points(preds[,3],fitted(obs)[,3],type="p",pch=1,col="red")
points(preds[,4],fitted(obs)[,4],type="p",pch=1,col="brown")
points(preds[,5],fitted(obs)[,5],type="p",pch=1,col="black")
lines(c(0,1),c(0,1),type="l",col="gray",lty=3) # plot the ideal diagonal line


# CALIBRATION CURVES BASED PER OUTCOME CATEGORY ON FLEXIBLE RECALIBRATION MODEL
# (MLR, 5 CATEGORIES)

wa1=smooth.spline(preds[,1],fitted(obs)[,1])
wa2=smooth.spline(preds[,2],fitted(obs)[,2])
wa3=smooth.spline(preds[,3],fitted(obs)[,3])
wa4=smooth.spline(preds[,4],fitted(obs)[,4])
wa5=smooth.spline(preds[,5],fitted(obs)[,5])
plot(wa1$x, wa1$y,type="l",col="green",ylab="Observed proportion",
     xlab="Estimated probability",xlim=0:1,ylim=0:1)
lines(wa2$x, wa2$y,col="orange")
lines(wa3$x, wa3$y,col="red")
lines(wa4$x, wa4$y,col="brown")
lines(wa5$x, wa5$y,col="black")
lines(c(0,1),c(0,1),type="l",col="gray",lty=3) # plot the ideal diagonal line


# CALIBRATION SCATTER PLOT PER OUTCOME DICHOTOMY BASED ON FLEXIBLE RECALIBRATION MODEL
# (MLR, 5 CATEGORIES)

plot(preds[,2]+preds[,3]+preds[,4]+preds[,5],
```

```
     fitted(obs)[,2]+fitted(obs)[,3]+fitted(obs)[,4]+fitted(obs)[,5],
     type="p",pch=1,col="orange",lwd=1,ylab="Observed proportion",
     xlab="Estimated probability",xlim=0:1,ylim=0:1)
points(preds[,3]+preds[,4]+preds[,5],fitted(obs)[,3]+fitted(obs)[,4]+fitted(obs)[,5],
       type="p",pch=1,col="red")
points(preds[,4]+preds[,5],fitted(obs)[,4]+fitted(obs)[,5],type="p",pch=1,col="brown")
points(preds[,5],fitted(obs)[,5],type="p",pch=1,col="black")
lines(c(0,1),c(0,1),type="l",col="gray",lty=3) # plot the ideal diagonal line


# CALIBRATION CURVES BASED PER OUTCOME CATEGORY ON FLEXIBLE RECALIBRATION MODEL
# (MLR, 5 CATEGORIES)

wa2=smooth.spline(preds[,2]+preds[,3]+preds[,4]+preds[,5],
                    fitted(obs)[,2]+fitted(obs)[,3]+fitted(obs)[,4]+fitted(obs)[,5])
wa3=smooth.spline(preds[,3]+preds[,4]+preds[,5],fitted(obs)[,3]+fitted(obs)[,4]+fitted(obs)[,5])
wa4=smooth.spline(preds[,4]+preds[,5],fitted(obs)[,4]+fitted(obs)[,5])
wa5=smooth.spline(preds[,5],fitted(obs)[,5])
plot(wa2$x, wa2$y,type="l",col="orange", ylab="Observed proportion",xlab="Estimated probability",
     xlim=0:1,ylim=0:1)
lines(wa3$x, wa3$y,col="red")
lines(wa4$x, wa4$y,col="brown")
lines(wa5$x, wa5$y,col="black")
lines(ref,ref,type="l",col="gray",lty=3) # plot the ideal diagonal line


# ORDINAL C-STATISTIC (ORC)

orc <- function(out,preds,k){
  library(DescTools) # Cstat
  Ec=preds
  for (i in (1:k)){
    Ec[,i]=i*Ec[,i]
  }
  E=rowSums(Ec)
  pwc = rep(NA,k*(k-1)*0.5)
  for (i in (2:k)){
    for (j in (1:(i-1))){
      pwc[((i-1)*(i-2)*0.5)+j]=Cstat(x=E[out==(j) | out==i],resp=out[out==(j) | out==i]==i) # c
statistic 1 vs 2
    }
  }
  mean(pwc)
}
mlrc = orc(dataset$y,mlrpred,5) # MLR ONLY, 5 CATEGORIES
```

## 3. Comparison of flexible recalibration models

The flexible recalibration model that we used in this work was based on the MLR framework (Equation 18). This may disadvantage the resulting calibration plots, and derived measures such as the ECI, for prediction models that were based on another type of model. It is perhaps impossible to propose a recalibration model that is fully model agnostic, but we can compare different setups to evaluate their impact on the results. The recalibration model should in any case not make a proportional odds assumption. We evaluated 6 alternative flexible recalibration models, by varying whether the model has an MLR or CR-NP setup and whether the $K - 1$ linear predictors compare every category with a reference category, compare every dichotomy, or compare a category with their complement (i.e. all other categories combined):

MLR, reference LP: $\log\left(\frac{P(Y=k)}{P(Y=1)}\right) = a_{flex,k} + \sum_{j=2}^{K} s_{k,j}\left(\hat{Z}_j\right),\ k = 2,\dots,K,\ \hat{Z}_j = \log\left(\hat{P}_j/\hat{P}_1\right)$

CR-NP, reference LP: $\log\left(\frac{P(Y>k)}{P(Y\geq k)}\right) = a_{flex,k}^{\sim} + \sum_{j=2}^{K} s_{k,j}^{\sim}\left(\hat{Z}_j\right),\ k = 1,\dots,K-1,\ \hat{Z}_j = \log\left(\hat{P}_j/\hat{P}_1\right)$

MLR, dichotomy LP: $\log\left(\frac{P(Y=k)}{P(Y=1)}\right) = a_{flex,k}' + \sum_{j=2}^{K} s_{k,j}'\left(\widehat{D}_j\right),\ k = 2,\dots,K,\ \widehat{D}_j = \operatorname{logit}\left(\hat{V}_j\right)$

CR-NP, dichotomy LP: $\log\left(\frac{P(Y>k)}{P(Y\geq k)}\right) = a_{flex,k}'' + \sum_{j=2}^{K} s_{k,j}''\left(\widehat{D}_j\right),\ k = 1,\dots,K-1,\ \widehat{D}_j = \operatorname{logit}\left(\hat{V}_j\right)$

MLR, category LP: $\log\left(\frac{P(Y=k)}{P(Y=1)}\right) = a_{flex,k}^{*} + \sum_{j=1}^{K-1} s_{k,j}^{*}\left(\hat{C}_j\right),\ k = 2,\dots,K,\ \hat{C}_j = \log\left(\hat{P}_j/\left(1-\hat{P}_j\right)\right)$

CR-NP, category LP: $\log\left(\frac{P(Y>k)}{P(Y\geq k)}\right) = a_{flex,k}^{**} + \sum_{j=1}^{K-1} s_{k,j}^{**}\left(\hat{C}_j\right),\ k = 1,\dots,K-1,\ \hat{C}_j = \log\left(\hat{P}_j/\left(1-\hat{P}_j\right)\right)$

Th first option is equal to Equation 18 from the main paper. Note that MLR equals the adjacent category approach without proportional odds (AC-NP). We did not include CL-NP models, because these may lead to invalid models. Indeed, when we tried to fit flexible recalibration models of the CL-NP type, we nearly always received error messages (for each of the three types of linear predictors). We applied these models to all simulation scenarios (using the large sample datasets) and the case study. We summarized calibration using the ECI (Tables A3.1-3). Differences between results for the six recalibration models were small, although linear predictors for a category vs its complement appear less appealing. Further, we constructed calibration scatter plots. This yielded 516 plots (480 for the simulation study and 36 for the case study). We only show plots for the MLR and CL-PO models from the case study. These plots confirm that differences between the six recalibration models were small. If the approach used in the main paper favors MLR models, the advantage is marginal.

Table A3.1. ECI values on large sample simulated datasets (n=200,000) using MLR truth.

| Simulation scenario | Model | MLR Ref LP | CR-NP Ref LP | MLR Dich LP | CR-NP Dich LP | MLR Cat LP | CR-NP Cat LP |
|---|---|---|---|---|---|---|---|
| MLR 1 | MLR | 0.0000 | 0.0001 | 0.0000 | 0.0001 | 0.0001 | 0.0001 |
| | CL-PO | 0.0057 | 0.0052 | 0.0060 | 0.0055 | 0.0052 | 0.0056 |
| | AC-PO | 0.0000 | 0.0001 | 0.0000 | 0.0001 | 0.0001 | 0.0001 |
| | SLM | 0.0000 | 0.0001 | 0.0000 | 0.0001 | 0.0001 | 0.0001 |
| MLR 2 | MLR | 0.0000 | 0.0000 | 0.0001 | 0.0000 | 0.0000 | 0.0000 |
| | CL-PO | 0.0098 | 0.0090 | 0.0099 | 0.0096 | 0.0097 | 0.0098 |
| | AC-PO | 0.0000 | 0.0000 | 0.0002 | 0.0000 | 0.0000 | 0.0000 |
| | SLM | 0.0000 | 0.0000 | 0.0001 | 0.0000 | 0.0000 | 0.0000 |
| MLR 3 | MLR | 0.0000 | 0.0009 | 0.0008 | 0.0007 | 0.0012 | 0.0004 |
| | CL-PO | 0.0493 | 0.0486 | 0.0490 | 0.0484 | 0.0483 | 0.0485 |
| | AC-PO | 0.0457 | 0.0455 | 0.0462 | 0.0459 | 0.0455 | 0.0460 |
| | SLM | 0.0000 | 0.0000 | 0.0000 | 0.0001 | 0.0000 | 0.0000 |
| MLR 4 | MLR | 0.0000 | 0.0004 | 0.0008 | 0.0004 | 0.0008 | 0.0003 |
| | CL-PO | 0.0324 | 0.0325 | 0.0326 | 0.0322 | 0.0326 | 0.0321 |
| | AC-PO | 0.0587 | 0.0583 | 0.0588 | 0.0584 | 0.0585 | 0.0584 |
| | SLM | 0.0000 | 0.0000 | 0.0000 | 0.0000 | 0.0000 | 0.0000 |
| MLR 5 | MLR | 0.0000 | 0.0019 | 0.0101 | 0.0006 | 0.0042 | 0.0001 |
| | CL-PO | 0.0128 | 0.0126 | 0.0130 | 0.0125 | 0.0127 | 0.0124 |
| | AC-PO | 0.0178 | 0.0175 | 0.0182 | 0.0176 | 0.0175 | 0.0176 |
| | SLM | 0.0000 | 0.0000 | 0.0000 | 0.0001 | 0.0000 | 0.0000 |
| MLR 6 | MLR | 0.0000 | 0.0032 | 0.0165 | 0.0004 | 0.0061 | 0.0002 |
| | CL-PO | 0.1185 | 0.1154 | 0.1184 | 0.1184 | 0.1188 | 0.1187 |
| | AC-PO | 0.0961 | 0.0957 | 0.0926 | 0.0963 | 0.0960 | 0.0963 |
| | SLM | 0.0000 | 0.0000 | 0.0001 | 0.0001 | 0.0000 | 0.0001 |
| MLR 7 | MLR | 0.0000 | 0.0018 | 0.0074 | 0.0009 | 0.0042 | 0.0002 |
| | CL-PO | 0.1560 | 0.1502 | 0.1560 | 0.1560 | 0.1556 | 0.1561 |
| | AC-PO | 0.1163 | 0.1164 | 0.1117 | 0.1169 | 0.1166 | 0.1169 |
| | SLM | 0.0000 | 0.0001 | 0.0003 | 0.0001 | 0.0001 | 0.0001 |
| MLR 8 | MLR | 0.0000 | 0.0017 | 0.0096 | 0.0011 | 0.0043 | 0.0002 |
| | CL-PO | 0.1845 | 0.1772 | 0.1841 | 0.1845 | 0.1838 | 0.1846 |
| | AC-PO | 0.1584 | 0.1598 | 0.1551 | 0.1601 | 0.1600 | 0.1601 |
| | SLM | 0.0000 | 0.0001 | 0.0001 | 0.0001 | 0.0001 | 0.0001 |
| MLR 9 | MLR | 0.0000 | 0.0003 | 0.0002 | 0.0002 | 0.0003 | 0.0001 |
| | CL-PO | 0.1060 | 0.1073 | 0.0986 | 0.1062 | 0.1082 | 0.1063 |
| | AC-PO | 0.1448 | 0.1408 | 0.1435 | 0.1423 | 0.1407 | 0.1426 |
| | SLM | 0.0480 | 0.0587 | 0.0558 | 0.0416 | 0.0588 | 0.0464 |
| MLR 10 | MLR | 0.0000 | 0.0001 | 0.0001 | 0.0002 | 0.0003 | 0.0002 |
| | CL-PO | 0.2362 | 0.2531 | 0.2275 | 0.2267 | 0.2526 | 0.2226 |
| | AC-PO | 0.2178 | 0.2298 | 0.2145 | 0.2023 | 0.2318 | 0.1714 |
| | SLM | 0.1482 | 0.1512 | 0.1478 | 0.1094 | 0.1475 | 0.1307 |
| MLR 11 | MLR | 0.0000 | 0.0004 | 0.0008 | 0.0004 | 0.0008 | 0.0002 |
| | CL-PO | 0.0324 | 0.0325 | 0.0326 | 0.0322 | 0.0326 | 0.0321 |
| | AC-PO | 0.0587 | 0.0583 | 0.0588 | 0.0584 | 0.0585 | 0.0584 |
| | SLM | 0.0000 | 0.0000 | 0.0000 | 0.0000 | 0.0000 | 0.0000 |
| | | | | | | | |
| Mean | MLR | 0.0000 | 0.0010 | 0.0042 | 0.0005 | 0.0020 | 0.0002 |
| | CL-PO | 0.0858 | 0.0858 | 0.0843 | 0.0847 | 0.0873 | 0.0844 |
| | AC-PO | 0.0831 | 0.0838 | 0.0818 | 0.0817 | 0.0841 | 0.0789 |
| | SLM | 0.0178 | 0.0191 | 0.0186 | 0.0138 | 0.0188 | 0.0161 |

ECI, estimated calibration index; MLR, multinomial logistic regression; CL-PO, cumulative logit model with proportional odds; AC-PO, adjacent category logit model with proportional odds; SLM, stereotype logit model; CR-NP, continuation ration logit model without proportional odds; Ref LP, linear predictors using a reference category; Dich LP, linear predictors based on dichotomize the outcome; Cat LP, linear predictors based on one category vs all other categories.

Table A3.2. ECI values on large sample simulated datasets (n=200,000) using CL-PO truth.

| Simulation scenario | Model | MLR Ref LP | CR-NP Ref LP | MLR Dich LP | CR-NP Dich LP | MLR Cat LP | CR-NP Cat LP |
|---|---|---|---|---|---|---|---|
| CLPO 1 | MLR | 0.0057 | 0.0068 | 0.0055 | 0.0073 | 0.0072 | 0.0070 |
| | CL-PO | 0.0000 | 0.0000 | 0.0001 | 0.0001 | 0.0001 | 0.0001 |
| | AC-PO | 0.0058 | 0.0068 | 0.0055 | 0.0072 | 0.0071 | 0.0069 |
| | SLM | 0.0058 | 0.0068 | 0.0055 | 0.0072 | 0.0070 | 0.0069 |
| CLPO 2 | MLR | 0.0054 | 0.0064 | 0.0057 | 0.0067 | 0.0068 | 0.0065 |
| | CL-PO | 0.0001 | 0.0001 | 0.0001 | 0.0001 | 0.0001 | 0.0001 |
| | AC-PO | 0.0122 | 0.0135 | 0.0128 | 0.0137 | 0.0132 | 0.0135 |
| | SLM | 0.0057 | 0.0063 | 0.0057 | 0.0066 | 0.0064 | 0.0064 |
| CLPO 3 | MLR | 0.0038 | 0.0042 | 0.0039 | 0.0045 | 0.0042 | 0.0044 |
| | CL-PO | 0.0001 | 0.0000 | 0.0001 | 0.0002 | 0.0001 | 0.0003 |
| | AC-PO | 0.0153 | 0.0156 | 0.0160 | 0.0159 | 0.0155 | 0.0158 |
| | SLM | 0.0038 | 0.0042 | 0.0024 | 0.0045 | 0.0042 | 0.0044 |
| CLPO 4 | MLR | 0.0069 | 0.0088 | 0.0067 | 0.0091 | 0.0090 | 0.0091 |
| | CL-PO | 0.0000 | 0.0001 | 0.0000 | 0.0000 | 0.0002 | 0.0000 |
| | AC-PO | 0.0172 | 0.0190 | 0.0169 | 0.0189 | 0.0187 | 0.0190 |
| | SLM | 0.0076 | 0.0088 | 0.0069 | 0.0090 | 0.0087 | 0.0090 |
| CLPO 5 | MLR | 0.0036 | 0.0055 | 0.0041 | 0.0056 | 0.0055 | 0.0056 |
| | CL-PO | 0.0001 | 0.0002 | 0.0001 | 0.0001 | 0.0003 | 0.0001 |
| | AC-PO | 0.0153 | 0.0181 | 0.0169 | 0.0183 | 0.0183 | 0.0182 |
| | SLM | 0.0039 | 0.0055 | 0.0038 | 0.0056 | 0.0055 | 0.0055 |
| CLPO 6 | MLR | 0.0030 | 0.0043 | 0.0027 | 0.0046 | 0.0044 | 0.0045 |
| | CL-PO | 0.0000 | 0.0002 | 0.0000 | 0.0001 | 0.0003 | 0.0001 |
| | AC-PO | 0.0244 | 0.0259 | 0.0251 | 0.0258 | 0.0259 | 0.0258 |
| | SLM | 0.0031 | 0.0045 | 0.0029 | 0.0045 | 0.0045 | 0.0044 |
| CLPO 7 | MLR | 0.0046 | 0.0050 | 0.0043 | 0.0050 | 0.0050 | 0.0051 |
| | CL-PO | 0.0001 | 0.0001 | 0.0001 | 0.0001 | 0.0001 | 0.0001 |
| | AC-PO | 0.0126 | 0.0130 | 0.0123 | 0.0130 | 0.0127 | 0.0130 |
| | SLM | 0.0047 | 0.0052 | 0.0042 | 0.0050 | 0.0051 | 0.0050 |
| CLPO 8 | MLR | 0.0075 | 0.0081 | 0.0073 | 0.0079 | 0.0082 | 0.0078 |
| | CL-PO | 0.0001 | 0.0001 | 0.0001 | 0.0001 | 0.0001 | 0.0001 |
| | AC-PO | 0.0179 | 0.0181 | 0.0171 | 0.0181 | 0.0180 | 0.0181 |
| | SLM | 0.0080 | 0.0082 | 0.0078 | 0.0083 | 0.0082 | 0.0082 |
| CLPO 9 | MLR | 0.0057 | 0.0068 | 0.0054 | 0.0073 | 0.0073 | 0.0070 |
| | CL-PO | 0.0000 | 0.0001 | 0.0001 | 0.0001 | 0.0001 | 0.0001 |
| | AC-PO | 0.0058 | 0.0068 | 0.0055 | 0.0072 | 0.0071 | 0.0069 |
| | SLM | 0.0058 | 0.0068 | 0.0055 | 0.0072 | 0.0070 | 0.0069 |
| | | | | | | | |
| Mean | MLR | 0.0051 | 0.0062 | 0.0051 | 0.0064 | 0.0064 | 0.0063 |
| | CL-PO | 0.0001 | 0.0001 | 0.0001 | 0.0001 | 0.0002 | 0.0001 |
| | AC-PO | 0.0141 | 0.0152 | 0.0142 | 0.0153 | 0.0152 | 0.0152 |
| | SLM | 0.0054 | 0.0063 | 0.0050 | 0.0064 | 0.0063 | 0.0063 |

ECI, estimated calibration index; MLR, multinomial logistic regression; CL-PO, cumulative logit model with proportional odds; AC-PO, adjacent category logit model with proportional odds; SLM, stereotype logit model; CR-NP, continuation ration logit model without proportional odds; Ref LP, linear predictors using a reference category; Dich LP, linear predictors based on dichotomize the outcome; Cat LP, linear predictors based on one category vs all other categories.

Table A3.3. ECI values for the case study (n=4884, apparent validation).

| Model | MLR Ref LP | CR-NP Ref LP | MLR Dich LP | CR-NP Dich LP | MLR Cat LP | CR-NP Cat LP |
|---|---|---|---|---|---|---|
| MLR | 0.0052 | 0.0078 | 0.0047 | 0.0068 | 0.0152 | 0.0124 |
| CL-PO | 0.0295 | 0.0328 | 0.0286 | 0.0314 | 0.0319 | 0.0335 |
| AC-PO | 0.1412 | 0.1490 | 0.1462 | 0.1479 | 0.1474 | 0.1513 |
| CR-PO | 0.1939 | 0.2008 | 0.1844 | 0.1826 | 0.1868 | 0.1896 |
| CR-NP | 0.0052 | 0.0056 | 0.0095 | 0.0067 | 18.3* | 0.0154 |
| SLM | 0.0044 | 0.0061 | 0.0090 | 0.0069 | 0.0084 | 0.0060 |

* Warning that “fitted probabilities numerically 0 or 1 occurred”.

ECI, estimated calibration index; MLR, multinomial logistic regression; CL-PO, cumulative logit model with proportional odds; AC-PO, adjacent category logit model with proportional odds; SLM, stereotype logit model; CR-PO, continuation ration logit model with proportional odds; CR-NP, continuation ration logit model without proportional odds; Ref LP, linear predictors using a reference category; Dich LP, linear predictors based on dichotomize the outcome; Cat LP, linear predictors based on one category vs all other categories.

Figure A3.1. Calibration scatter plots based on six different flexible recalibration models for the MLR model in the case study. Green: category 1 (no coronary artery disease); orange: category 2 (non-obstructive stenosis); red: category 3 (1-vessel disease); brown: category 4 (2-vessel disease); black: category 5 (3-vessel disease).

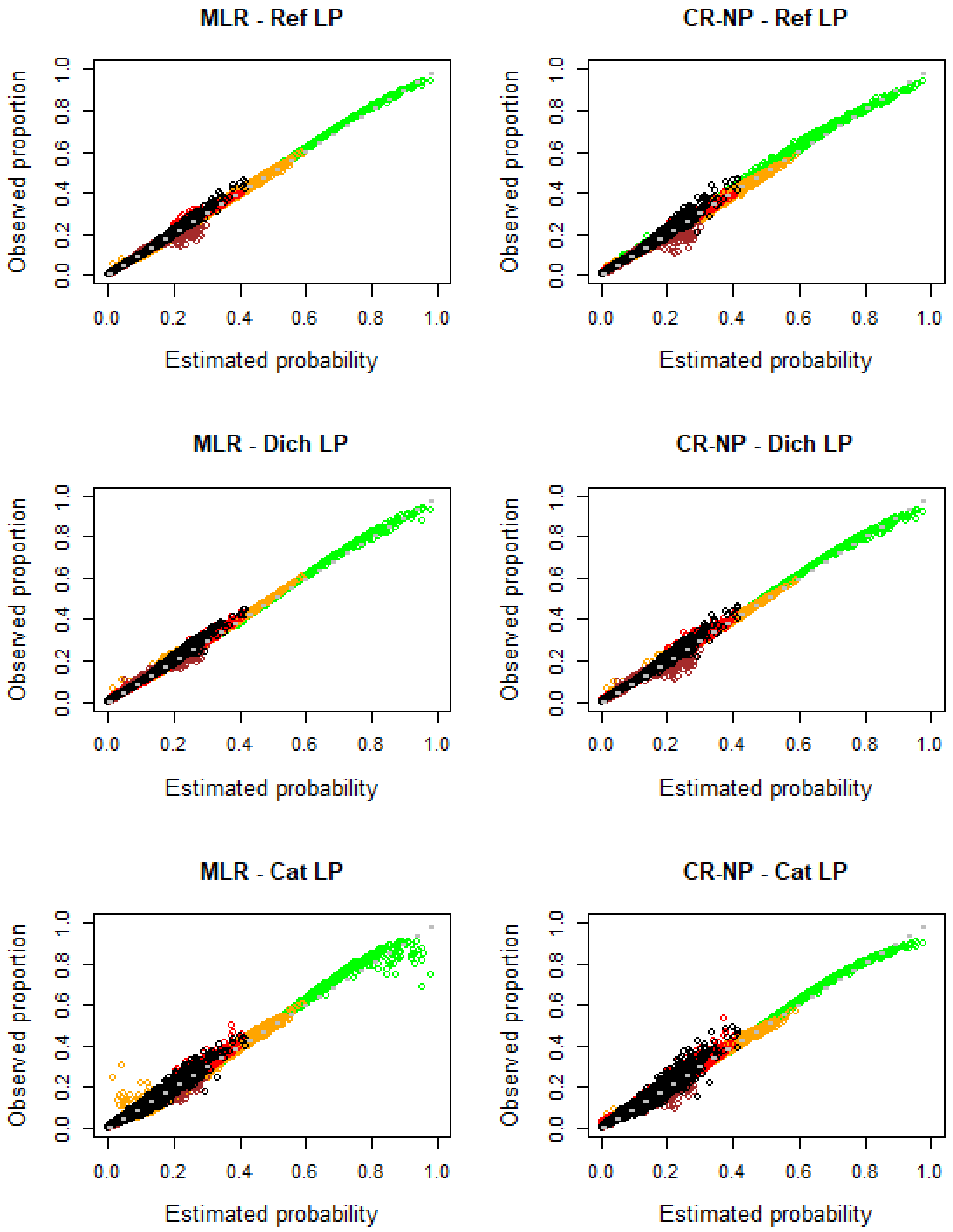

Figure A3.2. Calibration scatter plots based on six different flexible recalibration models for the CL-PO model in the case study. Green: category 1 (no coronary artery disease); orange: category 2 (non-obstructive stenosis); red: category 3 (1-vessel disease); brown: category 4 (2-vessel disease); black: category 5 (3-vessel disease).

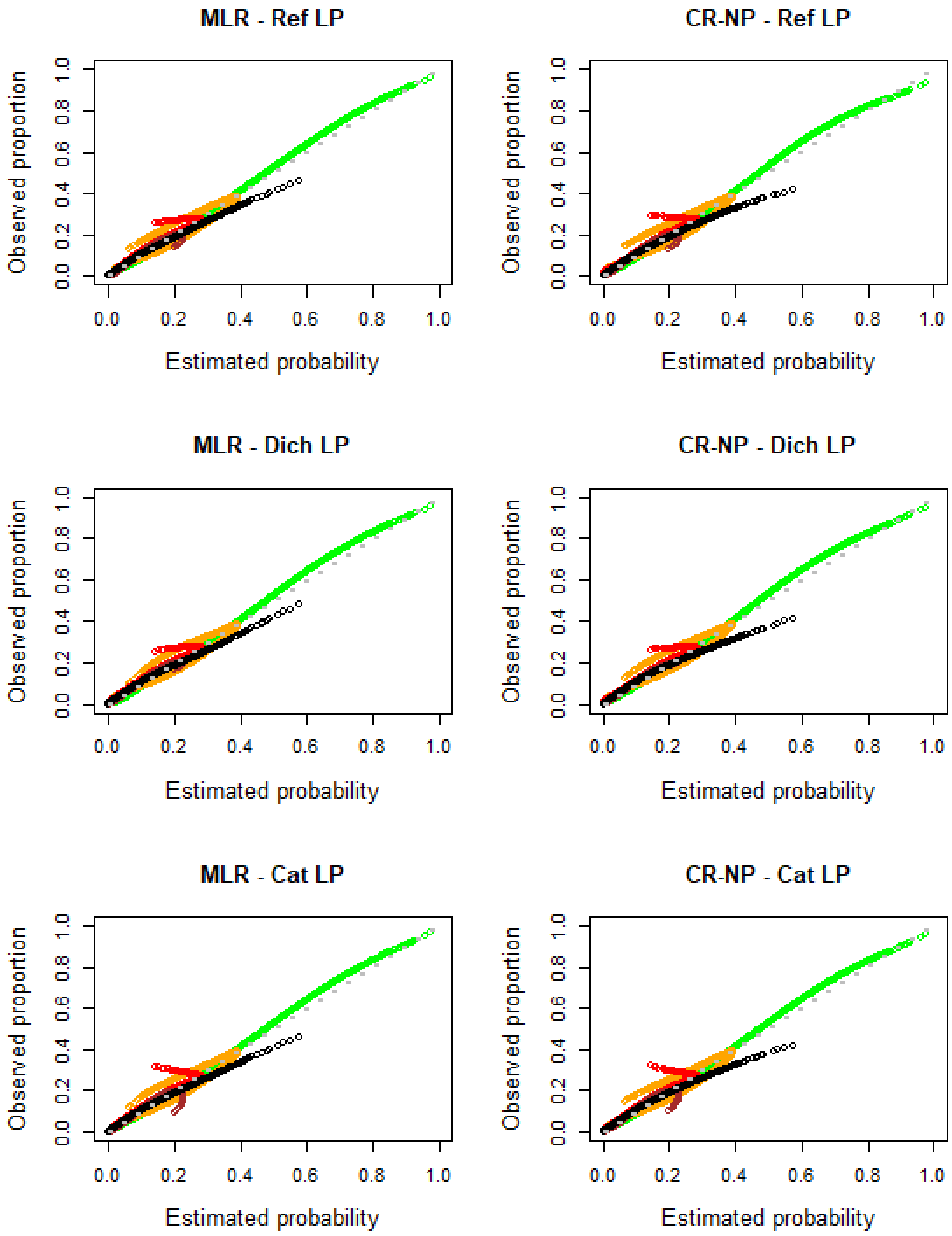

## 4. Supplementary tables

Table S1. Details of the simulation scenarios when the true model has an MLR form.

| **Scenario** | ***Q*** | ***K*** | **ORC** | **Outcome distribution** | **Means of $X_{p,k}$*** |
|---|---|---|---|---|---|
| 1 | 4 continuous | 3 | 0.74 | $\left(\frac{1}{3},\frac{1}{3},\frac{1}{3}\right)$ | $\boldsymbol{\mu}_{\mathbf{1k}} = (0.0, 0.4, 0.8)$<br>$\boldsymbol{\mu}_{\mathbf{2k}} = (0.0, 0.3, 0.6)$<br>$\boldsymbol{\mu}_{\mathbf{3k}} = (0.0, 0.4, 0.8)$<br>$\boldsymbol{\mu}_{\mathbf{4k}} = (0.0, 0.3, 0.6)$ |
| 2 | 4 continuous | 3 | 0.74 | $(0.55, 0.30, 0.15)$ | $\boldsymbol{\mu}_{\mathbf{1k}} = (0.0, 0.4, 0.8)$<br>$\boldsymbol{\mu}_{\mathbf{2k}} = (0.0, 0.3, 0.6)$<br>$\boldsymbol{\mu}_{\mathbf{3k}} = (0.0, 0.4, 0.8)$<br>$\boldsymbol{\mu}_{\mathbf{4k}} = (0.0, 0.3, 0.6)$ |
| 3 | 4 continuous | 3 | 0.74 | $\left(\frac{1}{3},\frac{1}{3},\frac{1}{3}\right)$ | $\boldsymbol{\mu}_{1k} = (0.0, 0.7, 0.8)$<br>$\boldsymbol{\mu}_{2k} = (0.0, 0.6, 0.6)$<br>$\boldsymbol{\mu}_{3k} = (0.0, 0.5, 0.8)$<br>$\boldsymbol{\mu}_{4k} = (0.0, 0.1, 0.6)$ |
| 4 | 4 continuous | 3 | 0.74 | $(0.55, 0.30, 0.15)$ | $\boldsymbol{\mu}_{1k} = (0.0, 0.7, 0.8)$<br>$\boldsymbol{\mu}_{2k} = (0.0, 0.6, 0.6)$<br>$\boldsymbol{\mu}_{3k} = (0.0, 0.5, 0.8)$<br>$\boldsymbol{\mu}_{4k} = (0.0, 0.1, 0.6)$ |
| 5 | 4 continuous | 3 | 0.74 | $(0.55, 0.30, 0.15)$ | $\boldsymbol{\mu}_{1k} = (0.0, 0.7, 0.8)$<br>$\boldsymbol{\mu}_{2k} = (0.0, 0.7, 0.6)$<br>$\boldsymbol{\mu}_{3k} = (0.0, 0.0, 1.0)$<br>$\boldsymbol{\mu}_{4k} = (0.3, 0.0, 0.3)$ |
| 6 | 3 continuous | 4 | 0.74 | $(0.40, 0.25, 0.20, 0.15)$ | $\boldsymbol{\mu}_{1k} = (0.0, 0.0, 1.0, 1.0)$<br>$\boldsymbol{\mu}_{2k} = (0.0, 0.8, 0.8, 0.9)$<br>$\boldsymbol{\mu}_{3k} = (0.2, 0.0, 0.9, 1.0)$ |
| 7 | 3 continuous | 4 | 0.66 | $(0.40, 0.25, 0.20, 0.15)$ | $\boldsymbol{\mu}_{1k} = (0.0, 0.0, 0.6, 0.6)$<br>$\boldsymbol{\mu}_{2k} = (0.0, 0.4, 0.4, 0.5)$<br>$\boldsymbol{\mu}_{3k} = (0.1, 0.0, 0.6, 0.7)$ |
| 8 | 3 continuous | 4 | 0.66 | $(0.45, 0.30, 0.20, 0.05)$ | $\boldsymbol{\mu}_{1k} = (0.0, 0.0, 0.6, 0.6)$<br>$\boldsymbol{\mu}_{2k} = (0.0, 0.4, 0.4, 0.5)$<br>$\boldsymbol{\mu}_{3k} = (0.1, 0.0, 0.6, 0.7)$ |
| 9 | 4 binary | 3 | 0.74 | $(0.55, 0.30, 0.15)$ | $\boldsymbol{\mu}_{1k} = (0.20, 0.55, 0.58)$<br>$\boldsymbol{\mu}_{2k} = (0.20, 0.50, 0.50)$<br>$\boldsymbol{\mu}_{3k} = (0.20, 0.45, 0.58)$<br>$\boldsymbol{\mu}_{4k} = (0.20, 0.25, 0.50)$ |
| 10 | 3 binary | 4 | 0.74 | $(0.40, 0.25, 0.20, 0.15)$ | $\boldsymbol{\mu}_{1k} = (0.20, 0.20, 0.65, 0.65)$<br>$\boldsymbol{\mu}_{2k} = (0.20, 0.40, 0.40, 0.60)$<br>$\boldsymbol{\mu}_{3k} = (0.25, 0.20, 0.60, 0.70)$ |
| 11 | 8 continuous (4 true + 4 noise) | 3 | 0.74 | $(0.55, 0.30, 0.15)$ | $\boldsymbol{\mu}_{1k} = (0.0, 0.7, 0.8)$<br>$\boldsymbol{\mu}_{2k} = (0.0, 0.6, 0.6)$<br>$\boldsymbol{\mu}_{3k} = (0.0, 0.5, 0.8)$<br>$\boldsymbol{\mu}_{4k} = (0.0, 0.1, 0.6)$<br>$\boldsymbol{\mu}_{5k} = (0.0, 0.0, 0.0)$<br>$\boldsymbol{\mu}_{6k} = (0.0, 0.0, 0.0)$<br>$\boldsymbol{\mu}_{7k} = (0.0, 0.0, 0.0)$<br>$\boldsymbol{\mu}_{8k} = (0.0, 0.0, 0.0)$ |

* For binary predictors, the means refer to the prevalences of the predictor for each outcome category.
MLR, multinomial logistic regression; ORC, ordinal C statistic.

Table S2. Details of the simulation scenarios when the true model has a CL-PO form.

| Scenario | Q | K | ORC | Outcome distribution | True model parameters |
|---|---|---|---|---|---|
| 1 | 4 continuous | 3 | 0.74 | $\left(\frac{1}{3},\frac{1}{3},\frac{1}{3}\right)$ | $\boldsymbol{\alpha_k} = [-0.18,1.55]^T$<br>$\boldsymbol{\beta} = [-0.55,-0.41,-0.55,-0.41]^T$ |
| 2 | 4 continuous | 3 | 0.74 | $(0.55, 0.30, 0.15)$ | $\boldsymbol{\alpha_k} = [0.92,2.80]^T$<br>$\boldsymbol{\beta} = [-0.53,-0.39,-0.53,-0.39]^T$ |
| 3 | 4 continuous | 3 | 0.74 | $(0.70, 0.25, 0.05)$ | $\boldsymbol{\alpha_k} = [1.73,4.15]^T$<br>$\boldsymbol{\beta} = [-0.53,-0.39,-0.53,-0.39]^T$ |
| 4 | 3 continuous | 4 | 0.74 | $(0.55, 0.30, 0.15)$ | $\boldsymbol{\alpha_k} = [-0.12,1.22,2.62]^T$<br>$\boldsymbol{\beta} = [-0.54,-0.47,-0.51]^T$ |
| 5 | 3 continuous | 4 | 0.66 | $(0.55, 0.30, 0.15)$ | $\boldsymbol{\alpha_k} = [-0.05,1.10,2.35]^T$<br>$\boldsymbol{\beta} = [-0.54,-0.47,-0.51]^T$ |
| 6 | 3 continuous | 4 | 0.66 | $(0.70, 0.25, 0.05)$ | $\boldsymbol{\alpha_k} = [-0.18,1.55]^T$<br>$\boldsymbol{\beta} = [-0.55,-0.41,-0.55,-0.41]^T$ |
| 7 | 4 binary | 3 | 0.74 | $(0.55, 0.30, 0.15)$ | $\boldsymbol{\alpha_k} = [-0.18,1.55]^T$<br>$\boldsymbol{\beta} = [-0.55,-0.41,-0.55,-0.41]^T$ |
| 8 | 3 binary | 4 | 0.74 | $(0.55, 0.30, 0.15)$ | $\boldsymbol{\alpha_k} = [-0.18,1.55]^T$<br>$\boldsymbol{\beta} = [-0.55,-0.41,-0.55,-0.41]^T$ |
| 9 | 8 continuous (4 true + 4 noise) | 3 | 0.74 | $\left(\frac{1}{3},\frac{1}{3},\frac{1}{3}\right)$ | $\boldsymbol{\alpha_k} = [-0.18,1.55]^T$<br>$\boldsymbol{\beta} = [-0.55,-0.41,-0.55,-0.41, 0,0,0,0]^T$ |

* For binary predictors, the means refer to the prevalences of the predictor for each outcome category.
CL-PO, cumulative logit model with proportional odds; ORC, ordinal C statistic.

Table S3. Large sample estimates of the model coefficients for scenarios 1-11 under MLR truth.

| **Model** <br> **Parameter** | **Scen. 1** | **Scen. 2** | **Scen. 3** | **Scen. 4** | **Scen. 5** | **Scen. 6** | **Scen. 7** | **Scen. 8** | **Scen. 9** | **Scen. 10** | **Scen. 11[a]** |
|---|---|---|---|---|---|---|---|---|---|---|---|
| MLR | | | | | | | | | | | |
| Int, k=2 vs k=1 | -0.25 | -0.86 | -0.56 | -1.17 | -1.06 | -0.77 | -0.55 | -0.48 | -2.09 | -0.70 | -1.17 |
| Int, k=3 vs k=1 | -1.00 | -2.30 | -1.00 | -2.30 | -2.30 | -1.91 | -1.13 | -1.26 | -3.55 | -2.44 | -2.30 |
| Int, k=4 vs k=1 | na | na | na | na | na | -2.36 | -1.53 | -2.73 | na | -3.45 | na |
| $X_1$, k=2 vs k=1 | 0.40 | 0.40 | 0.70 | 0.70 | 0.70 | 0.00 | 0.00 | 0.00 | 1.59 | 0.03 | 0.70 |
| $X_1$, k=3 vs k=1 | 0.80 | 0.79 | 0.80 | 0.79 | 0.79 | 1.00 | 0.60 | 0.59 | 1.74 | 2.03 | 0.79 |
| $X_1$, k=4 vs k=1 | na | na | na | na | na | 0.99 | 0.59 | 0.61 | na | 2.06 | na |
| $X_2$, k=2 vs k=1 | 0.29 | 0.30 | 0.59 | 0.60 | 0.70 | 0.80 | 0.40 | 0.40 | 1.38 | 0.97 | 0.60 |
| $X_2$, k=3 vs k=1 | 0.59 | 0.59 | 0.59 | 0.59 | 0.59 | 0.80 | 0.40 | 0.40 | 1.38 | 0.96 | 0.59 |
| $X_2$, k=4 vs k=1 | na | na | na | na | na | 0.89 | 0.49 | 0.46 | na | 1.77 | na |
| $X_3$, k=2 vs k=1 | 0.40 | 0.40 | 0.50 | 0.50 | 0.00 | -0.20 | -0.10 | -0.10 | 1.20 | -0.28 | 0.50 |
| $X_3$, k=3 vs k=1 | 0.80 | 0.80 | 0.80 | 0.79 | 1.00 | 0.70 | 0.50 | 0.50 | 1.71 | 1.50 | 0.79 |
| $X_3$, k=4 vs k=1 | na | na | na | na | na | 0.79 | 0.59 | 0.59 | na | 1.95 | na |
| $X_4$, k=2 vs k=1 | 0.30 | 0.30 | 0.10 | 0.10 | -0.30 | na | na | na | 0.28 | na | 0.10 |
| $X_4$, k=3 vs k=1 | 0.60 | 0.59 | 0.60 | 0.59 | -0.01 | na | na | na | 1.38 | na | 0.59 |
| CL-PO | | | | | | | | | | | |
| Int, k≥2 vs k=1 | 0.18 | -0.63 | 0.07 | -0.81 | -0.68 | -0.14 | 0.20 | 0.05 | -1.73 | -0.69 | -0.81 |
| Int, k≥3 vs k≤2 | -1.55 | -2.47 | -1.65 | -2.72 | -2.55 | -1.47 | -0.94 | -1.36 | -3.72 | -2.02 | -2.72 |
| Int, k≥4 vs k≤3 | na | na | na | na | na | -2.87 | -2.18 | -3.31 | na | -3.49 | na |
| $X_1$ | 0.55 | 0.53 | 0.53 | 0.62 | 0.63 | 0.59 | 0.40 | 0.36 | 1.36 | 1.43 | 0.62 |
| $X_2$ | 0.41 | 0.39 | 0.39 | 0.49 | 0.54 | 0.60 | 0.34 | 0.34 | 1.09 | 1.00 | 0.49 |
| $X_3$ | 0.55 | 0.53 | 0.54 | 0.55 | 0.48 | 0.43 | 0.37 | 0.29 | 1.20 | 1.17 | 0.55 |
| $X_4$ | 0.41 | 0.39 | 0.42 | 0.32 | -0.11 | na | na | na | 0.84 | na | 0.32 |
| AC-PO | | | | | | | | | | | |
| Int, k=2 vs k=1 | -0.26 | -0.86 | -0.31 | -0.95 | -0.87 | -0.68 | -0.54 | -0.48 | -1.62 | -1.01 | -0.95 |
| Int, k=3 vs k=2 | -0.75 | -1.44 | -0.80 | -1.62 | -1.48 | -0.76 | -0.42 | -0.62 | -2.39 | -1.08 | -1.62 |
| Int, k=4 vs k=3 | na | na | na | na | na | -1.16 | -0.63 | -1.73 | na | -1.52 | na |
| $X_1$ | 0.40 | 0.40 | 0.39 | 0.45 | 0.46 | 0.36 | 0.23 | 0.24 | 1.00 | 0.81 | 0.45 |
| $X_2$ | 0.30 | 0.30 | 0.28 | 0.35 | 0.38 | 0.33 | 0.17 | 0.20 | 0.80 | 0.57 | 0.35 |
| $X_3$ | 0.40 | 0.40 | 0.39 | 0.42 | 0.40 | 0.28 | 0.21 | 0.20 | 0.91 | 0.70 | 0.42 |
| $X_4$ | 0.30 | 0.30 | 0.31 | 0.26 | -0.07 | na | na | na | 0.67 | na | 0.26 |
| SLM | | | | | | | | | | | |
| Int, k=2 vs k=1 | -0.25 | -0.86 | -0.51 | -1.15 | -1.00 | -0.53 | -0.48 | -0.41 | -2.13 | -0.57 | -1.15 |
| Int, k=3 vs k=1 | -1.00 | -2.30 | -1.00 | -2.26 | -2.11 | -1.91 | -1.13 | -1.25 | -3.50 | -2.45 | -2.26 |
| Int, k=4 vs k=1 | na | na | na | na | na | -2.35 | -1.51 | -2.72 | na | -3.38 | na |
| $X_1$ | 0.40 | 0.40 | 0.59 | 0.64 | 0.62 | 0.17 | 0.05 | 0.05 | 1.48 | 0.18 | 0.64 |
| $X_2$ | 0.30 | 0.30 | 0.45 | 0.52 | 0.54 | 0.11 | 0.02 | 0.02 | 1.23 | 0.09 | 0.52 |
| $X_3$ | 0.40 | 0.40 | 0.55 | 0.55 | 0.40 | 0.13 | 0.05 | 0.04 | 1.27 | 0.16 | 0.55 |
| $X_4$ | 0.30 | 0.30 | 0.37 | 0.28 | -0.14 | na | na | na | 0.75 | na | 0.28 |
| □, k=2 vs k=1 | 1.00 | 1.00 | 1.00 | 1.00 | 1.00 | 1.00 | 1.00 | 1.00 | 1.00 | 1.00 | 1.00 |
| □, k=3 vs k=1 | 2.01 | 1.98 | 1.41 | 1.33 | 1.41 | 5.98 | 12.20 | 13.16 | 1.28 | 10.67 | 1.33 |
| □, k=4 vs k=1 | na | na | na | na | na | 6.35 | 13.43 | 14.50 | na | 13.06 | na |

MLR, multinomial logistic regression; CL-PO, cumulative logit model with proportional odds; AC-PO, adjacent category logit model with proportional odds; SLM, stereotype logit model; na, not applicable.

[a] Coefficients for the noise variables were all 0.00 for CL-PO, AC-PO, and SLM. For MLR, 6 out of 8 coefficients were 0.00, one was -0.01, and one was 0.01.

Table S4. Apparent performance based on a large dataset of n=200,000 for simulation scenarios 5 to 11 under MLR truth.

| | CALIBRATION INTERCEPTS AND SLOPES | | | | | | | | | | SINGLE NUMBER METRICS | | |
|---|---|---|---|---|---|---|---|---|---|---|---|---|---|
| | Per outcome category | | | | Per outcome dichotomy | | | Model-specific | | | | | |
| **MODEL** | **Y=1** | **Y=2** | **Y=3** | **Y=4** | **Y>1** | **Y>2** | **Y>3** | **LP1** | **LP2** | **LP3** | **ECI** | **rMSPE** | **ORC** |
| **MLR truth scenario 5: K=3, Q=4, imbalanced outcome, highly non-equidistant means, ORC 0.74** | | | | | | | | | | | | | |
| MLR | 0.00 / 1.00 | 0.00 / 1.00 | 0.00 / 1.00 | na | 0.00 / 1.00 | 0.00 / 1.00 | na | 0.00 / 1.00 | 0.00 / 1.00 | na | 0.000 | 0.002 | 0.740 |
| CL-PO | -0.01 / 1.06 | 0.00 / 1.02 | 0.01 / 0.94 | na | 0.01 / 1.06 | 0.01 / 0.94 | na | 0.00 / 1.00 | 0.00 / 1.00 | na | 0.013 | 0.100 | 0.730 |
| AC-PO | 0.00 / 1.09 | 0.00 / 1.26 | 0.00 / 0.87 | na | 0.00 / 1.09 | 0.00 / 0.87 | na | 0.00 / 1.00 | 0.00 / 1.00 | na | 0.018 | 0.103 | 0.734 |
| SLM | 0.00 / 1.00 | 0.00 / 1.00 | 0.00 / 1.00 | na | 0.00 / 1.00 | 0.00 / 1.00 | na | 0.00 / 1.00 | 0.00 / 1.00 | na | 0.000 | 0.095 | 0.724 |
| **MLR truth scenario 6: K=4, Q=3, imbalanced outcome, highly non-equidistant means, ORC 0.74** | | | | | | | | | | | | | |
| MLR | 0.00 / 1.00 | 0.00 / 1.00 | 0.00 / 1.00 | 0.00 / 1.00 | 0.00 / 1.00 | 0.00 / 1.00 | 0.00 / 1.00 | 0.00 / 1.00 | 0.00 / 1.00 | 0.00 / 1.00 | 0.000 | 0.002 | 0.741 |
| CL-PO | 0.03 / 0.94 | -0.04 / 0.66 | -0.03 / 1.66 | 0.04 / 0.89 | -0.03 / 0.94 | 0.00 / 1.37 | 0.04 / 0.89 | 0.00 / 1.00 | 0.00 / 1.00 | 0.00 / 1.00 | 0.119 | 0.098 | 0.735 |
| AC-PO | 0.00 / 0.94 | 0.00 / 1.17 | 0.00 / 1.73 | 0.00 / 0.79 | 0.00 / 0.94 | 0.00 / 1.26 | 0.00 / 0.79 | 0.00 / 1.00 | 0.00 / 1.00 | 0.00 / 1.00 | 0.096 | 0.096 | 0.737 |
| SLM | 0.00 / 1.00 | 0.00 / 1.01 | 0.00 / 1.00 | 0.00 / 1.00 | 0.00 / 1.00 | 0.00 / 1.00 | 0.00 / 1.00 | 0.00 / 1.00 | 0.00 / 1.00 | 0.00 / 1.00 | 0.000 | 0.088 | 0.737 |
| **MLR truth scenario 7: K=4, Q=3, imbalanced outcome, highly non-equidistant means, ORC 0.66** | | | | | | | | | | | | | |
| MLR | 0.00 / 1.00 | 0.00 / 1.00 | 0.00 / 1.00 | 0.00 / 1.00 | 0.00 / 1.00 | 0.00 / 1.00 | 0.00 / 1.00 | 0.00 / 1.00 | 0.00 / 1.00 | 0.00 / 1.00 | 0.000 | 0.002 | 0.663 |
| CL-PO | 0.02 / 0.86 | -0.01 / 0.30 | -0.03 / 1.63 | 0.01 / 1.01 | -0.02 / 0.86 | -0.01 / 1.33 | 0.01 / 1.01 | 0.00 / 1.00 | 0.00 / 1.00 | 0.00 / 1.00 | 0.156 | 0.058 | 0.662 |
| AC-PO | 0.00 / 0.87 | 0.00 / 1.48 | 0.00 / 1.68 | 0.00 / 0.88 | 0.00 / 0.87 | 0.00 / 1.23 | 0.00 / 0.88 | 0.00 / 1.00 | 0.00 / 1.00 | 0.00 / 1.00 | 0.116 | 0.055 | 0.663 |
| SLM | 0.00 / 1.00 | 0.00 / 1.00 | 0.00 / 1.00 | 0.00 / 1.00 | 0.00 / 1.00 | 0.00 / 1.00 | 0.00 / 1.00 | 0.00 / 1.00 | 0.00 / 1.00 | 0.00 / 1.00 | 0.000 | 0.046 | 0.663 |
| **MLR truth scenario 8: K=4, Q=3, highly imbalanced outcome, highly non-equidistant means, ORC 0.66** | | | | | | | | | | | | | |
| MLR | 0.00 / 1.00 | 0.00 / 1.00 | 0.00 / 1.00 | 0.00 / 1.00 | 0.00 / 1.00 | 0.00 / 1.00 | 0.00 / 1.00 | 0.00 / 1.00 | 0.00 / 1.00 | 0.00 / 1.00 | 0.000 | 0.002 | 0.661 |
| CL-PO | 0.02 / 0.81 | -0.01 / -0.08 | -0.03 / 1.46 | 0.01 / 1.12 | -0.02 / 0.81 | -0.02 / 1.39 | 0.01 / 1.12 | 0.00 / 1.00 | 0.00 / 1.00 | 0.00 / 1.00 | 0.185 | 0.064 | 0.659 |
| AC-PO | 0.00 / 0.83 | 0.00 / -0.01 | 0.00 / 1.47 | 0.00 / 0.85 | 0.00 / 0.83 | 0.00 / 1.28 | 0.00 / 0.85 | 0.00 / 1.00 | 0.00 / 1.00 | 0.00 / 1.00 | 0.158 | 0.061 | 0.661 |
| SLM | 0.00 / 1.00 | 0.00 / 1.00 | 0.00 / 1.00 | 0.00 / 1.00 | 0.00 / 1.00 | 0.00 / 1.00 | 0.00 / 1.00 | 0.00 / 1.00 | 0.00 / 1.00 | 0.00 / 1.00 | 0.000 | 0.052 | 0.662 |
| **MLR truth scenario 9: K=3, Q=4 binary, imbalanced outcome, non-equidistant means, ORC 0.74** | | | | | | | | | | | | | |
| MLR | 0.00 / 1.00 | 0.00 / 1.00 | 0.00 / 1.00 | na | 0.00 / 1.00 | 0.00 / 1.00 | na | 0.00 / 1.00 | 0.00 / 1.00 | na | 0.000 | 0.002 | 0.745 |
| CL-PO | -0.03 / 1.15 | 0.01 / 1.20 | 0.04 / 0.81 | na | 0.03 / 1.15 | 0.04 / 0.81 | na | 0.00 / 1.00 | 0.00 / 1.00 | na | 0.106 | 0.066 | 0.742 |
| AC-PO | 0.00 / 1.22 | 0.00 / 1.58 | 0.00 / 0.74 | na | 0.00 / 1.22 | 0.00 / 0.74 | na | 0.00 / 1.00 | 0.00 / 1.00 | na | 0.145 | 0.074 | 0.742 |
| SLM | 0.00 / 1.00 | 0.00 / 1.01 | 0.00 / 0.99 | na | 0.00 / 1.00 | 0.00 / 0.99 | na | 0.00 / 1.00 | 0.00 / 1.00 | na | 0.048 | 0.051 | 0.742 |
| **MLR truth scenario 10: K=4, Q=3 binary, imbalanced outcome, highly non-equidistant means, ORC 0.74** | | | | | | | | | | | | | |
| MLR | 0.00 / 1.00 | 0.00 / 1.00 | 0.00 / 1.00 | 0.00 / 1.00 | 0.00 / 1.00 | 0.00 / 1.00 | 0.00 / 1.00 | 0.00 / 1.00 | 0.00 / 1.00 | 0.00 / 1.00 | 0.000 | 0.002 | 0.742 |
| CL-PO | 0.06 / 0.81 | -0.02 / 0.90 | -0.06 / 1.51 | 0.02 / 1.03 | -0.06 / 0.81 | -0.04 / 1.34 | 0.02 / 1.03 | 0.00 / 1.00 | 0.00 / 1.00 | 0.00 / 1.00 | 0.236 | 0.077 | 0.742 |
| AC-PO | 0.00 / 0.83 | 0.00 / 1.60 | 0.00 / 1.53 | 0.00 / 0.93 | 0.00 / 0.83 | 0.00 / 1.24 | 0.00 / 0.93 | 0.00 / 1.00 | 0.00 / 1.00 | 0.00 / 1.00 | 0.218 | 0.071 | 0.742 |
| SLM | 0.00 / 0.99 | 0.00 / 1.05 | 0.00 / 1.01 | 0.00 / 0.99 | 0.00 / 0.99 | 0.00 / 1.00 | 0.00 / 0.99 | 0.00 / 1.00 | 0.00 / 1.00 | 0.00 / 1.00 | 0.148 | 0.058 | 0.742 |
| **MLR truth scenario 11: K=3, Q=8 continuous (4 true + 4 noise), imbalanced outcome, non-equidistant means, ORC 0.74** | | | | | | | | | | | | | |
| MLR | 0.00 / 1.00 | 0.00 / 1.00 | 0.00 / 1.00 | na | 0.00 / 1.00 | 0.00 / 1.00 | na | 0.00 / 1.00 | 0.00 / 1.00 | na | 0.000 | 0.003 | 0.737 |
| CL-PO | -0.02 / 1.11 | 0.01 / 1.13 | 0.03 / 0.85 | na | 0.02 / 1.11 | 0.03 / 0.85 | na | 0.00 / 1.00 | 0.00 / 1.00 | na | 0.032 | 0.058 | 0.735 |
| AC-PO | 0.00 / 1.17 | 0.00 / 1.47 | 0.00 / 0.77 | na | 0.00 / 1.17 | 0.00 / 0.77 | na | 0.00 / 1.00 | 0.00 / 1.00 | na | 0.059 | 0.064 | 0.736 |
| SLM | 0.00 / 1.00 | 0.00 / 1.00 | 0.00 / 1.00 | na | 0.00 / 1.00 | 0.00 / 1.00 | na | 0.00 / 1.00 | 0.00 / 1.00 | na | 0.000 | 0.047 | 0.733 |

MLR, multinomial logistic regression; CL-PO, cumulative logit model with proportional odds; AC-PO, adjacent category logit model with proportional odds; SLM, stereotype logit model; LP, linear predictor; ECI, estimated calibration index; rMSPE, root mean squared prediction error; ORC, ordinal C statistic; CAD, coronary artery disease; na, not applicable.

Table S5. Apparent performance based on a large dataset of n=200,000 for simulation scenarios 4 to 9 under CL-PO truth.

| | **CALIBRATION INTERCEPTS AND SLOPES** | | | | | | | | | | **SINGLE NUMBER METRICS** | | |
|---|---|---|---|---|---|---|---|---|---|---|---|---|---|
| | **Per outcome category** | | | | **Per outcome dichotomy** | | | **Model-specific** | | | | | |
| **MODEL** | **Y=1** | **Y=2** | **Y=3** | **Y=4** | **Y>1** | **Y>2** | **Y>3** | **LP1** | **LP2** | **LP3** | **ECI** | **rMSPE** | **ORC** |
| **CL-PO truth scenario 4: K=4, Q=3, imbalanced outcome, ORC 0.74** | | | | | | | | | | | | | |
| MLR | 0.00 / 0.99 | 0.00 / 1.34 | 0.00 / 1.05 | 0.00 / 0.98 | 0.00 / 0.99 | 0.00 / 1.00 | 0.00 / 0.98 | 0.00 / 1.00 | 0.00 / 1.00 | 0.00 / 1.00 | 0.007 | 0.014 | 0.741 |
| CL-PO | 0.00 / 1.00 | 0.00 / 1.00 | 0.00 / 1.00 | 0.00 / 1.00 | 0.00 / 1.00 | 0.00 / 1.00 | 0.00 / 1.00 | 0.00 / 1.00 | 0.00 / 1.00 | 0.00 / 1.00 | 0.000 | 0.001 | 0.741 |
| AC-PO | 0.00 / 1.09 | 0.00 / 1.28 | 0.00 / 1.03 | 0.00 / 0.92 | 0.00 / 1.09 | 0.00 / 0.95 | 0.00 / 0.92 | 0.00 / 1.00 | 0.00 / 1.00 | 0.00 / 1.00 | 0.017 | 0.020 | 0.741 |
| SLM | 0.00 / 0.99 | 0.00 / 1.34 | 0.00 / 1.05 | 0.00 / 0.98 | 0.00 / 0.99 | 0.00 / 1.00 | 0.00 / 0.98 | 0.00 / 1.00 | 0.00 / 1.00 | 0.00 / 1.00 | 0.008 | 0.014 | 0.741 |
| **CL-PO truth scenario 5: K=4, Q=3, imbalanced outcome, ORC 0.66** | | | | | | | | | | | | | |
| MLR | 0.00 / 1.00 | 0.00 / 1.39 | 0.00 / 1.02 | 0.00 / 0.99 | 0.00 / 1.00 | 0.00 / 1.00 | 0.00 / 0.99 | 0.00 / 1.00 | 0.00 / 1.00 | 0.00 / 1.00 | 0.004 | 0.008 | 0.661 |
| CL-PO | 0.00 / 1.00 | 0.00 / 0.98 | 0.00 / 1.00 | 0.00 / 1.00 | 0.00 / 1.00 | 0.00 / 1.00 | 0.00 / 1.00 | 0.00 / 1.00 | 0.00 / 1.00 | 0.00 / 1.00 | 0.000 | 0.001 | 0.661 |
| AC-PO | 0.00 / 1.09 | 0.00 / 1.20 | 0.00 / 1.06 | 0.00 / 0.91 | 0.00 / 1.09 | 0.00 / 0.97 | 0.00 / 0.91 | 0.00 / 1.00 | 0.00 / 1.00 | 0.00 / 1.00 | 0.015 | 0.014 | 0.661 |
| SLM | 0.00 / 1.00 | 0.00 / 1.39 | 0.00 / 1.02 | 0.00 / 0.99 | 0.00 / 1.00 | 0.00 / 1.00 | 0.00 / 0.99 | 0.00 / 1.00 | 0.00 / 1.00 | 0.00 / 1.00 | 0.004 | 0.008 | 0.661 |
| **CL-PO truth scenario 6: K=4, Q=3, highly imbalanced outcome, ORC 0.66** | | | | | | | | | | | | | |
| MLR | 0.00 / 1.00 | 0.00 / 1.15 | 0.00 / 1.00 | 0.00 / 0.98 | 0.00 / 1.00 | 0.00 / 1.00 | 0.00 / 0.98 | 0.00 / 1.00 | 0.00 / 1.00 | 0.00 / 1.00 | 0.003 | 0.007 | 0.660 |
| CL-PO | 0.00 / 1.00 | 0.00 / 1.01 | 0.00 / 1.00 | 0.00 / 0.99 | 0.00 / 1.00 | 0.00 / 1.00 | 0.00 / 0.99 | 0.00 / 1.00 | 0.00 / 1.00 | 0.00 / 1.00 | 0.000 | 0.001 | 0.660 |
| AC-PO | 0.00 / 1.11 | 0.00 / 1.65 | 0.00 / 1.02 | 0.00 / 0.77 | 0.00 / 1.11 | 0.00 / 0.95 | 0.00 / 0.77 | 0.00 / 1.00 | 0.00 / 1.00 | 0.00 / 1.00 | 0.024 | 0.014 | 0.660 |
| SLM | 0.00 / 1.00 | 0.00 / 1.15 | 0.00 / 1.00 | 0.00 / 0.98 | 0.00 / 1.00 | 0.00 / 1.00 | 0.00 / 0.98 | 0.00 / 1.00 | 0.00 / 1.00 | 0.00 / 1.00 | 0.003 | 0.007 | 0.660 |
| **CL-PO truth scenario 7: K=3, Q=4 binary, imbalanced outcome, ORC 0.74** | | | | | | | | | | | | | |
| MLR | 0.00 / 1.00 | 0.00 / 1.07 | 0.00 / 0.99 | na | 0.00 / 1.00 | 0.00 / 0.99 | na | 0.00 / 1.00 | 0.00 / 1.00 | na | 0.005 | 0.013 | 0.742 |
| CL-PO | 0.00 / 1.00 | 0.00 / 1.01 | 0.00 / 0.99 | na | 0.00 / 1.00 | 0.00 / 0.99 | na | 0.00 / 1.00 | 0.00 / 1.00 | na | 0.000 | 0.002 | 0.742 |
| AC-PO | 0.00 / 1.07 | 0.00 / 1.32 | 0.00 / 0.88 | na | 0.00 / 1.07 | 0.00 / 0.88 | na | 0.00 / 1.00 | 0.00 / 1.00 | na | 0.013 | 0.018 | 0.742 |
| SLM | 0.00 / 1.00 | 0.00 / 1.07 | 0.00 / 0.99 | na | 0.00 / 1.00 | 0.00 / 0.99 | na | 0.00 / 1.00 | 0.00 / 1.00 | na | 0.005 | 0.013 | 0.742 |
| **CL-PO truth scenario 8: K=4, Q=3 binary, imbalanced outcome, ORC 0.74** | | | | | | | | | | | | | |
| MLR | 0.00 / 0.99 | 0.00 / 1.35 | 0.00 / 1.03 | 0.00 / 0.98 | 0.00 / 0.99 | 0.00 / 1.00 | 0.00 / 0.98 | 0.00 / 1.00 | 0.00 / 1.00 | 0.00 / 1.00 | 0.007 | 0.014 | 0.740 |
| CL-PO | 0.00 / 1.00 | 0.00 / 1.02 | 0.00 / 0.99 | 0.00 / 1.00 | 0.00 / 1.00 | 0.00 / 1.00 | 0.00 / 1.00 | 0.00 / 1.00 | 0.00 / 1.00 | 0.00 / 1.00 | 0.000 | 0.001 | 0.740 |
| AC-PO | 0.00 / 1.09 | 0.00 / 1.30 | 0.00 / 1.02 | 0.00 / 0.91 | 0.00 / 1.09 | 0.00 / 0.95 | 0.00 / 0.91 | 0.00 / 1.00 | 0.00 / 1.00 | 0.00 / 1.00 | 0.018 | 0.020 | 0.740 |
| SLM | 0.00 / 0.99 | 0.00 / 1.38 | 0.00 / 1.03 | 0.00 / 0.98 | 0.00 / 0.99 | 0.00 / 1.00 | 0.00 / 0.98 | 0.00 / 1.00 | 0.00 / 1.00 | 0.00 / 1.00 | 0.008 | 0.014 | 0.740 |
| **CL-PO truth scenario 9: K=3, Q=8 continuous (4 true + 4 noise), balanced outcome, ORC 0.74** | | | | | | | | | | | | | |
| MLR | 0.00 / 0.99 | 0.00 / 1.38 | 0.00 / 0.99 | na | 0.00 / 0.99 | 0.00 / 0.99 | na | 0.00 / 1.00 | 0.00 / 1.00 | na | 0.006 | 0.014 | 0.740 |
| CL-PO | 0.00 / 1.00 | 0.00 / 1.02 | 0.00 / 0.99 | na | 0.00 / 1.00 | 0.00 / 0.99 | na | 0.00 / 1.00 | 0.00 / 1.00 | na | 0.000 | 0.003 | 0.740 |
| AC-PO | 0.00 / 1.00 | 0.00 / 1.38 | 0.00 / 0.99 | na | 0.00 / 1.00 | 0.00 / 0.99 | na | 0.00 / 1.00 | 0.00 / 1.00 | na | 0.006 | 0.014 | 0.740 |
| SLM | 0.00 / 0.99 | 0.00 / 1.38 | 0.00 / 0.99 | na | 0.00 / 0.99 | 0.00 / 0.99 | na | 0.00 / 1.00 | 0.00 / 1.00 | na | 0.006 | 0.014 | 0.740 |

MLR, multinomial logistic regression; CL-PO, cumulative logit model with proportional odds; AC-PO, adjacent category logit model with proportional odds; SLM, stereotype logit model; LP, linear predictor; ECI, estimated calibration index; rMSPE, root mean squared prediction error; ORC, ordinal C statistic; CAD, coronary artery disease; na, not applicable.

Table S6. Large sample estimates of the model coefficients for scenarios 1-8 under CL-PO truth.

| **Model** Parameter | **Scen. 1** | **Scen. 2** | **Scen. 3** | **Scen. 4** | **Scen. 5** | **Scen. 6** | **Scen. 7** | **Scen. 8** | **Scen. 9[a]** |
|---|---|---|---|---|---|---|---|---|---|
| MLR | | | | | | | | | |
| Int, k=2 vs k=1 | -0.27 | -1.14 | -1.81 | -0.51 | -0.65 | -0.64 | -2.07 | -1.66 | -0.27 |
| Int, k=3 vs k=1 | -0.99 | -2.58 | -4.18 | -1.07 | -1.10 | -1.38 | -4.19 | -2.98 | -0.99 |
| Int, k=4 vs k=1 | na | na | na | -2.00 | -1.74 | -3.03 | na | -4.74 | na |
| $X_1$, k=2 vs k=1 | 0.41 | 0.45 | 0.50 | 0.37 | 0.36 | 0.40 | 1.19 | 1.18 | 0.41 |
| $X_1$, k=3 vs k=1 | 0.79 | 0.75 | 0.75 | 0.62 | 0.60 | 0.70 | 1.95 | 2.03 | 0.79 |
| $X_1$, k=4 vs k=1 | na | na | na | 0.90 | 0.83 | 0.87 | na | 2.89 | na |
| $X_2$, k=2 vs k=1 | 0.29 | 0.32 | 0.37 | 0.33 | 0.31 | 0.36 | 0.93 | 0.82 | 0.29 |
| $X_2$, k=3 vs k=1 | 0.58 | 0.55 | 0.56 | 0.54 | 0.52 | 0.60 | 1.56 | 1.38 | 0.58 |
| $X_2$, k=4 vs k=1 | na | na | na | 0.78 | 0.72 | 0.76 | na | 2.02 | na |
| $X_3$, k=2 vs k=1 | 0.39 | 0.45 | 0.50 | 0.36 | 0.34 | 0.39 | 1.03 | 1.06 | 0.39 |
| $X_3$, k=3 vs k=1 | 0.78 | 0.76 | 0.75 | 0.59 | 0.57 | 0.64 | 1.73 | 1.71 | 0.78 |
| $X_3$, k=4 vs k=1 | na | na | na | 0.85 | 0.79 | 0.81 | na | 2.42 | na |
| $X_4$, k=2 vs k=1 | 0.29 | 0.32 | 0.35 | na | na | na | 0.69 | na | 0.29 |
| $X_4$, k=3 vs k=1 | 0.58 | 0.54 | 0.55 | na | na | na | 1.17 | Na | 0.57 |
| CL-PO | | | | | | | | | |
| Int, k≥2 vs k=1 | 0.17 | -0.92 | -1.73 | 0.13 | 0.05 | -0.18 | -2.03 | -1.52 | 0.17 |
| Int, k≥3 vs k≤2 | -1.56 | -2.79 | -4.16 | -1.22 | -1.10 | -1.63 | -3.94 | -2.88 | -1.56 |
| Int, k≥4 vs k≤3 | na | na | na | -2.62 | -2.35 | -3.58 | na | -4.30 | na |
| $X_1$ | 0.55 | 0.53 | 0.53 | 0.54 | 0.54 | 0.54 | 1.39 | 1.75 | 0.55 |
| $X_2$ | 0.41 | 0.39 | 0.40 | 0.47 | 0.47 | 0.47 | 1.09 | 1.20 | 0.41 |
| $X_3$ | 0.55 | 0.54 | 0.53 | 0.51 | 0.51 | 0.50 | 1.22 | 1.46 | 0.55 |
| $X_4$ | 0.41 | 0.38 | 0.38 | na | na | na | 0.82 | na | 0.41 |
| AC-PO | | | | | | | | | |
| Int, k=2 vs k=1 | -0.26 | -1.06 | -1.69 | -0.48 | -0.61 | -0.58 | -1.87 | -1.41 | -0.26 |
| Int, k=3 vs k=2 | -0.73 | -1.66 | -2.96 | -0.61 | -0.48 | -0.77 | -2.54 | -1.52 | -0.73 |
| Int, k=4 vs k=3 | na | na | na | -1.01 | -0.72 | -1.91 | na | -1.96 | na |
| $X_1$ | 0.39 | 0.39 | 0.44 | 0.30 | 0.28 | 0.33 | 1.03 | 0.97 | 0.39 |
| $X_2$ | 0.29 | 0.28 | 0.33 | 0.26 | 0.24 | 0.28 | 0.81 | 0.68 | 0.29 |
| $X_3$ | 0.39 | 0.40 | 0.44 | 0.29 | 0.27 | 0.31 | 0.90 | 0.83 | 0.39 |
| $X_4$ | 0.29 | 0.28 | 0.32 | na | na | na | 0.61 | na | 0.29 |
| SLM | | | | | | | | | |
| Int, k=2 vs k=1 | -0.27 | -1.14 | -1.81 | -0.51 | -0.65 | -0.64 | -2.07 | -1.65 | -0.27 |
| Int, k=3 vs k=1 | -0.99 | -2.58 | -4.18 | -1.07 | -1.10 | -1.38 | -4.19 | -2.99 | -0.99 |
| Int, k=4 vs k=1 | na | na | na | -2.00 | -1.74 | -3.03 | na | -4.75 | na |
| $X_1$ | 0.40 | 0.45 | 0.50 | 0.38 | 0.36 | 0.41 | 1.17 | 1.21 | 0.40 |
| $X_2$ | 0.29 | 0.33 | 0.37 | 0.33 | 0.31 | 0.36 | 0.93 | 0.84 | 0.29 |
| $X_3$ | 0.40 | 0.45 | 0.50 | 0.36 | 0.34 | 0.38 | 1.03 | 1.02 | 0.40 |
| $X_4$ | 0.29 | 0.32 | 0.36 | na | na | na | 0.70 | na | 0.29 |
| $\phi$, k=2 vs k=1 | 1.00 | 1.00 | 1.00 | 1.00 | 1.00 | 1.00 | 1.00 | 1.00 | 1.00 |
| $\phi$, k=3 vs k=1 | 1.98 | 1.67 | 1.52 | 1.66 | 1.67 | 1.69 | 1.67 | 1.68 | 1.98 |
| $\phi$, k=4 vs k=1 | na | na | na | 2.39 | 2.33 | 2.12 | na | 2.40 | na |

MLR, multinomial logistic regression; CL-PO, cumulative logit model with proportional odds; AC-PO, adjacent category logit model with proportional odds; SLM, stereotype logit model; na, not applicable.

[a] For MLR, 5 out of 8 coefficients for the noise variables were 0.00, three were (-)0.01. For CL-PO, AC-PO, and SLM, 3 out of 4 coefficients for the noise variables were 0.00, one was -0.01.

Table S7. Descriptive statistics for the CARDIIGAN cohort (n=4888). The statistics are based on the imputed dataset. The amount of missing values that had to be imputed is stated as well.

| **Predictor** | **All patients (n=4888)** | **No CAD (n=1381, 28%)** | **Non-obstructive stenosis (n=1606, 33%)** | **1-vessel disease (n=997, 20%)** | **2-vessel disease (n=475, 10%)** | **3-vessel disease 429 (9%)** | **Missing, n (%)** |
|---|---|---|---|---|---|---|---|
| Age, years | 64 (10.8), range 18-89 | 59 (11.2) | 66 (9.7) | 65 (10.3) | 67 (9.9) | 67 (10.7) | 0 |
| HDL cholesterol | 57 (17.4), range 15-188 | 61 (19.0) | 57 (17.3) | 54 (15.5) | 52 (15.4) | 52 (15.9) | 312 |
| LDL cholesterol | 128 (37.7), range 21-341 | 126 (36.2) | 126 (35.7) | 130 (39.5) | 129 (38.2) | 134 (40.1) | 310 |
| Fibrinogen | 380 (120), range 97-1414 | 355 (104) | 379 (112) | 393 (133) | 396 (129) | 413 (139) | 119 |
| Male sex | 3028 (62%) | 621 (45%) | 968 (60%) | 729 (73%) | 373 (79%) | 337 (79%) | 0 |
| Chest pain | 2987 (61%) | 749 (54%) | 927 (58%) | 663 (66%) | 334 (70%) | 314 (73%) | 0 |
| Diabetes mellitus | 757 (15%) | 124 (9%) | 264 (16%) | 167 (17%) | 94 (20%) | 108 (25%) | 0 |
| Hypertension | 4090 (84%) | 1073 (78%) | 1376 (86%) | 833 (84%) | 428 (90%) | 380 (89%) | 0 |
| Dyslipidemia | 3591 (73%) | 947 (69%) | 1163 (72%) | 743 (75%) | 396 (83%) | 342 (80) | 0 |
| Ever smoked | 2317 (47%) | 600 (43%) | 746 (46%) | 501 (50%) | 244 (51%) | 226 (53%) | 640 |
| CRP > 1 mg/dl | 676 (14%) | 142 (10%) | 201 (13%) | 178 (18%) | 76 (16%) | 79 (18%) | 96 |

Results are presented as mean (standard deviation) or n (%).
CAD, coronary artery disease; CRP, c-reactive protein.

Table S8. P-values for the likelihood ratio test of the proportional odds assumption in the CL-PO model.

| **Predictor** | **p-value** |
|---|---|
| Age | <0.0001 |
| HDL cholesterol | 0.60 |
| LDL cholesterol | 0.010 |
| Log(Fibrinogen) | 0.22 |
| Male sex | 0.46 |
| Chest pain | 0.13 |
| Diabetes mellitus | 0.017 |
| Hypertension | 0.0024 |
| Dyslipidemia | 0.016 |
| Ever smoked | 0.99 |
| C-reactive protein > 1 mg/dl | 0.13 |

Table S9. Coefficients for the models in the case study.

| Model | Int | Age | Sex | Ch.pain | Diab | Hypert | Dyslip | Smok | HDL | LDL | Log(fib) | CRP>1 | Phi |
|---|---|---|---|---|---|---|---|---|---|---|---|---|---|
| MLR | | | | | | | | | | | | | |
| k=2 vs k=1 | -7.60 | 0.070 | 0.85 | 0.19 | 0.50 | 0.26 | 0.26 | 0.32 | -0.012 | 0.0019 | 0.46 | -0.053 | na |
| k=3 vs k=1 | -10.78 | 0.072 | 1.46 | 0.64 | 0.56 | 0.083 | 0.38 | 0.31 | -0.019 | 0.0057 | 0.76 | 0.32 | na |
| k=4 vs k=1 | -14.04 | 0.10 | 1.83 | 0.79 | 0.68 | 0.53 | 1.03 | 0.34 | -0.030 | 0.0044 | 0.81 | 0.12 | na |
| k=5 vs k=1 | -17.85 | 0.10 | 1.90 | 0.95 | 1.07 | 0.34 | 0.64 | 0.42 | -0.028 | 0.0098 | 1.27 | 0.13 | na |
| CL-PO | | | | | | | | | | | | | |
| k≥2 vs k=1 | -7.23 | 0.058 | 1.12 | 0.54 | 0.46 | 0.18 | 0.43 | 0.19 | -0.017 | 0.0046 | 0.56 | 0.15 | na |
| k≥3 vs k≤2 | -8.90 | | | | | | | | | | | | |
| k≥4 vs k≤3 | -10.08 | | | | | | | | | | | | |
| k=5 vs k≤4 | -11.02 | | | | | | | | | | | | |
| AC-PO | | | | | | | | | | | | | |
| k=2 vs k=1 | -3.78 | 0.026 | 0.54 | 0.27 | 0.23 | 0.10 | 0.22 | 0.087 | -0.0083 | 0.0024 | 0.29 | 0.053 | na |
| k=3 vs k=2 | -4.60 | | | | | | | | | | | | |
| k=4 vs k=3 | -5.03 | | | | | | | | | | | | |
| k=5 vs k=4 | -4.54 | | | | | | | | | | | | |
| CR-PO | | | | | | | | | | | | | |
| k>2 vs k≥1 | -5.56 | 0.041 | 0.84 | 0.41 | 0.38 | 0.15 | 0.34 | 0.15 | -0.013 | 0.0039 | 0.49 | 0.076 | na |
| k>3 vs k≥2 | -6.62 | | | | | | | | | | | | |
| k>4 vs k≥3 | -7.06 | | | | | | | | | | | | |
| k>5 vs k≥4 | -7.20 | | | | | | | | | | | | |
| CR-NP | | | | | | | | | | | | | |
| k>2 vs k≥1 | -8.85 | 0.076 | 1.21 | 0.44 | 0.59 | 0.24 | 0.40 | 0.33 | -0.017 | 0.0039 | 0.65 | 0.10 | na |
| k>3 vs k≥2 | -4.40 | 0.015 | 0.78 | 0.55 | 0.21 | -0.030 | 0.33 | 0.032 | -0.011 | 0.0040 | 0.42 | 0.29 | na |
| k>4 vs k≥3 | -4.47 | 0.028 | 0.39 | 0.21 | 0.32 | 0.37 | 0.50 | 0.078 | -0.011 | 0.0013 | 0.29 | -0.20 | na |
| k>5 vs k≥4 | -4.04 | 0.0054 | 0.11 | 0.14 | 0.43 | -0.16 | -0.37 | 0.080 | 0.0026 | 0.0059 | 0.47 | 0.056 | na |
| SLM | | | | | | | | | | | | | |
| k=2 vs k=1 | -7.31 | 0.053 | 0.98 | 0.45 | 0.44 | 0.18 | 0.38 | 0.20 | -0.014 | 0.0038 | 0.54 | 0.11 | 1.00 |
| k=3 vs k=1 | -10.83 | | | | | | | | | | | | 1.39 |
| k=4 vs k=1 | -15.33 | | | | | | | | | | | | 1.85 |
| k=5 vs k=1 | -16.83 | | | | | | | | | | | | 2.02 |

Int, intercept; ch.pain, chest pain; diab, diabetes; hypert, hypertension; dyslip, dyslipidemia; smok, ever smoked; HDL, HDL cholesterol; LDL, LDL cholesterol; log(fib), log of fibrinogen; CRP, c-reactive protein.

MLR, multinomial logistic regression; CL-PO, cumulative logit model with proportional odds; AC-PO, adjacent category logit model with proportional odds; CR-PO, continuation ratio logit model with proportional odds; CR-NP, continuation ration logit model without proportional odds; SLM, stereotype logit model; na, not applicable.

Table S10. Mean and range of predicted probabilities for the case study.

| Model | No CAD | Outcome category<br>Non-obstructive stenosis | One-vessel disease | Two-vessel disease | Three-vessel disease |
|---|---|---|---|---|---|
| MLR | 0.28 (0.01-0.98) | 0.33 (0.02-0.59) | 0.20 (<.01-0.41) | 0.10 (<.01-0.30) | 0.09 (<.01-0.42) |
| CL-PO | 0.29 (0.02-0.98) | 0.33 (0.02-0.39) | 0.20 (<.01-0.29) | 0.10 (<.01-0.23) | 0.09 (<.01-0.57) |
| AC-PO | 0.28 (0.01-0.87) | 0.33 (0.05-0.38) | 0.20 (0.01-0.26) | 0.10 (<.01-0.23) | 0.09 (<.01-0.58) |
| CR-PO | 0.28 (0.04-0.93) | 0.34 (0.07-0.40) | 0.20 (<.01-0.25) | 0.09 (<.01-0.17) | 0.09 (<.01-0.63) |
| CR-NP | 0.28 (0.01-0.99) | 0.33 (0.01-0.58) | 0.20 (<.01-0.40) | 0.10 (<.01-0.30) | 0.09 (<.01-0.40) |
| SLM | 0.28 (0.01-0.98) | 0.33 (0.02-0.38) | 0.20 (<.01-0.27) | 0.10 (<.01-0.29) | 0.09 (<.01-0.37) |

CAD, coronary artery disease; MLR, multinomial logistic regression; CL-PO, cumulative logit model with proportional odds; AC-PO, adjacent category logit model with proportional odds; CR-PO, continuation ratio logit model with proportional odds; CR-NP, continuation ration logit model without proportional odds; SLM, stereotype logit model

## 5. Supplementary figures

Figure S1. Flexible smoothed calibration curves per outcome category for simulation scenario 1 when the true model has the form of a multinomial logistic regression (green for category 1, orange for category 2, red for category 3). These curves are based on the dataset used to develop the model and are therefore apparent (or unvalidated) curves (n=200,000). For some models, lines overlap.

MLR, multinomial logistic regression; CL-PO, cumulative logit model with proportional odds; AC-PO, adjacent category logit model with proportional odds; SLM, stereotype logit model.

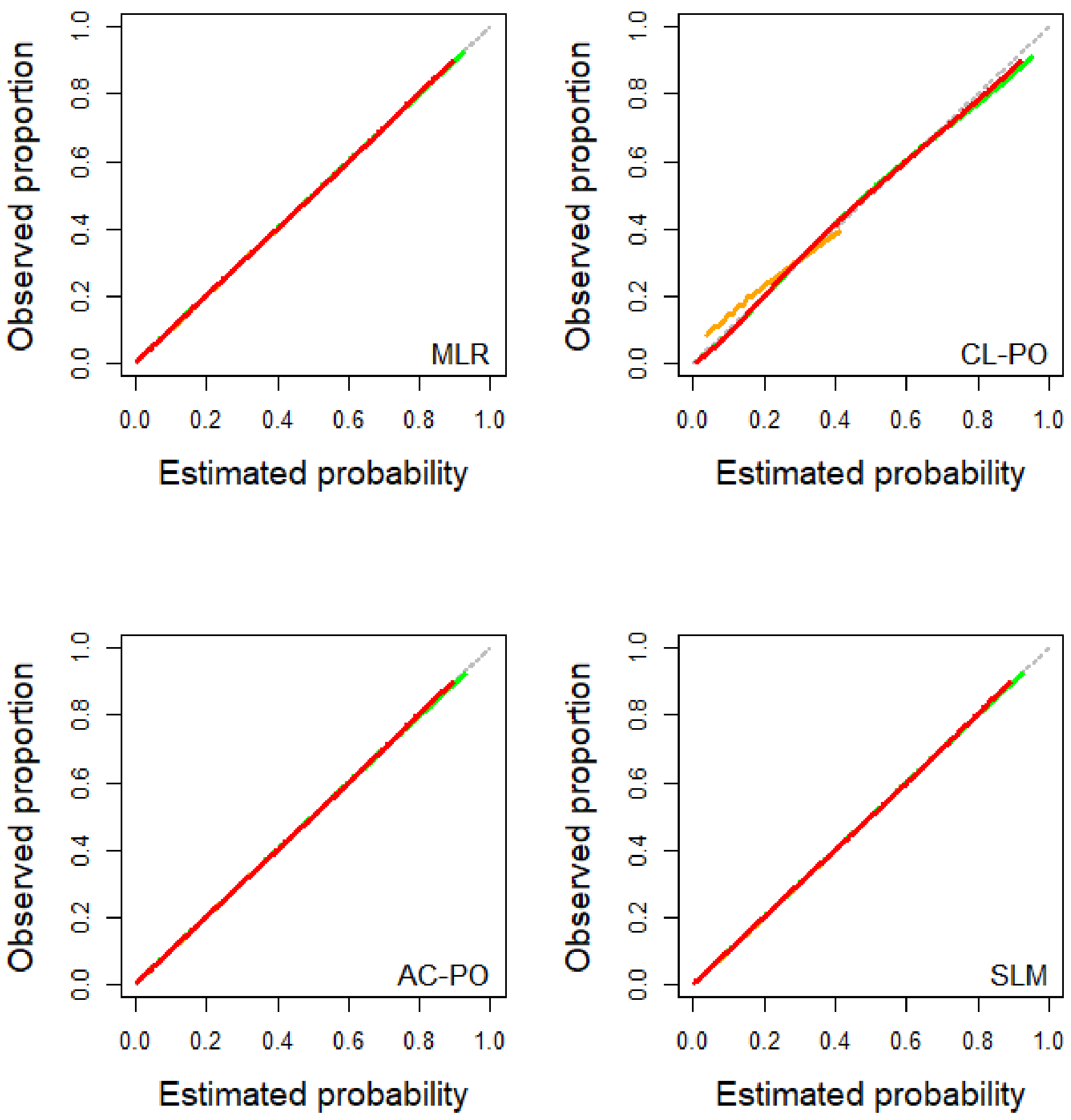

Figure S2. Flexible smoothed calibration curves per outcome category for simulation scenario 2 when the true model has the form of a multinomial logistic regression (green for category 1, orange for category 2, red for category 3). These curves are based on the dataset used to develop the model and are therefore apparent (or unvalidated) curves (n=200,000). For some models, lines overlap.

MLR, multinomial logistic regression; CL-PO, cumulative logit model with proportional odds; AC-PO, adjacent category logit model with proportional odds; SLM, stereotype logit model.

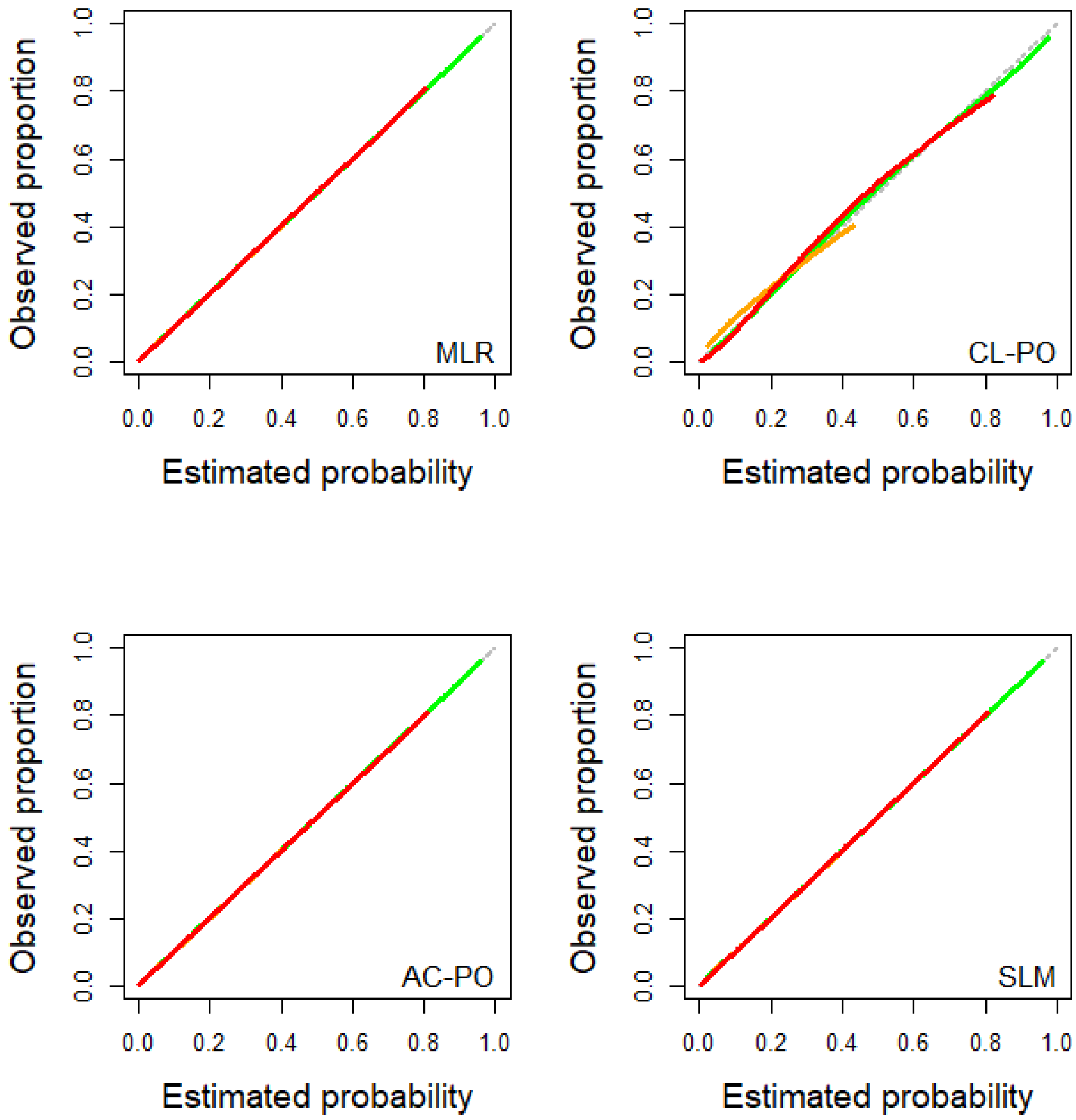

Figure S3. Flexible smoothed calibration curves per outcome category for simulation scenario 3 when the true model has the form of a multinomial logistic regression (green for category 1, orange for category 2, red for category 3). These curves are based on the dataset used to develop the model and are therefore apparent (or unvalidated) curves (n=200,000). For some models, lines overlap.

MLR, multinomial logistic regression; CL-PO, cumulative logit model with proportional odds; AC-PO, adjacent category logit model with proportional odds; SLM, stereotype logit model.

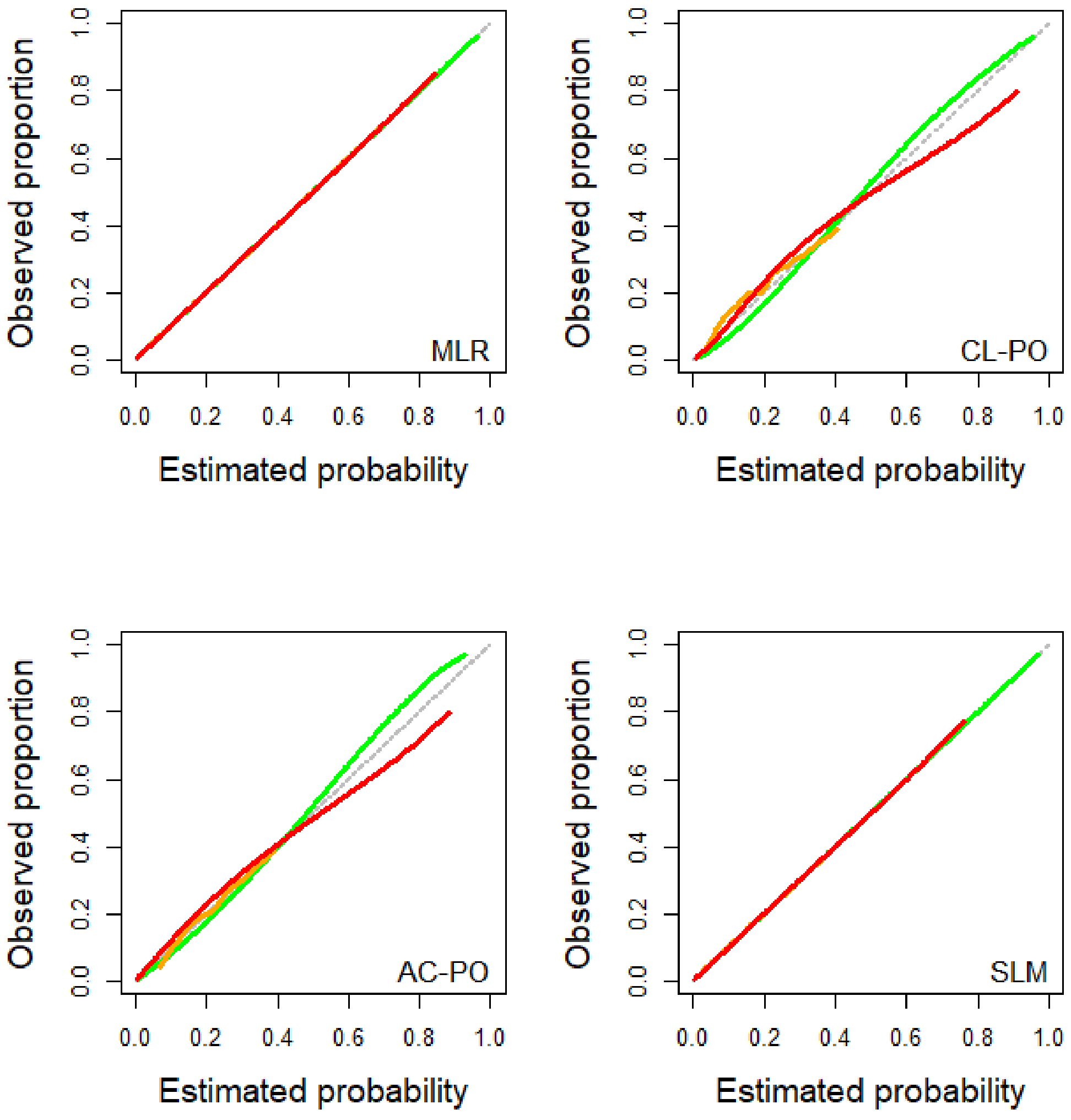

Figure S4. Flexible smoothed calibration curves per outcome category for simulation scenario 4 when the true model has the form of a multinomial logistic regression (green for category 1, orange for category 2, red for category 3). These curves are based on the dataset used to develop the model and are therefore apparent (or unvalidated) curves (n=200,000). For some models, lines overlap.

MLR, multinomial logistic regression; CL-PO, cumulative logit model with proportional odds; AC-PO, adjacent category logit model with proportional odds; SLM, stereotype logit model.

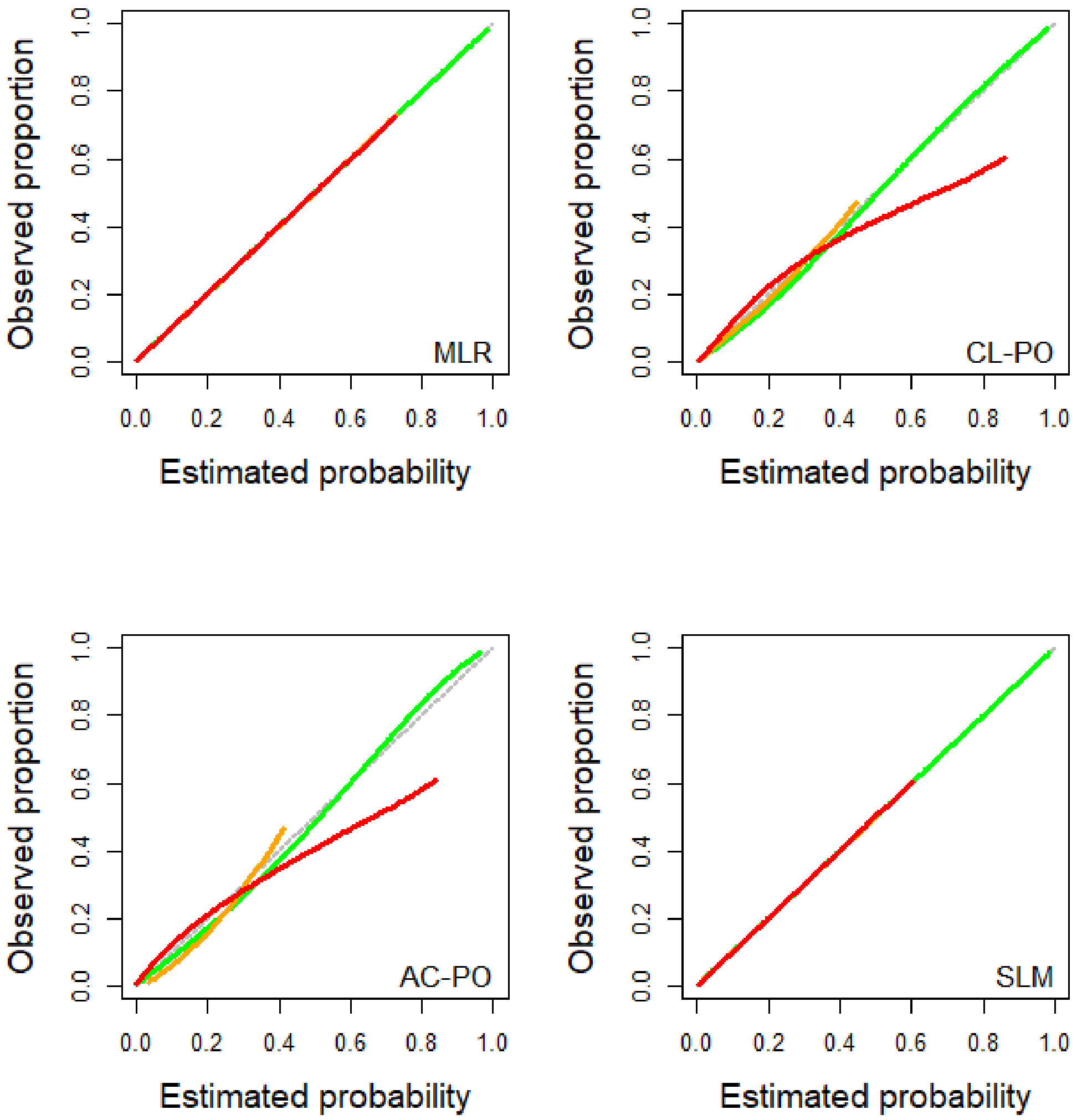

Figure S5. Scatter plots of true risks versus estimated risks for simulation scenario 5 when the true model has the form of a multinomial logistic regression. The plots are based on a random subset of 1,000 cases from all 200,000 cases.

MLR, multinomial logistic regression; CL-PO, cumulative logit model with proportional odds; AC-PO, adjacent category logit model with proportional odds; SLM, stereotype logit model.

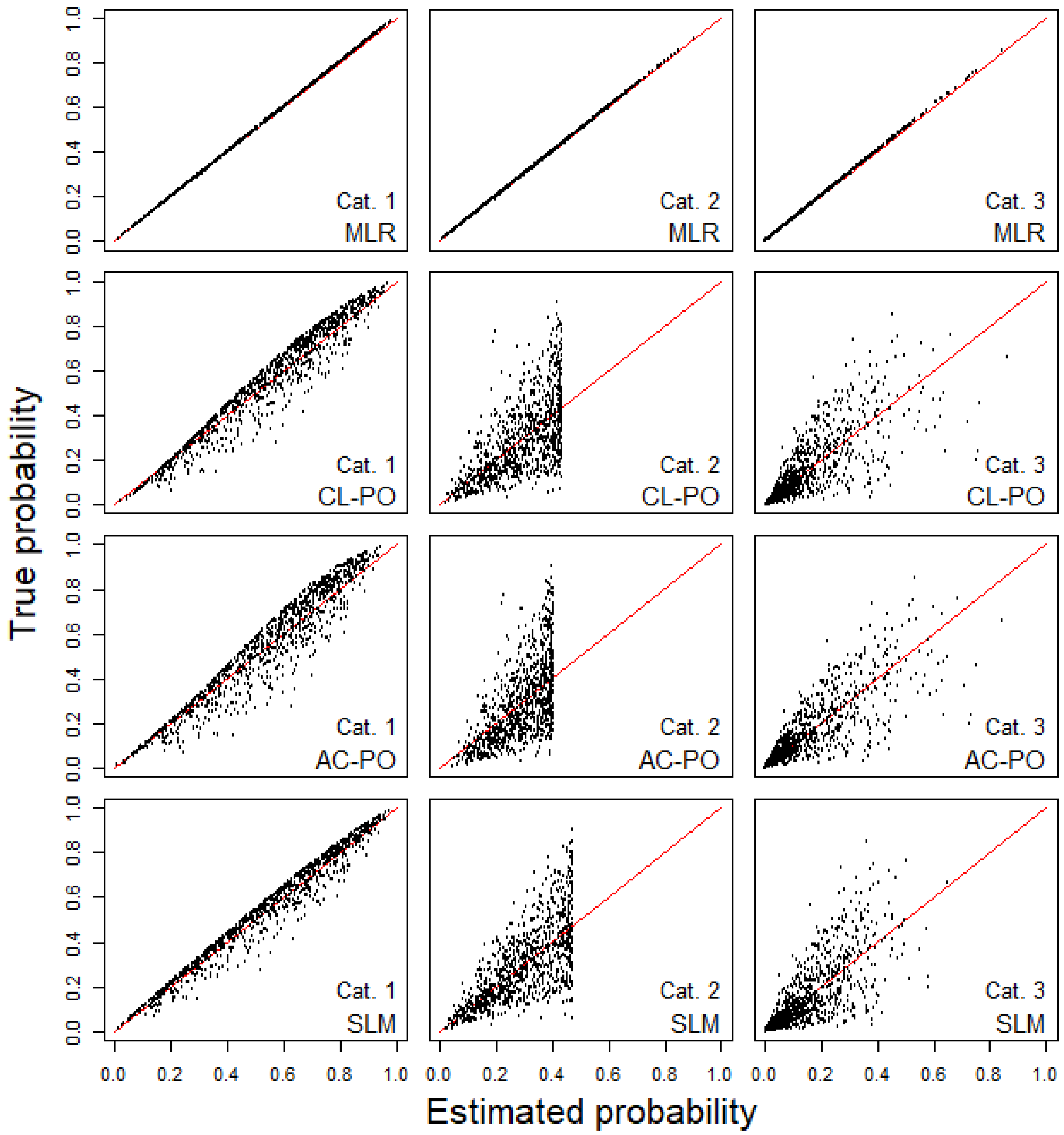

Figure S6. Scatter plots of true risks versus estimated risks for simulation scenario 6 when the true model has the form of a multinomial logistic regression. The plots are based on a random subset of 1,000 cases from all 200,000 cases.

MLR, multinomial logistic regression; CL-PO, cumulative logit model with proportional odds; AC-PO, adjacent category logit model with proportional odds; SLM, stereotype logit model.

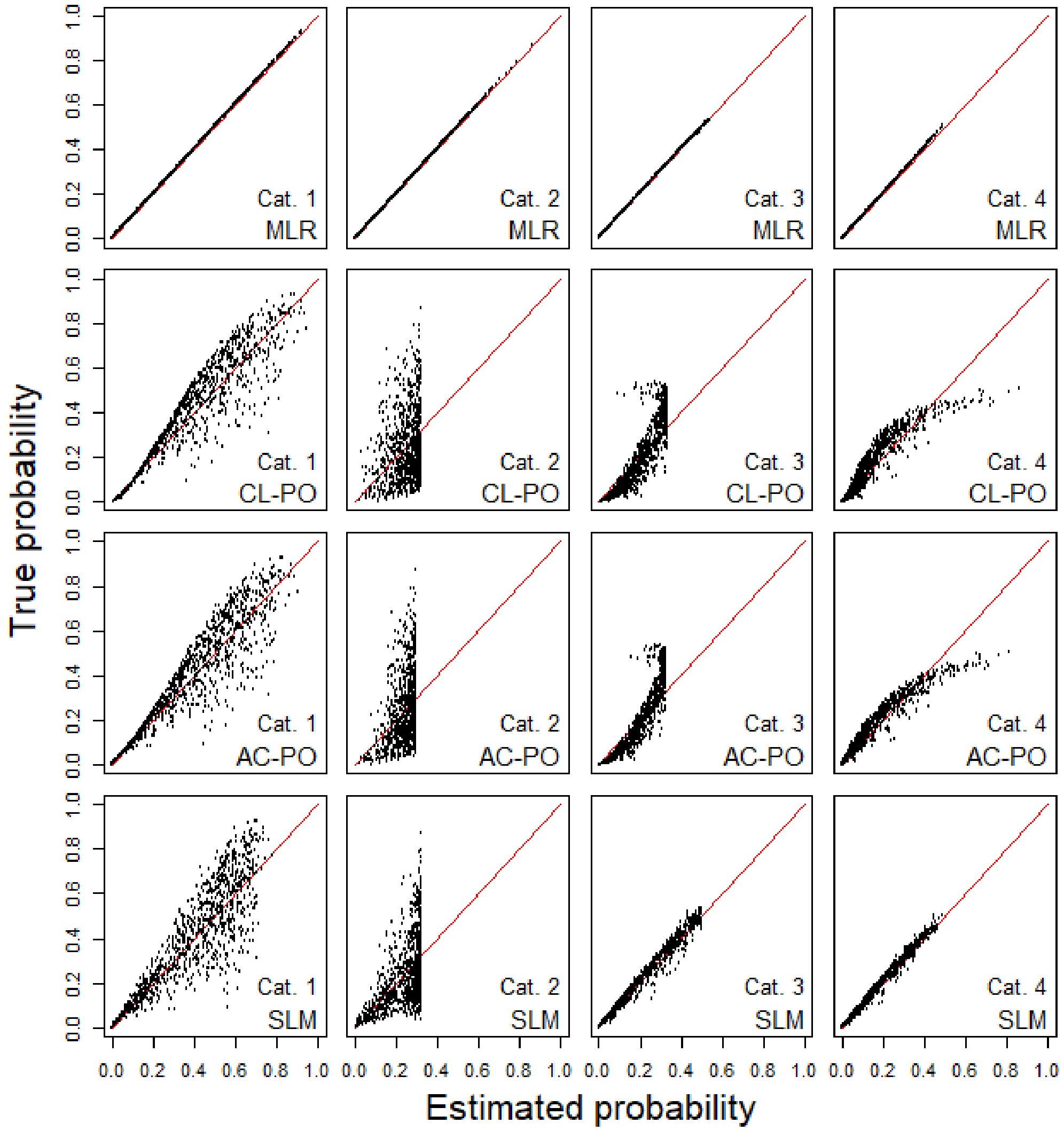

Figure S7. Scatter plots of true risks versus estimated risks for simulation scenario 7 when the true model has the form of a multinomial logistic regression. The plots are based on a random subset of 1,000 cases from all 200,000 cases.

MLR, multinomial logistic regression; CL-PO, cumulative logit model with proportional odds; AC-PO, adjacent category logit model with proportional odds; SLM, stereotype logit model.

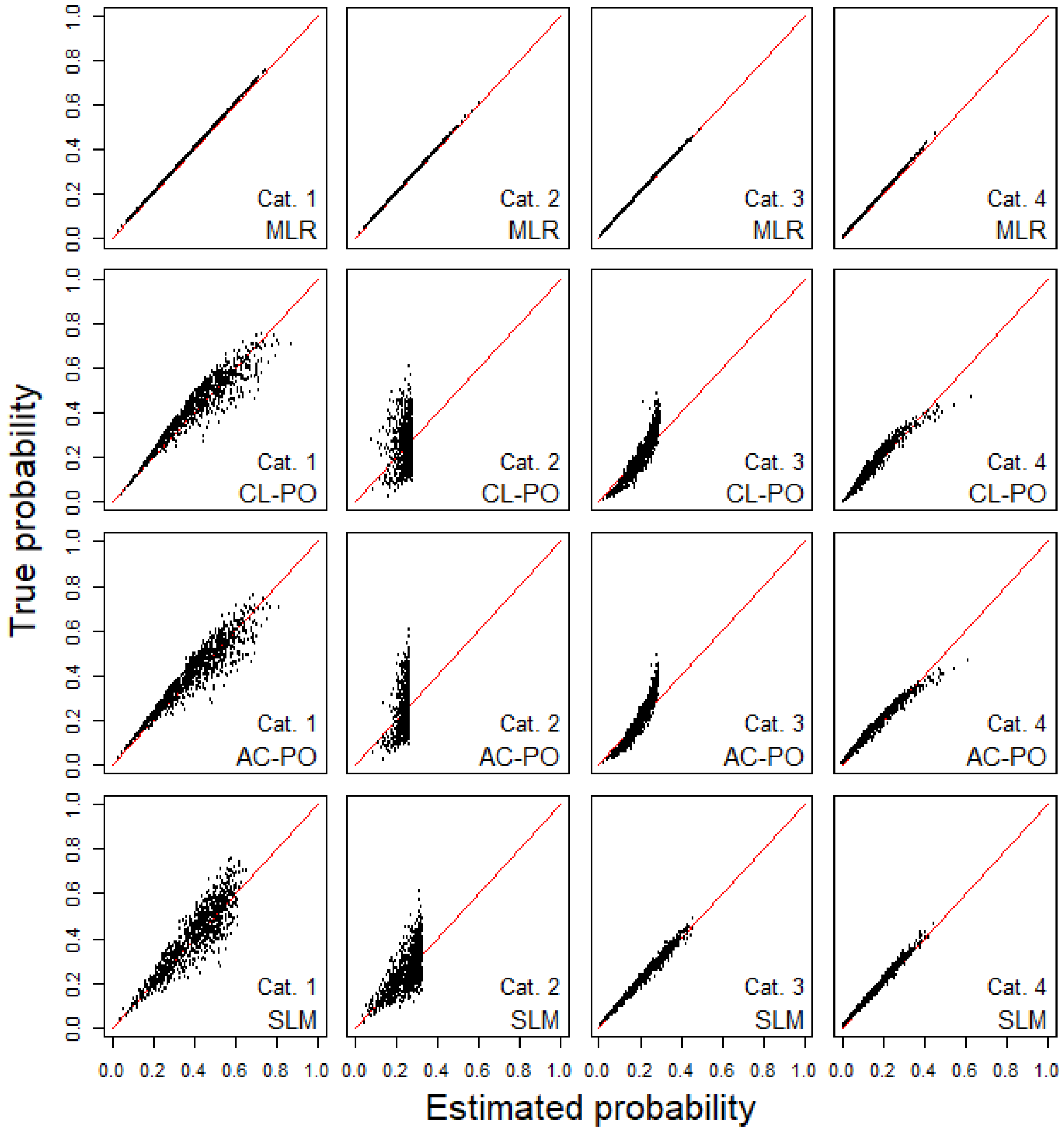

Figure S8. Scatter plots of true risks versus estimated risks for simulation scenario 8 when the true model has the form of a multinomial logistic regression. The plots are based on a random subset of 1,000 cases from all 200,000 cases.

MLR, multinomial logistic regression; CL-PO, cumulative logit model with proportional odds; AC-PO, adjacent category logit model with proportional odds; SLM, stereotype logit model.

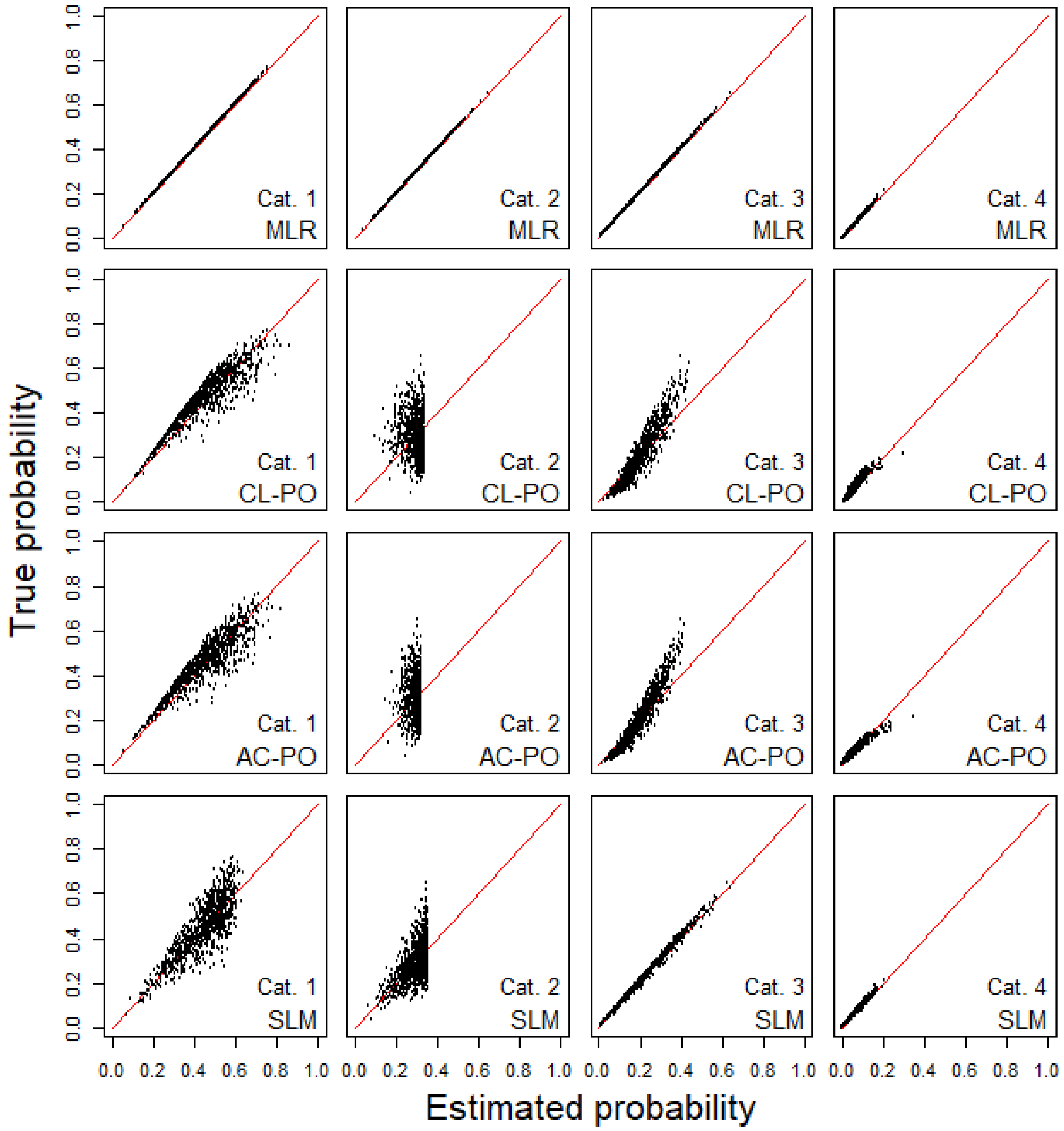

Figure S9. Scatter plots of true risks versus estimated risks for simulation scenario 9 when the true model has the form of a multinomial logistic regression.

MLR, multinomial logistic regression; CL-PO, cumulative logit model with proportional odds; AC-PO, adjacent category logit model with proportional odds; SLM, stereotype logit model.

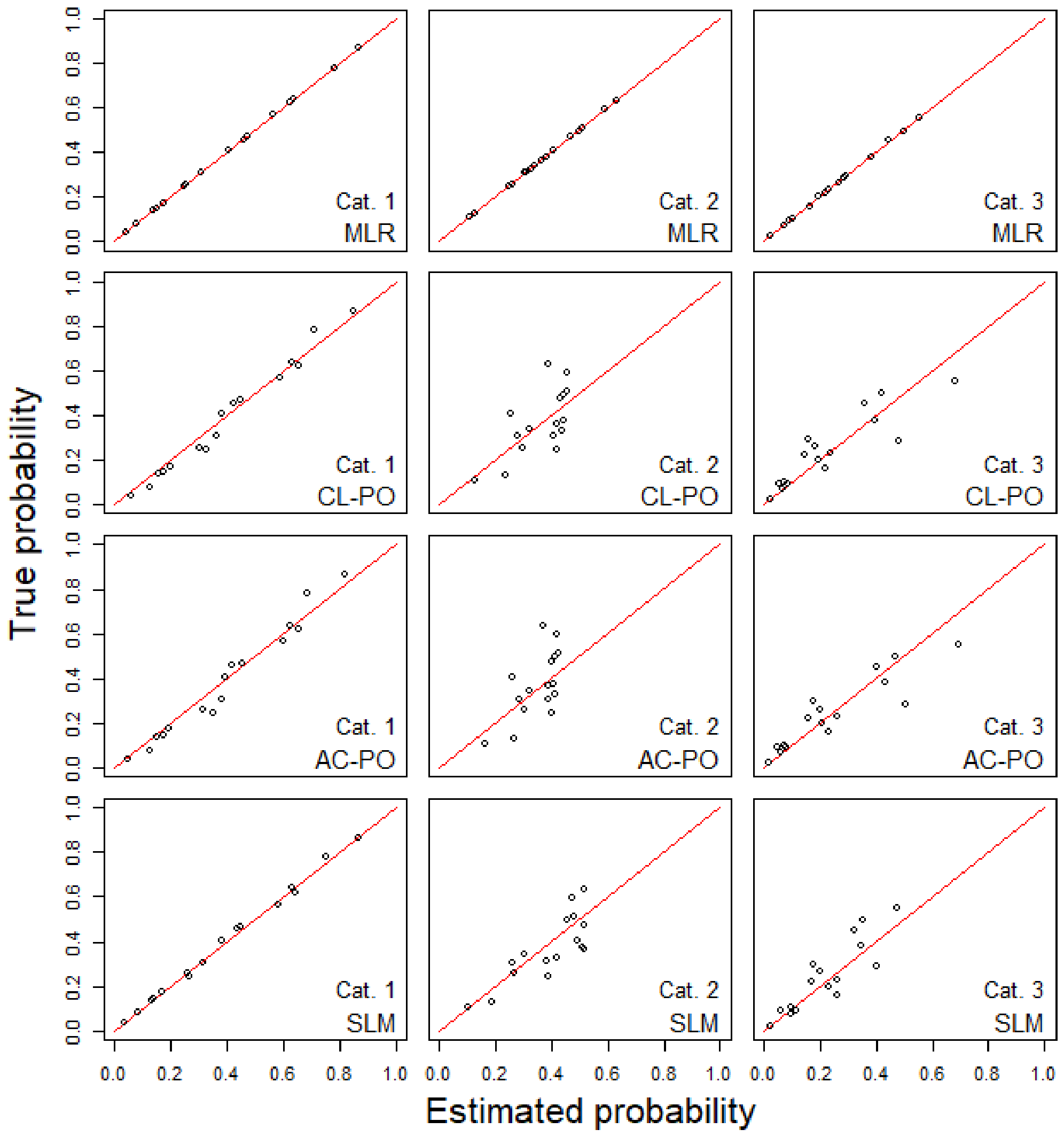

Figure S10. Scatter plots of true risks versus estimated risks for simulation scenario 10 when the true model has the form of a multinomial logistic regression.

MLR, multinomial logistic regression; CL-PO, cumulative logit model with proportional odds; AC-PO, adjacent category logit model with proportional odds; SLM, stereotype logit model.

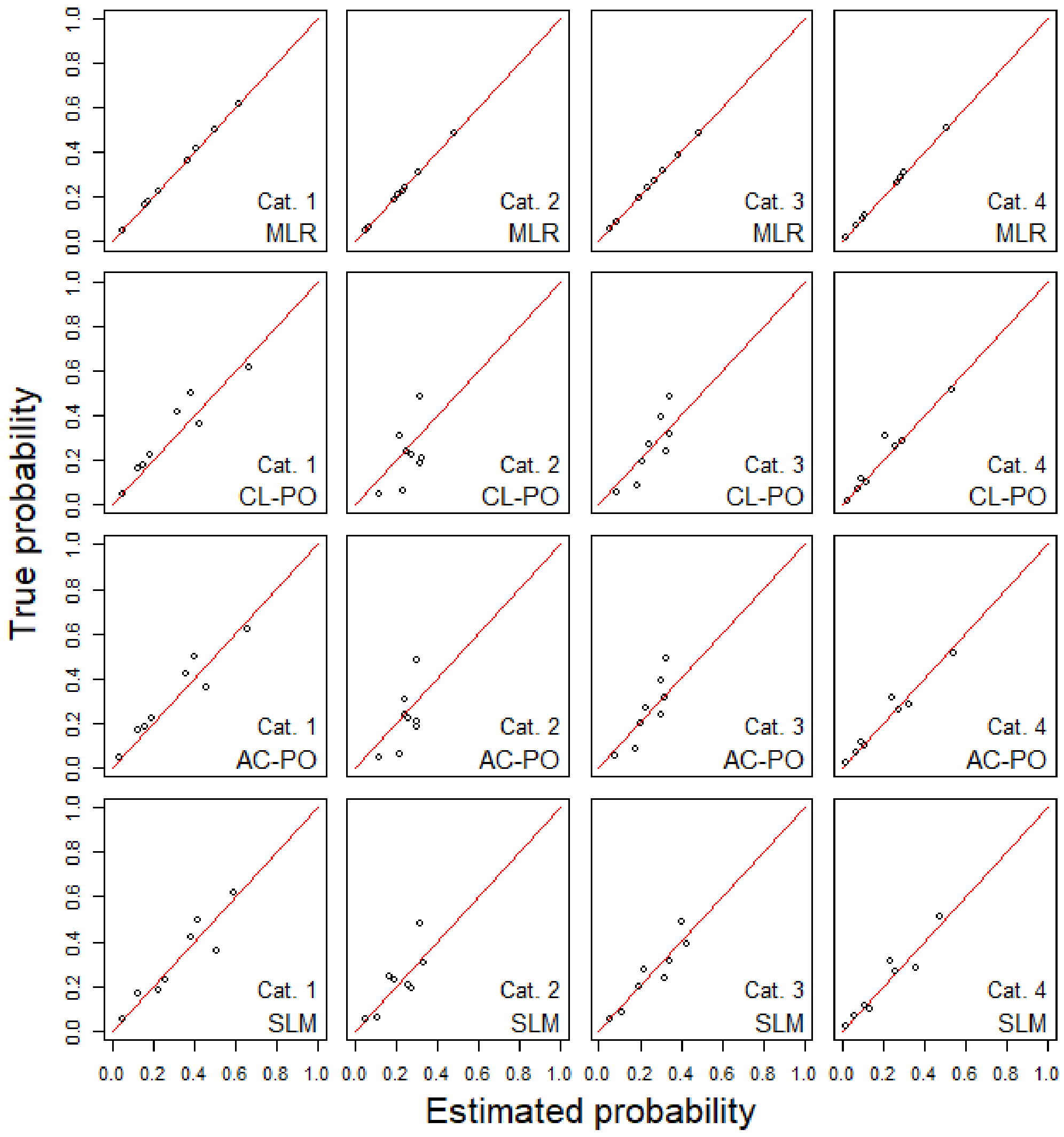

Figure S11. Scatter plots of true risks versus estimated risks for simulation scenario 11 when the true model has the form of a multinomial logistic regression. The plots are based on a random subset of 1,000 cases from all 200,000 cases.

MLR, multinomial logistic regression; CL-PO, cumulative logit model with proportional odds; AC-PO, adjacent category logit model with proportional odds; SLM, stereotype logit model.

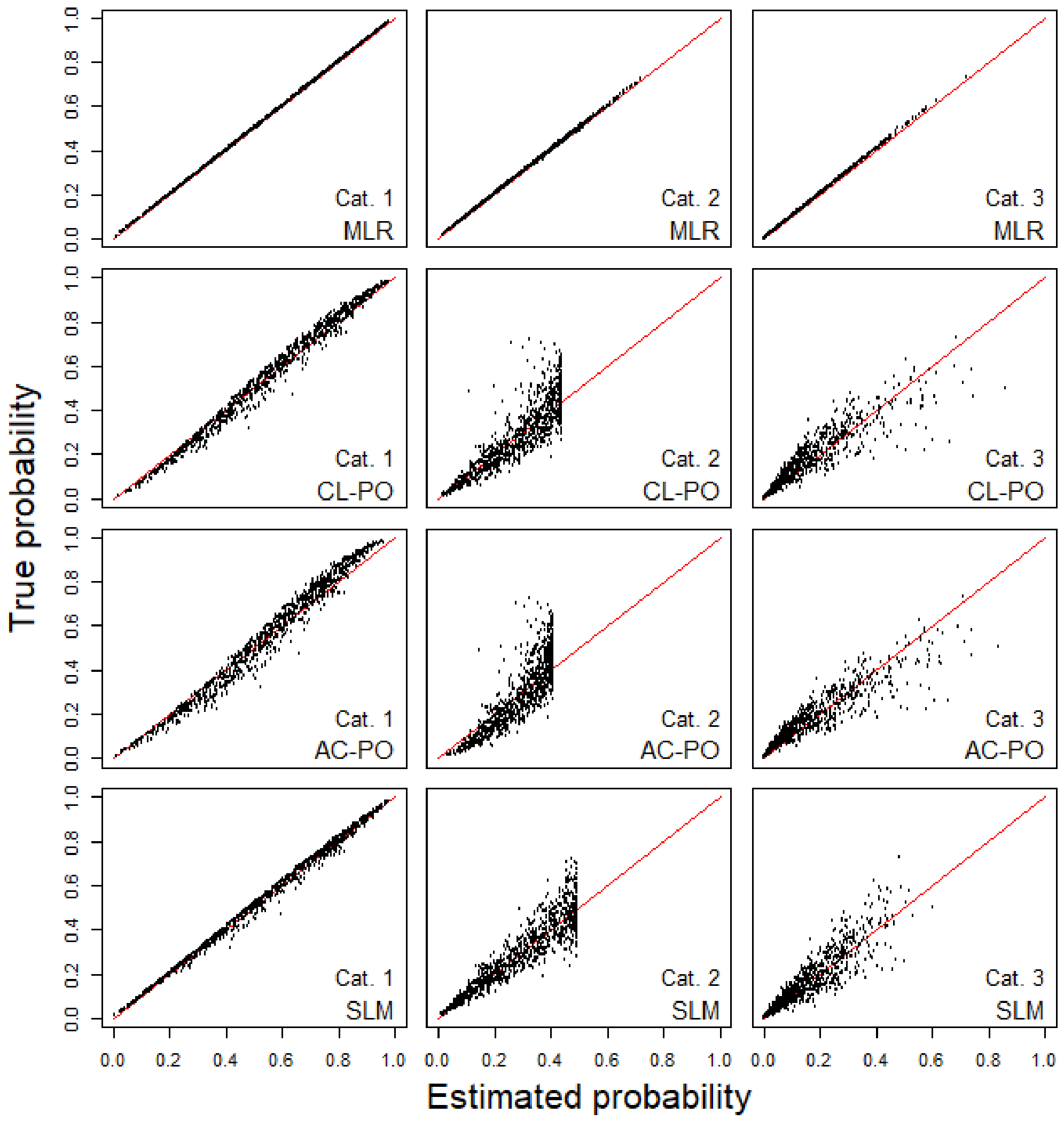

Figure S12. Flexible smoothed calibration curves per outcome category for simulation scenario 5 when the true model has the form of a multinomial logistic regression (green for category 1, orange for category 2, red for category 3). These curves are based on the dataset used to develop the model and are therefore apparent (or unvalidated) curves (n=200,000). For some models, lines overlap.

MLR, multinomial logistic regression; CL-PO, cumulative logit model with proportional odds; AC-PO, adjacent category logit model with proportional odds; SLM, stereotype logit model.

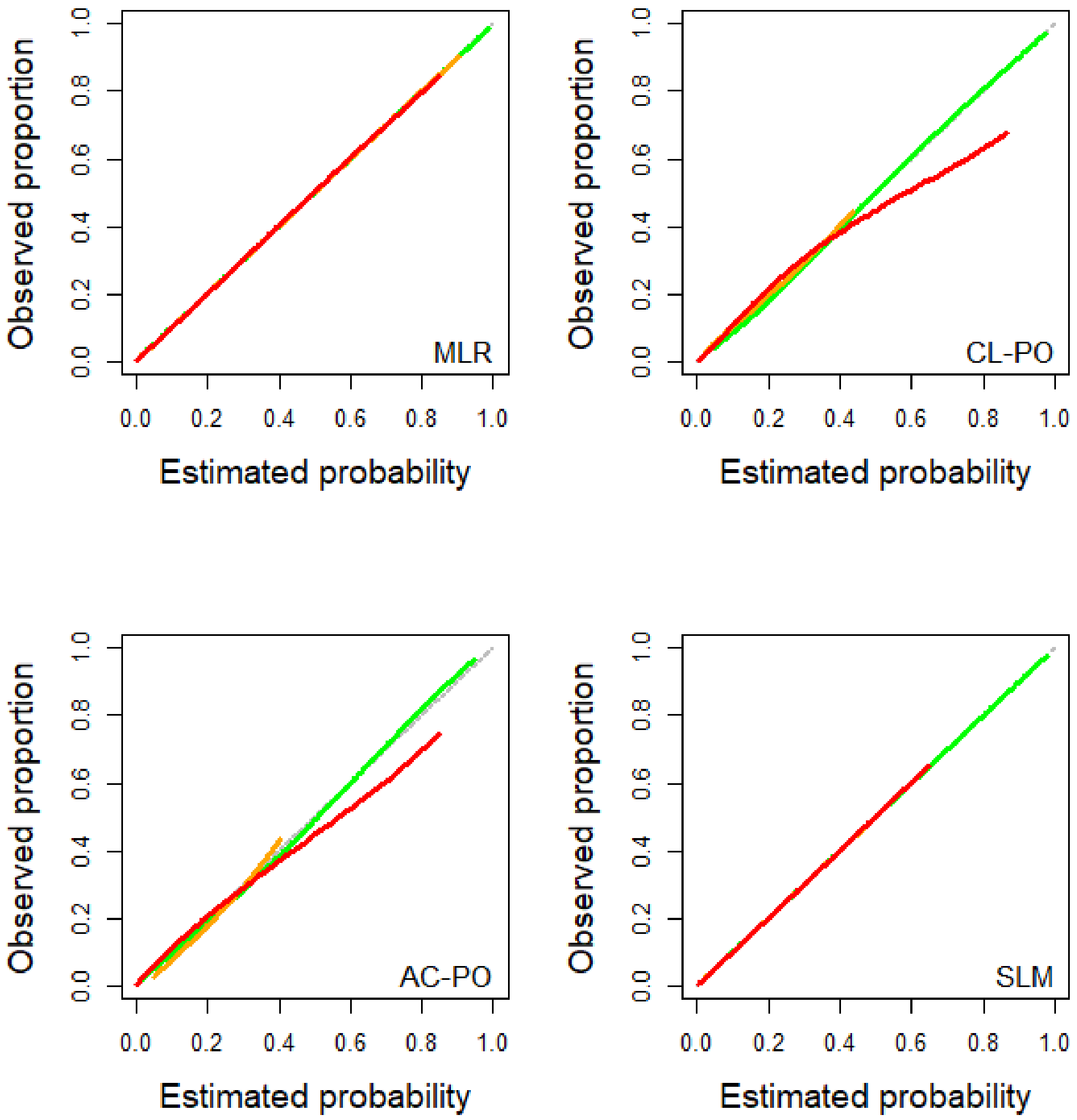

Figure S13. Flexible smoothed calibration curves per outcome category for simulation scenario 6 when the true model has the form of a multinomial logistic regression (green for category 1, orange for category 2, red for category 3, brown for category 4). These curves are based on the dataset used to develop the model and are therefore apparent (or unvalidated) curves (n=200,000). For some models, lines overlap.

MLR, multinomial logistic regression; CL-PO, cumulative logit model with proportional odds; AC-PO, adjacent category logit model with proportional odds; SLM, stereotype logit model.

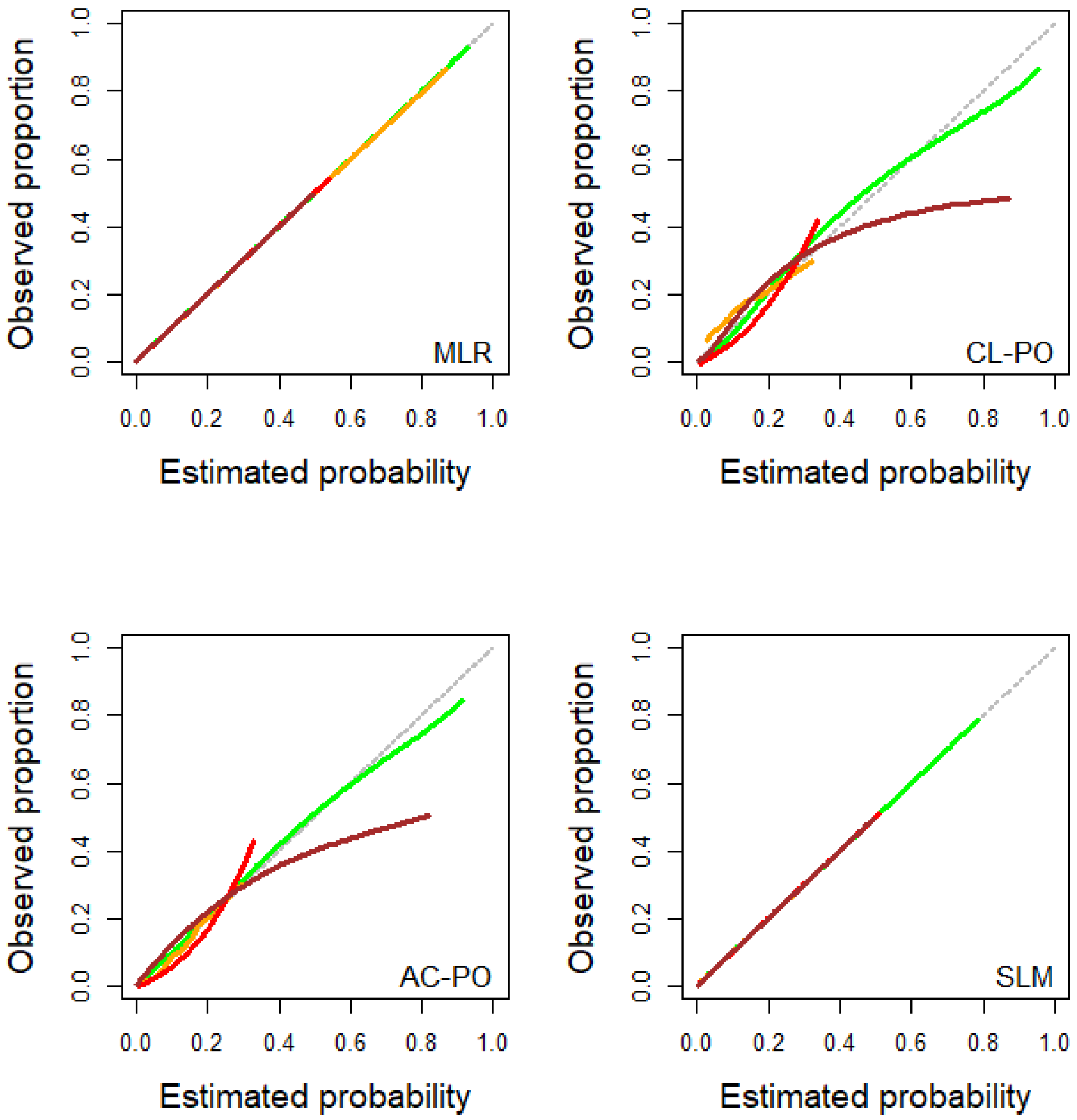

Figure S14. Flexible smoothed calibration curves per outcome category for simulation scenario 7 when the true model has the form of a multinomial logistic regression (green for category 1, orange for category 2, red for category 3, brown for category 4). These curves are based on the dataset used to develop the model and are therefore apparent (or unvalidated) curves (n=200,000). For some models, lines overlap.

MLR, multinomial logistic regression; CL-PO, cumulative logit model with proportional odds; AC-PO, adjacent category logit model with proportional odds; SLM, stereotype logit model.

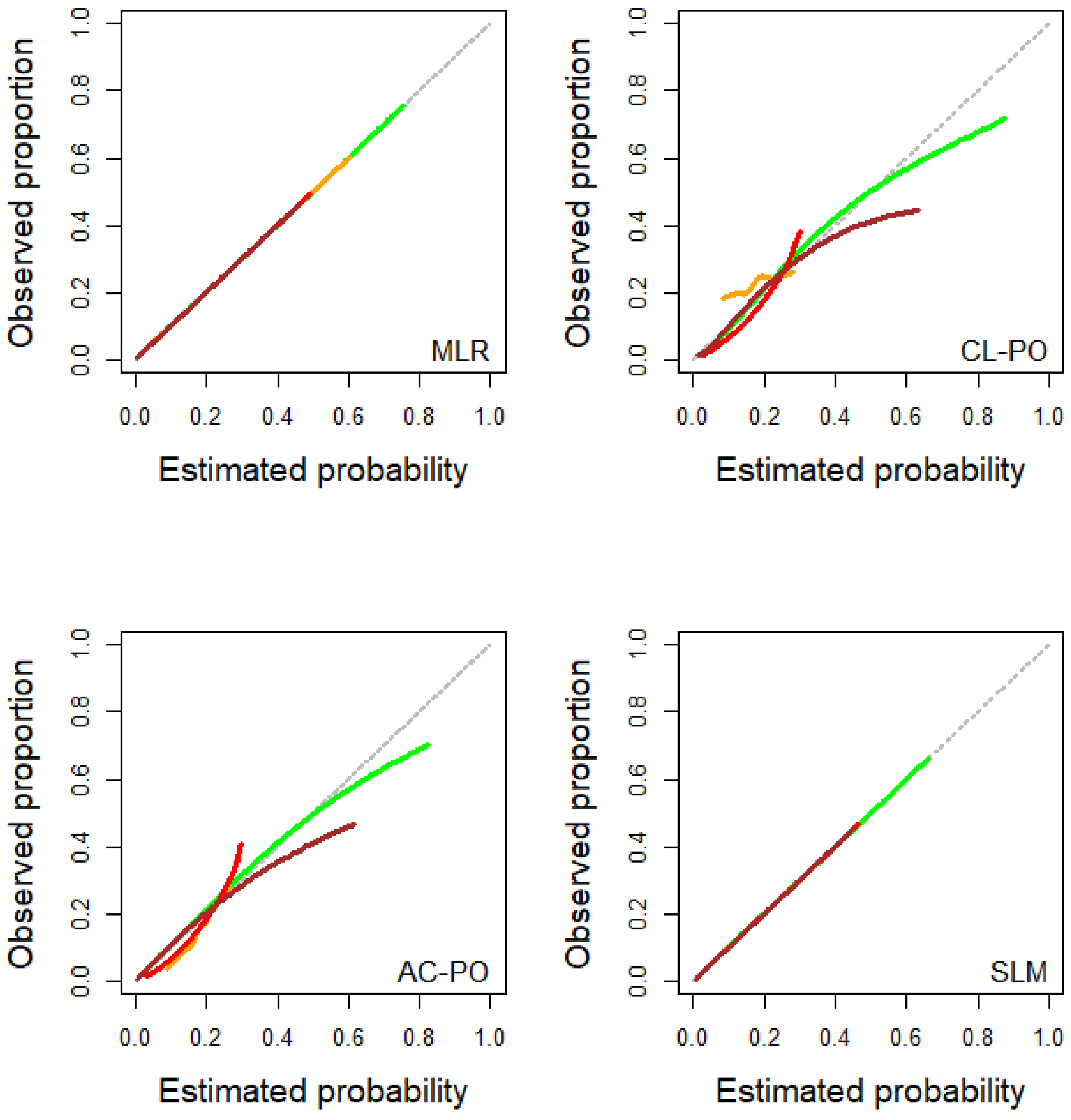

Figure S15. Flexible smoothed calibration curves per outcome category for simulation scenario 8 when the true model has the form of a multinomial logistic regression (green for category 1, orange for category 2, red for category 3, brown for category 4). These curves are based on the dataset used to develop the model and are therefore apparent (or unvalidated) curves (n=200,000). For some models, lines overlap.

MLR, multinomial logistic regression; CL-PO, cumulative logit model with proportional odds; AC-PO, adjacent category logit model with proportional odds; SLM, stereotype logit model.

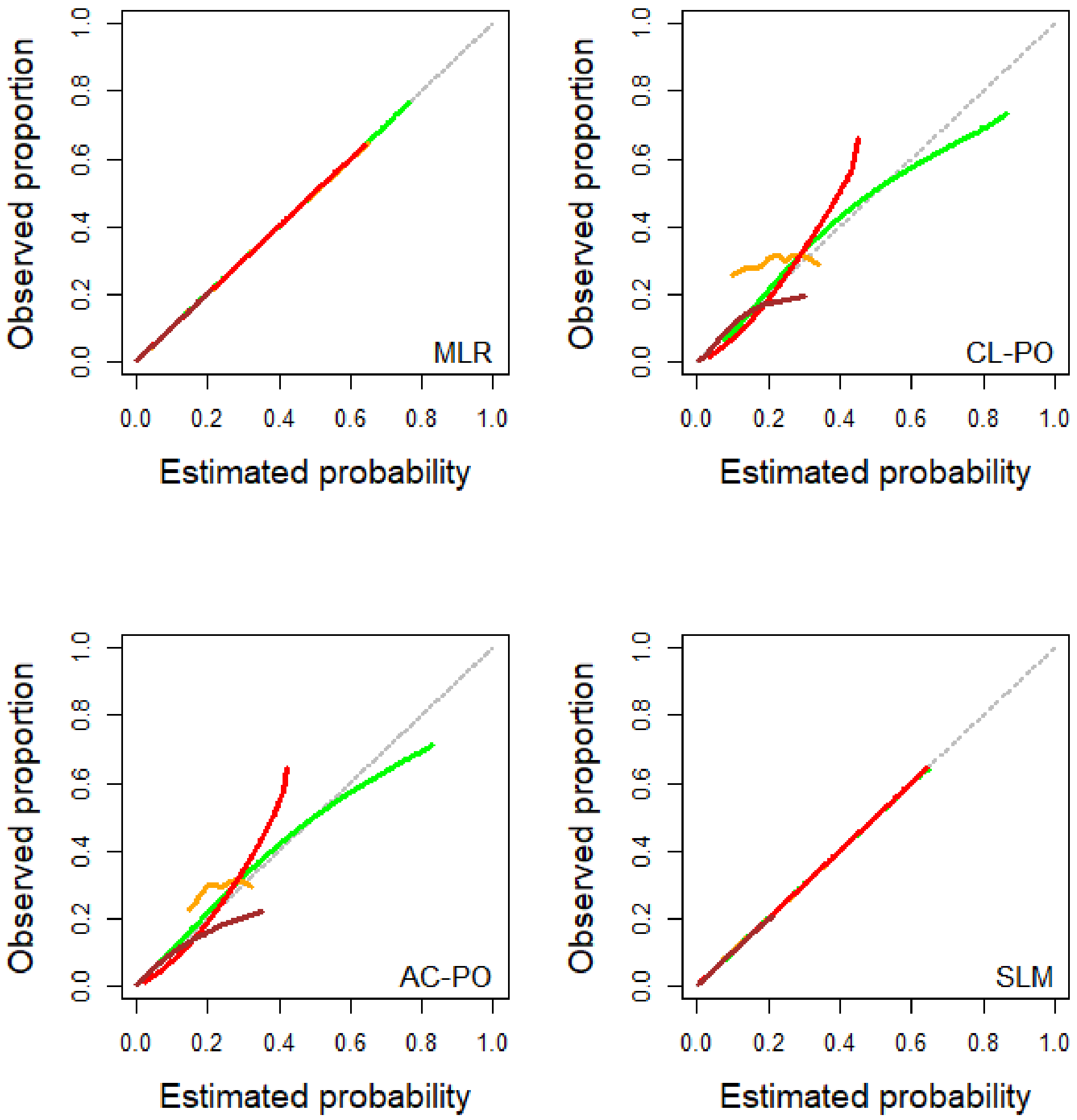

Figure S16. Calibration scatter plots per outcome category for simulation scenario 9 when the true model has the form of a multinomial logistic regression (green for category 1, orange for category 2, red for category 3). These graphs are based on the dataset used to develop the model and are therefore apparent (or unvalidated) curves (n=200,000). Because all predictors are binary, no flexible curves are shown.

MLR, multinomial logistic regression; CL-PO, cumulative logit model with proportional odds; AC-PO, adjacent category logit model with proportional odds; SLM, stereotype logit model.

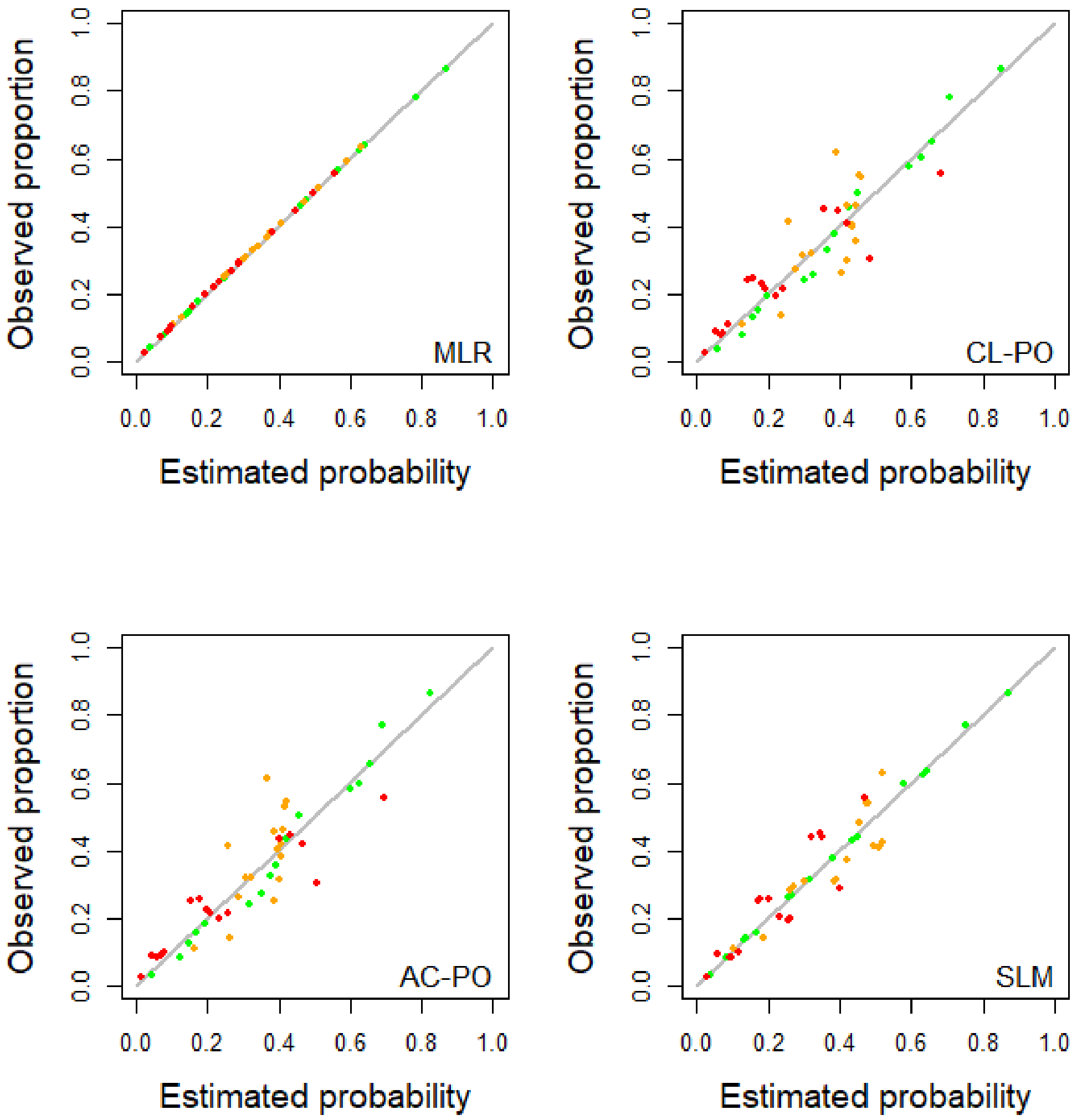

Figure S17. Calibration scatter plots per outcome category for simulation scenario 10 when the true model has the form of a multinomial logistic regression (green for category 1, orange for category 2, red for category 3, brown for category 4). These graphs are based on the dataset used to develop the model and are therefore apparent (or unvalidated) curves (n=200,000). Because all predictors are binary, no flexible curves are shown.

MLR, multinomial logistic regression; CL-PO, cumulative logit model with proportional odds; AC-PO, adjacent category logit model with proportional odds; SLM, stereotype logit model.

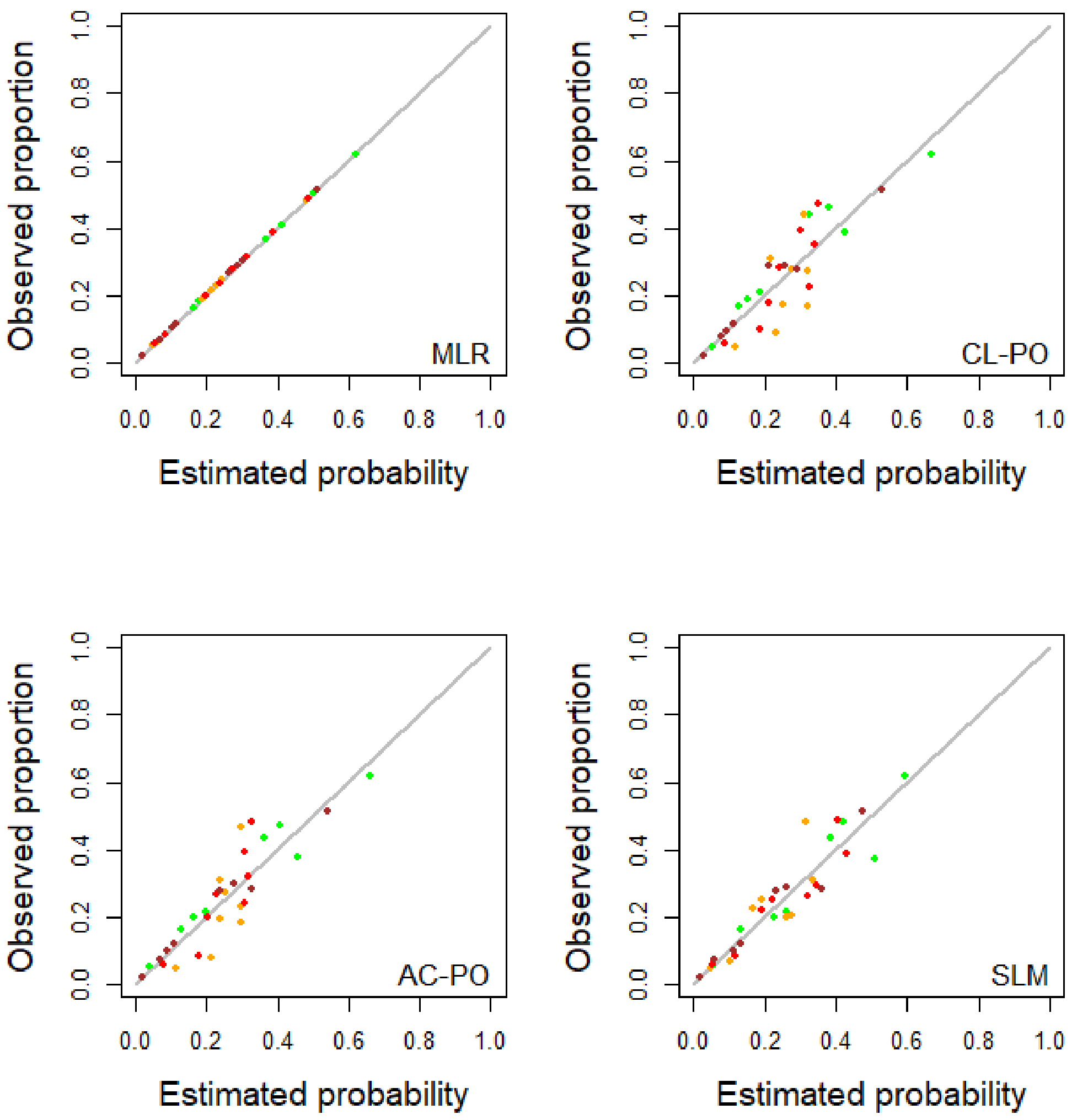

Figure S18. Flexible smoothed calibration curves per outcome category for simulation scenario 11 when the true model has the form of a multinomial logistic regression (green for category 1, orange for category 2, red for category 3). These curves are based on the dataset used to develop the model and are therefore apparent (or unvalidated) curves (n=200,000). For some models, lines overlap.

MLR, multinomial logistic regression; CL-PO, cumulative logit model with proportional odds; AC-PO, adjacent category logit model with proportional odds; SLM, stereotype logit model.

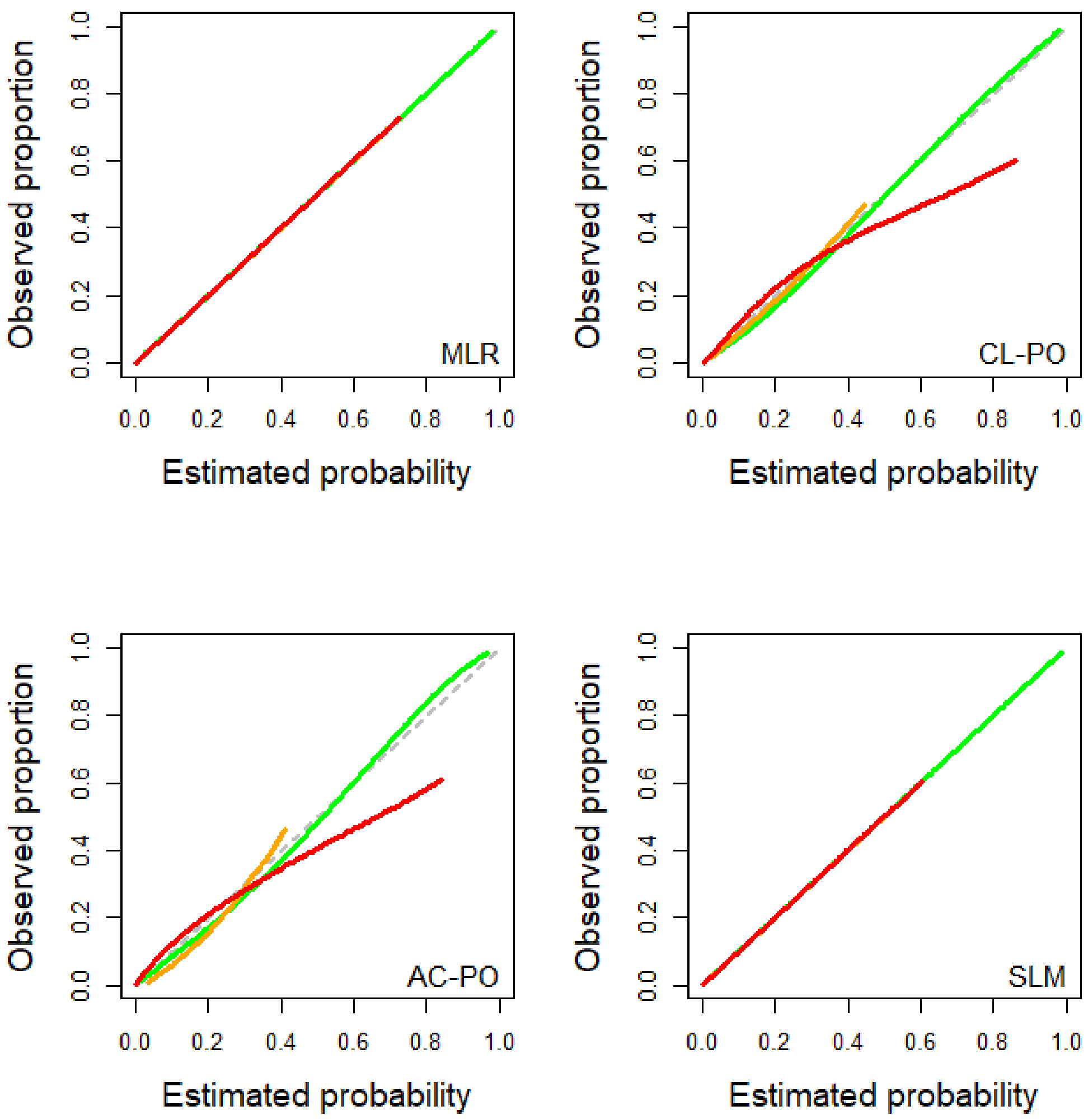

Figure S19. Scatter plots of true risks versus estimated risks for simulation scenario 4 when the true model has the form of a cumulative logit model with proportional odds. The plots are based on a random subset of 1,000 cases from all 200,000 cases.

MLR, multinomial logistic regression; CL-PO, cumulative logit model with proportional odds; AC-PO, adjacent category logit model with proportional odds; SLM, stereotype logit model.

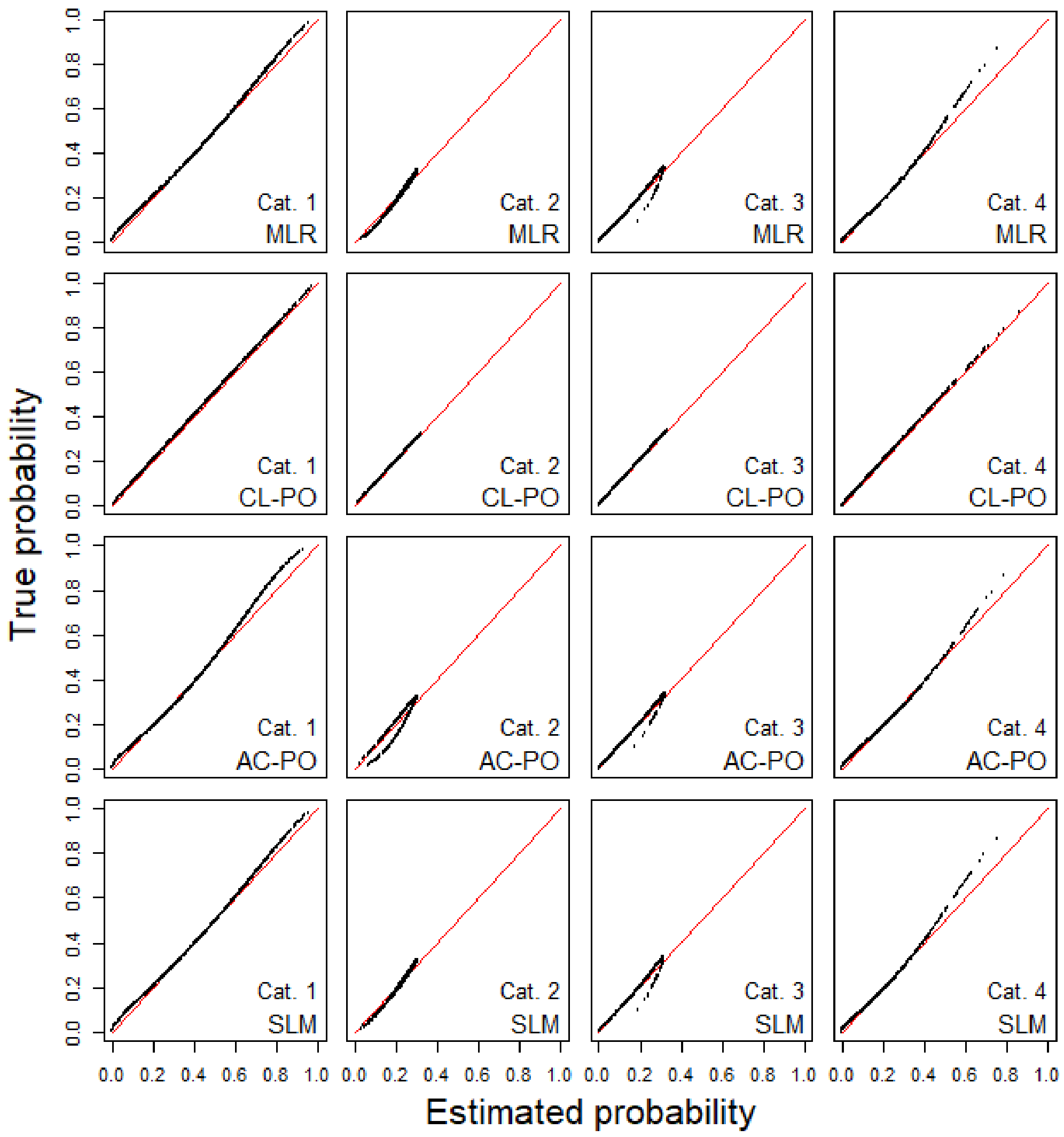

Figure S20. Scatter plots of true risks versus estimated risks for simulation scenario 5 when the true model has the form of a cumulative logit model with proportional odds. The plots are based on a random subset of 1,000 cases from all 200,000 cases.

MLR, multinomial logistic regression; CL-PO, cumulative logit model with proportional odds; AC-PO, adjacent category logit model with proportional odds; SLM, stereotype logit model.

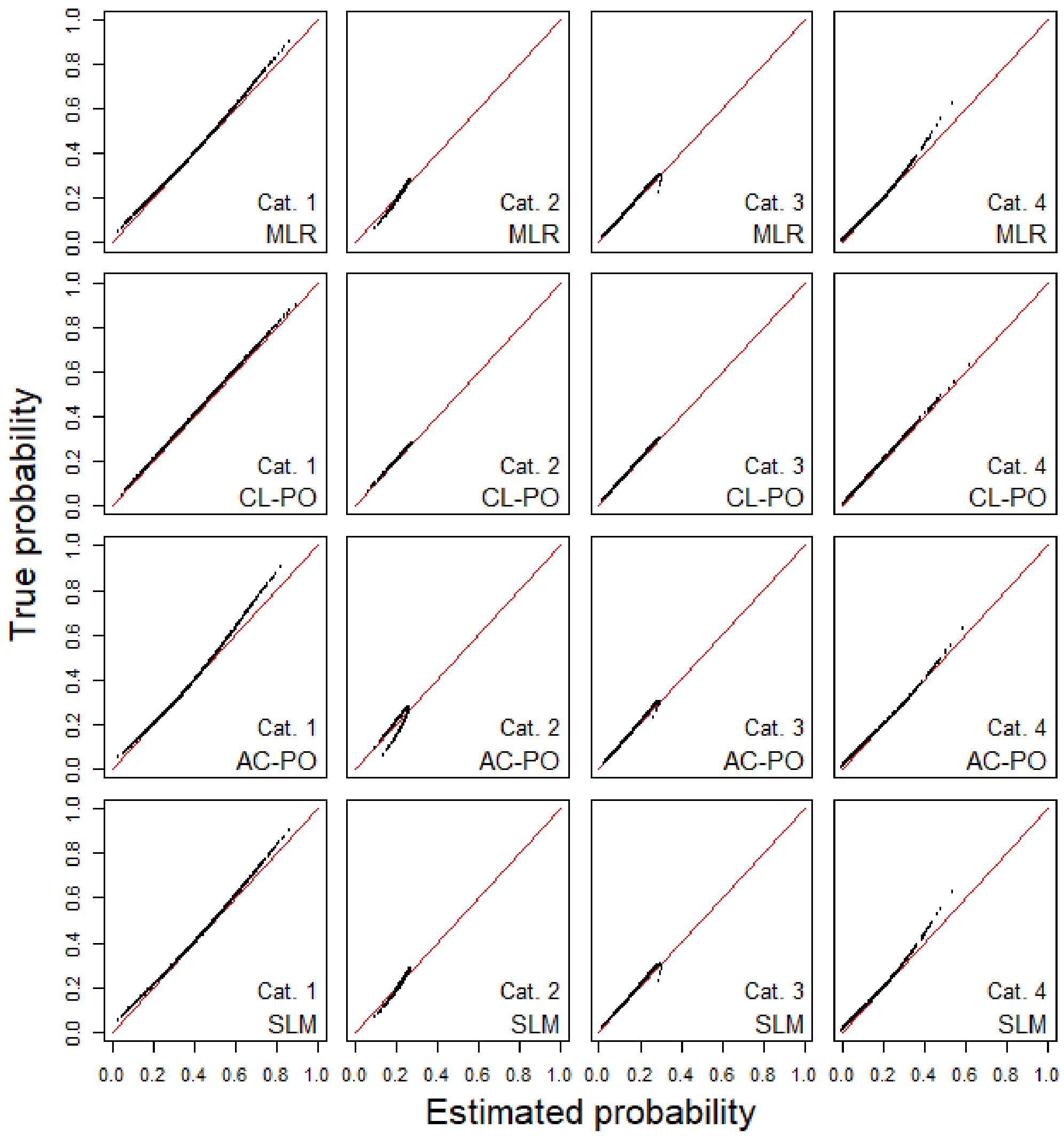

Figure S21. Scatter plots of true risks versus estimated risks for simulation scenario 6 when the true model has the form of a cumulative logit model with proportional odds. The plots are based on a random subset of 1,000 cases from all 200,000 cases.

MLR, multinomial logistic regression; CL-PO, cumulative logit model with proportional odds; AC-PO, adjacent category logit model with proportional odds; SLM, stereotype logit model.

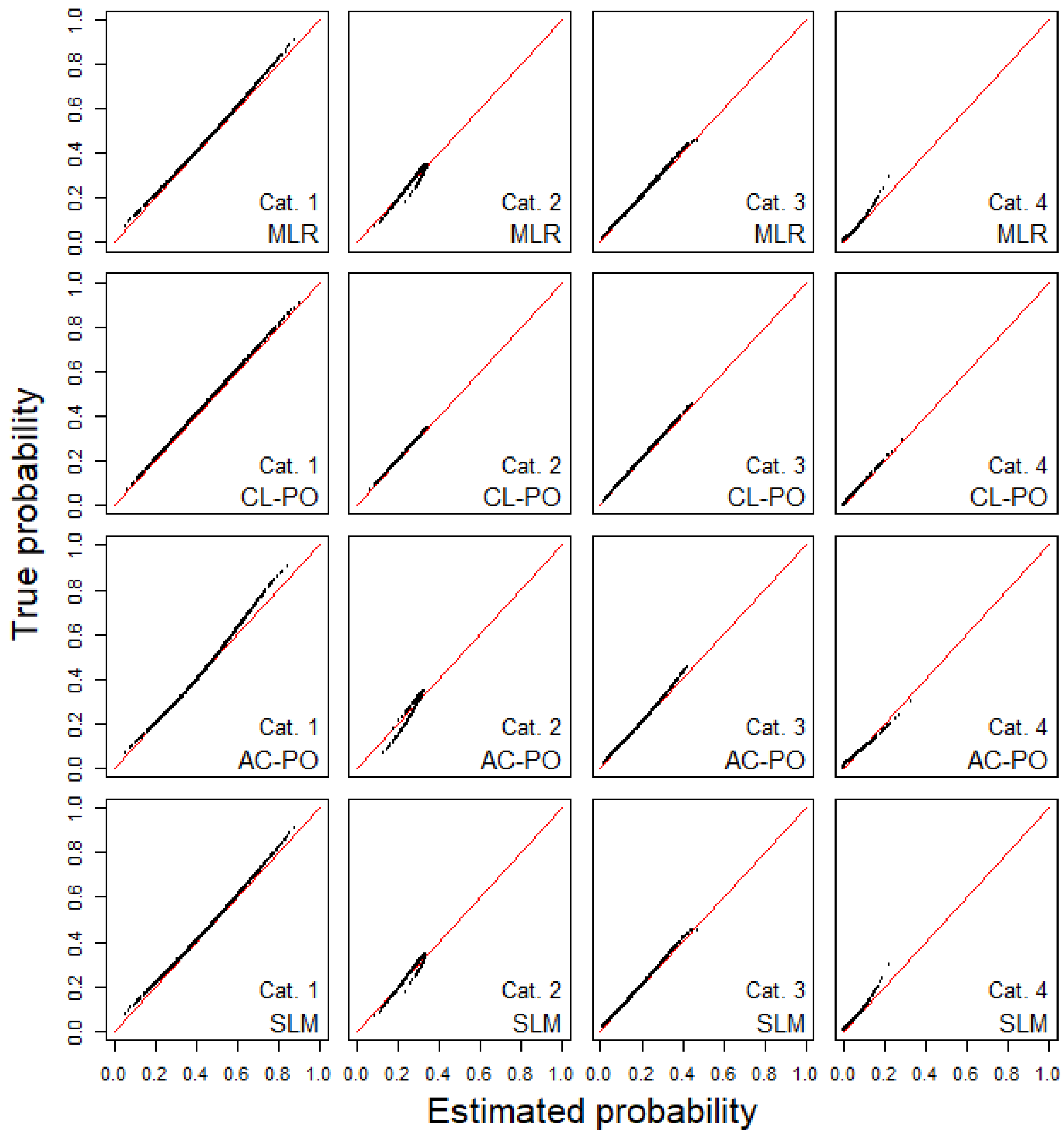

Figure S22. Scatter plots of true risks versus estimated risks for simulation scenario 7 when the true model has the form of a cumulative logit model with proportional odds.

MLR, multinomial logistic regression; CL-PO, cumulative logit model with proportional odds; AC-PO, adjacent category logit model with proportional odds; SLM, stereotype logit model.

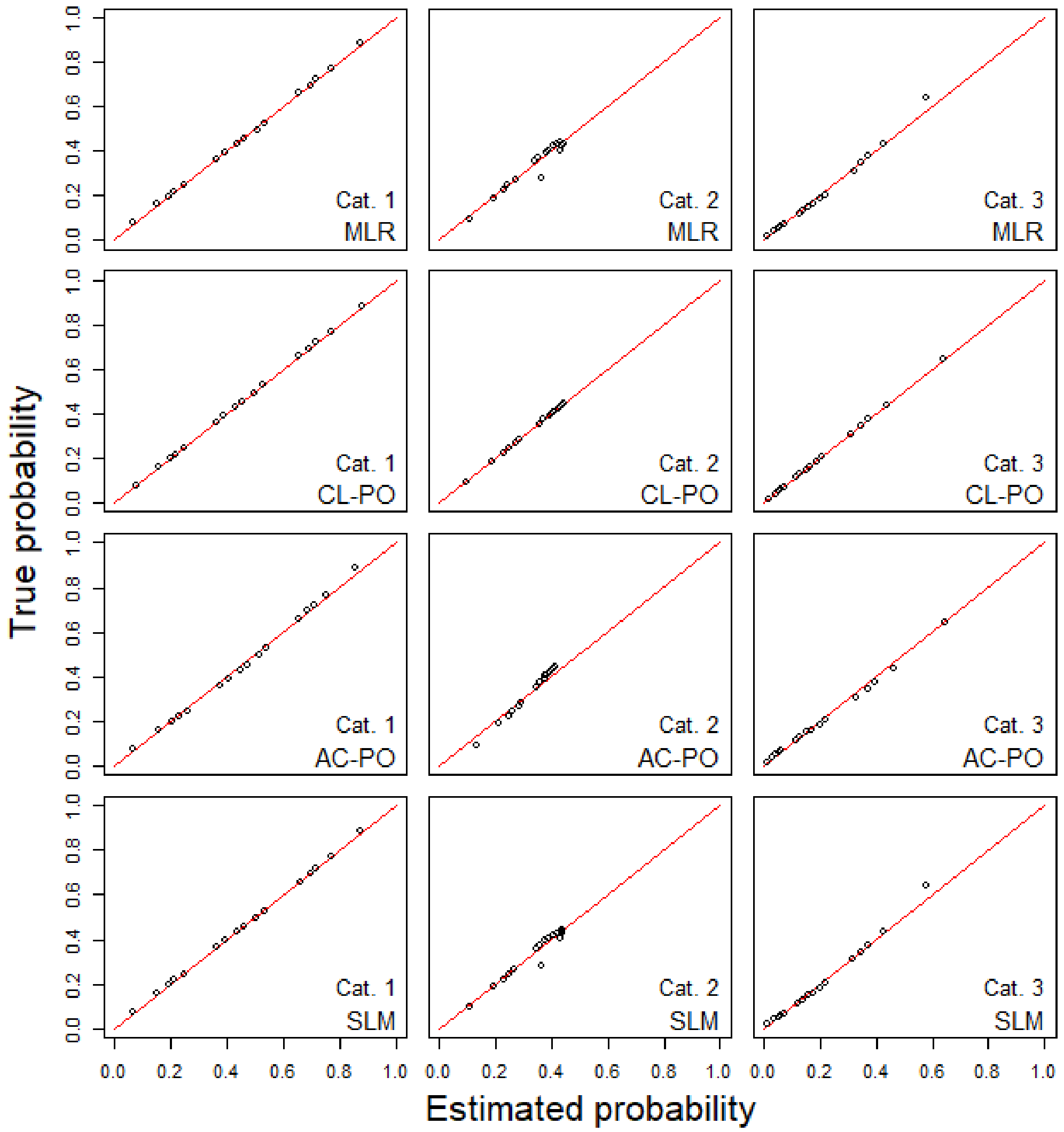

Figure S23. Scatter plots of true risks versus estimated risks for simulation scenario 8 when the true model has the form of a cumulative logit model with proportional odds.

MLR, multinomial logistic regression; CL-PO, cumulative logit model with proportional odds; AC-PO, adjacent category logit model with proportional odds; SLM, stereotype logit model.

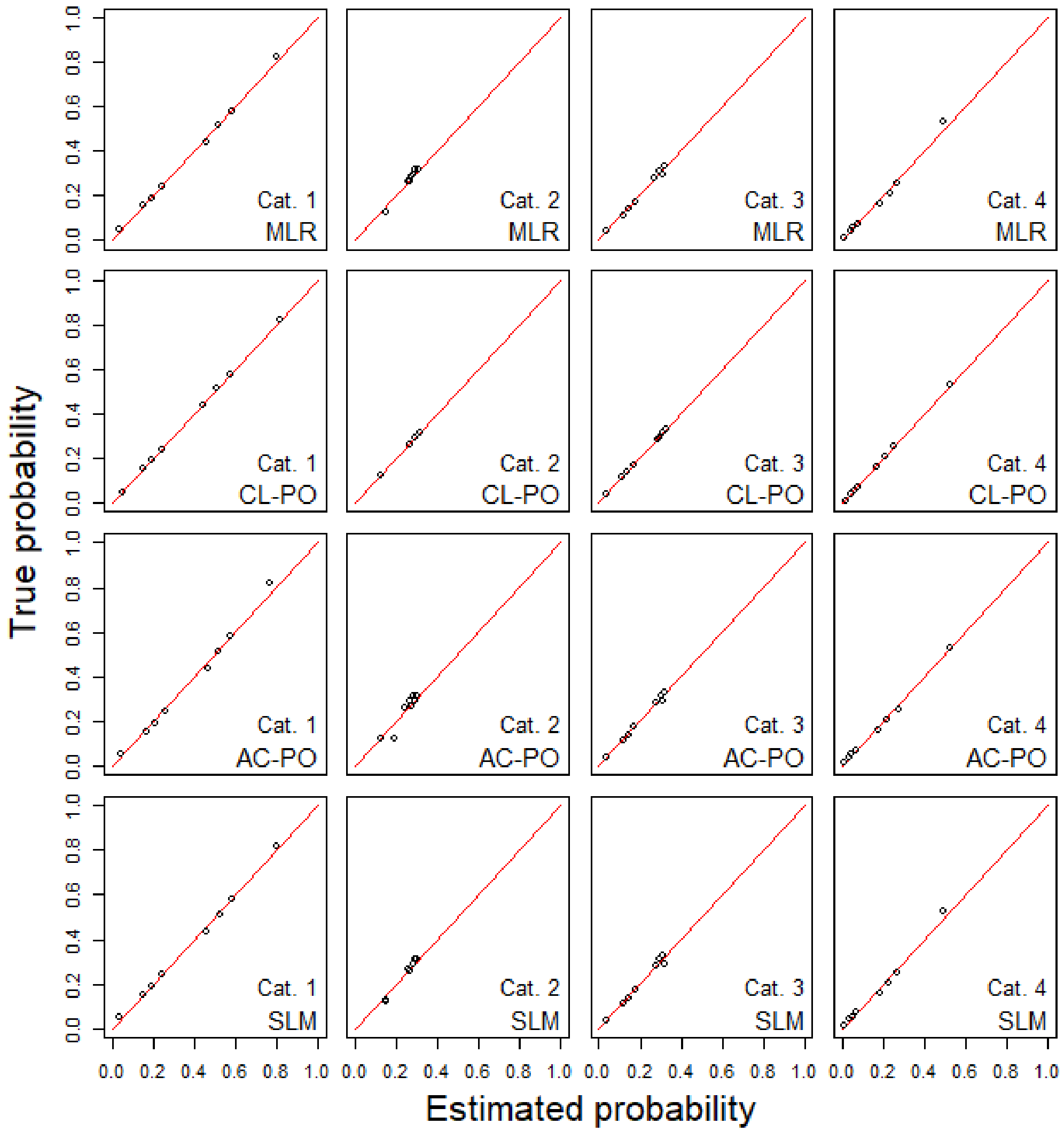

Figure S24. Scatter plots of true risks versus estimated risks for simulation scenario 9 when the true model has the form of a cumulative logit model with proportional odds. The plots are based on a random subset of 1,000 cases from all 200,000 cases.

MLR, multinomial logistic regression; CL-PO, cumulative logit model with proportional odds; AC-PO, adjacent category logit model with proportional odds; SLM, stereotype logit model.

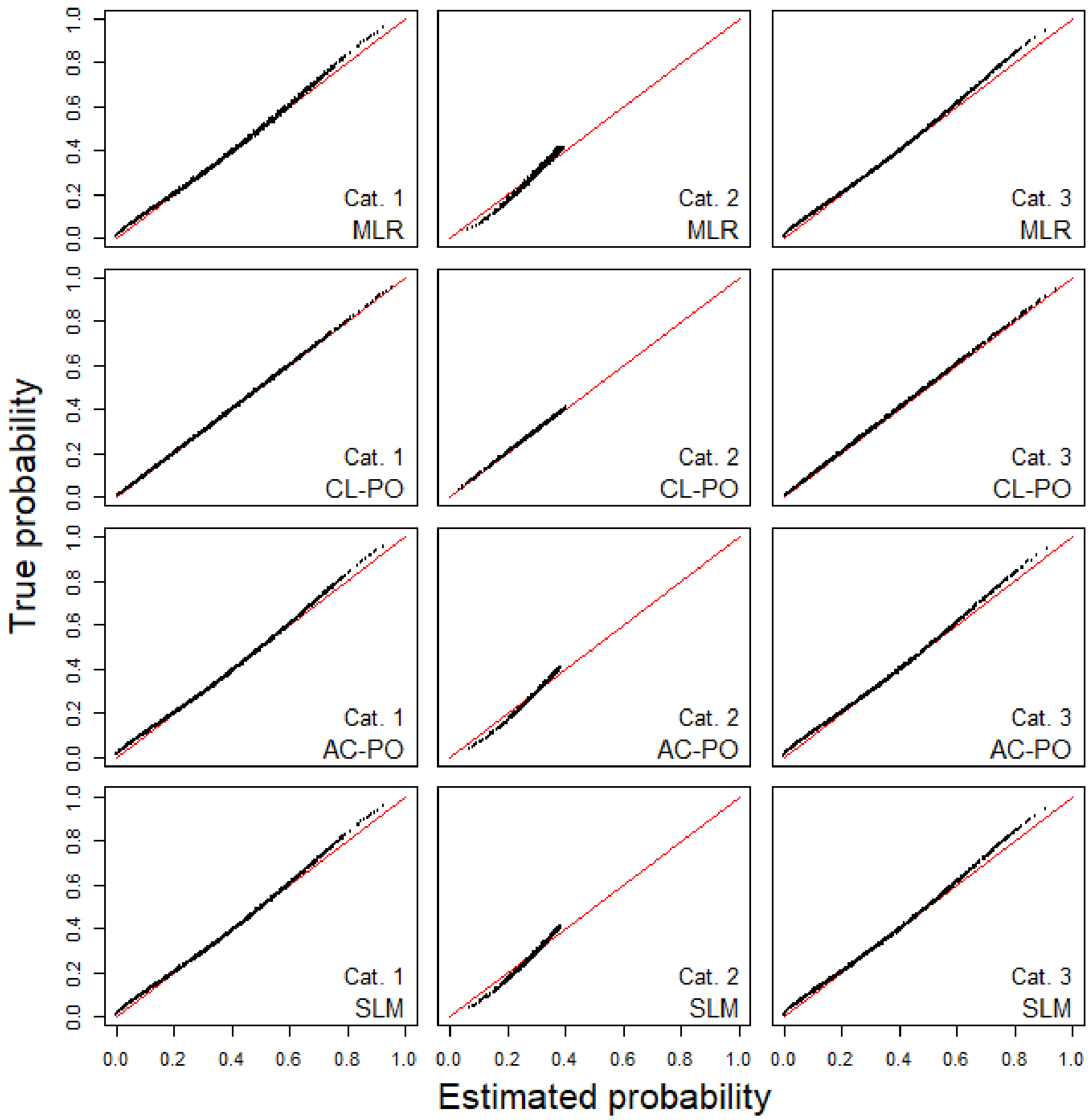

Figure S25. Flexible smoothed calibration curves per outcome category for simulation scenario 1 when the true model has the form of a cumulative logit model with proportional odds (green for category 1, orange for category 2, red for category 3). These curves are based on the dataset used to develop the model and are therefore apparent (or unvalidated) curves (n=200,000). For some models, lines overlap.

MLR, multinomial logistic regression; CL-PO, cumulative logit model with proportional odds; AC-PO, adjacent category logit model with proportional odds; SLM, stereotype logit model.

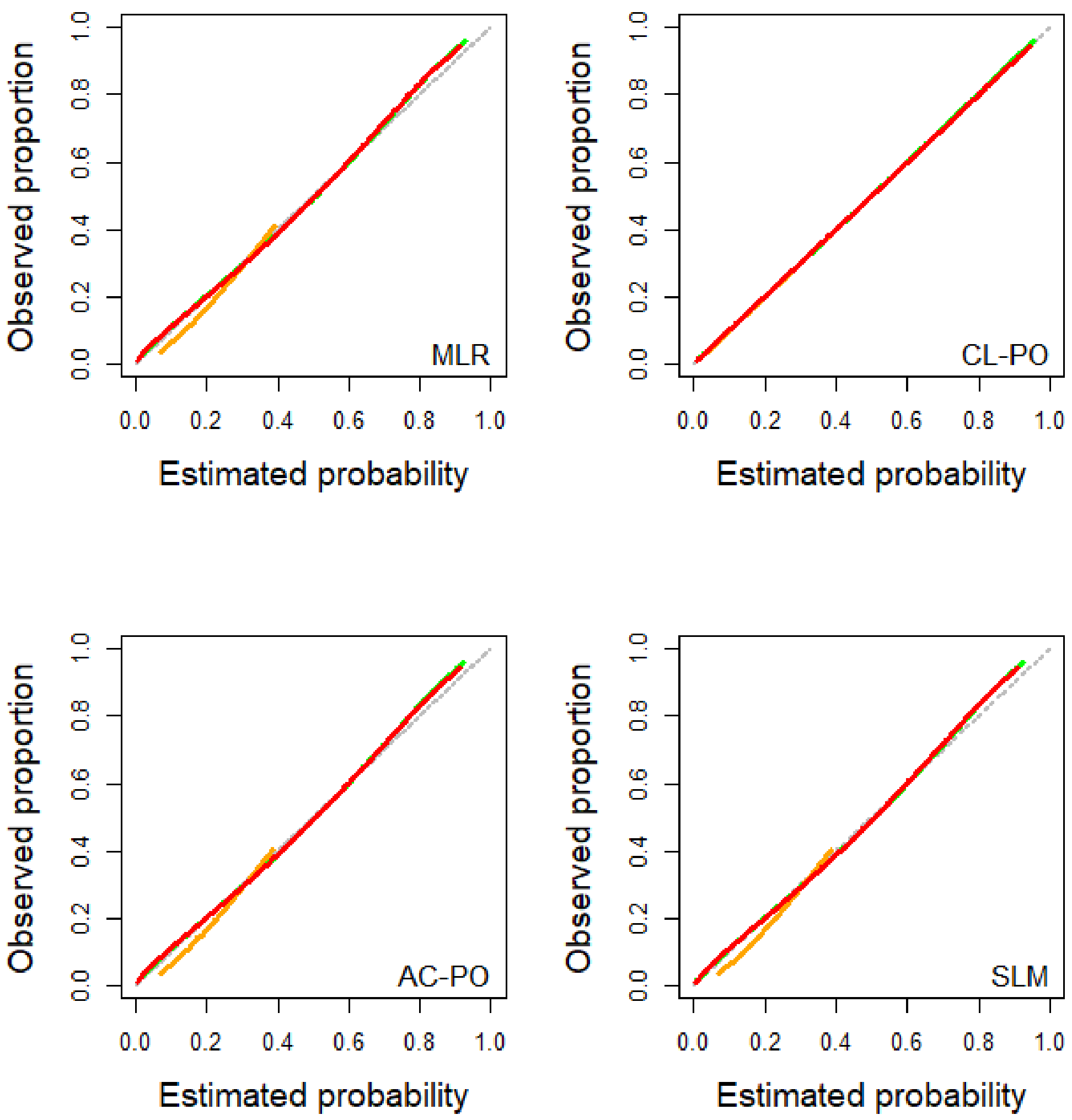

Figure S26. Flexible smoothed calibration curves per outcome category for simulation scenario 2 when the true model has the form of a cumulative logit model with proportional odds (green for category 1, orange for category 2, red for category 3). These curves are based on the dataset used to develop the model and are therefore apparent (or unvalidated) curves (n=200,000). For some models, lines overlap.

MLR, multinomial logistic regression; CL-PO, cumulative logit model with proportional odds; AC-PO, adjacent category logit model with proportional odds; SLM, stereotype logit model.

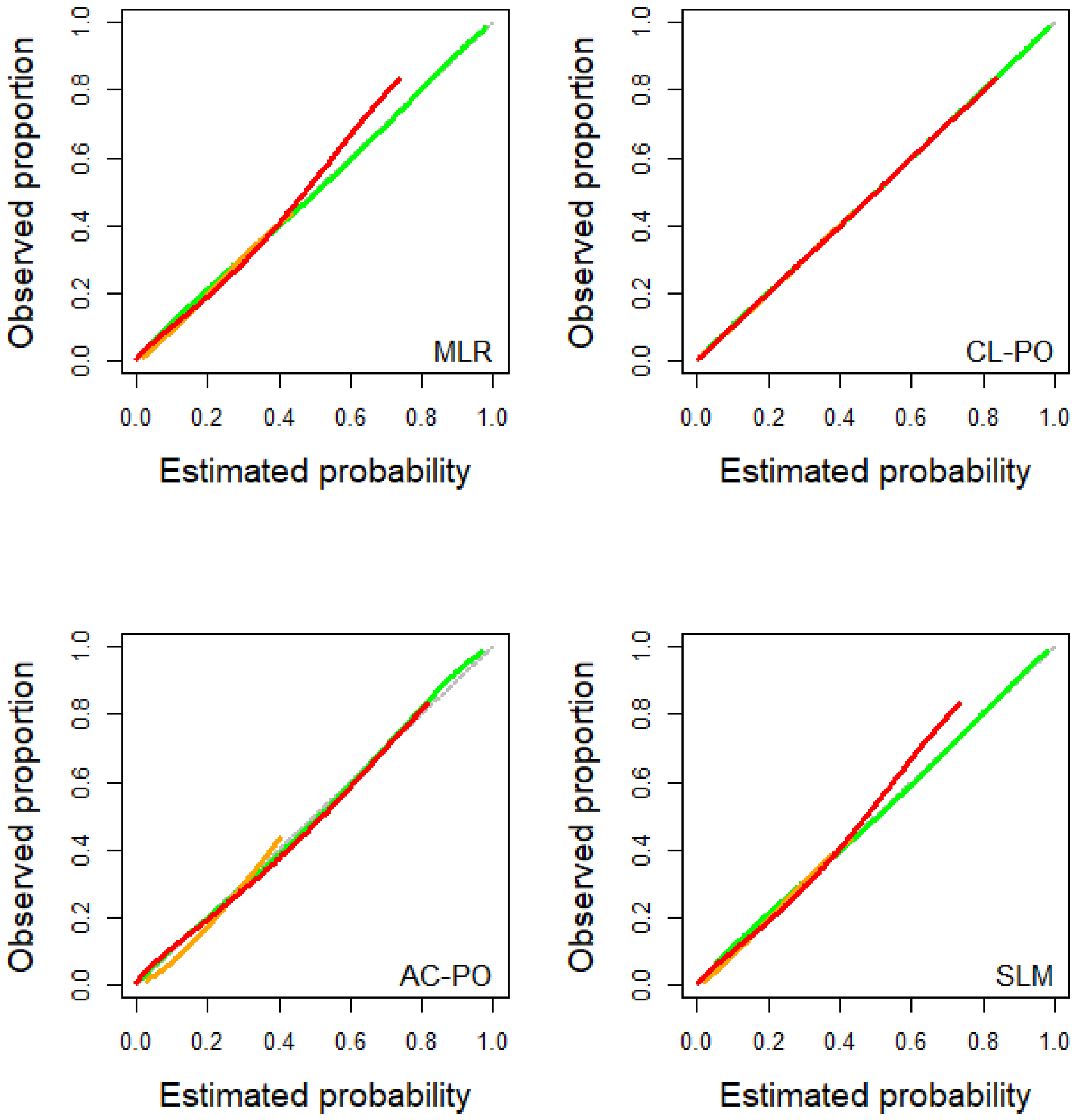

Figure S27. Flexible smoothed calibration curves per outcome category for simulation scenario 3 when the true model has the form of a cumulative logit model with proportional odds (green for category 1, orange for category 2, red for category 3). These curves are based on the dataset used to develop the model and are therefore apparent (or unvalidated) curves (n=200,000). For some models, lines overlap.

MLR, multinomial logistic regression; CL-PO, cumulative logit model with proportional odds; AC-PO, adjacent category logit model with proportional odds; SLM, stereotype logit model.

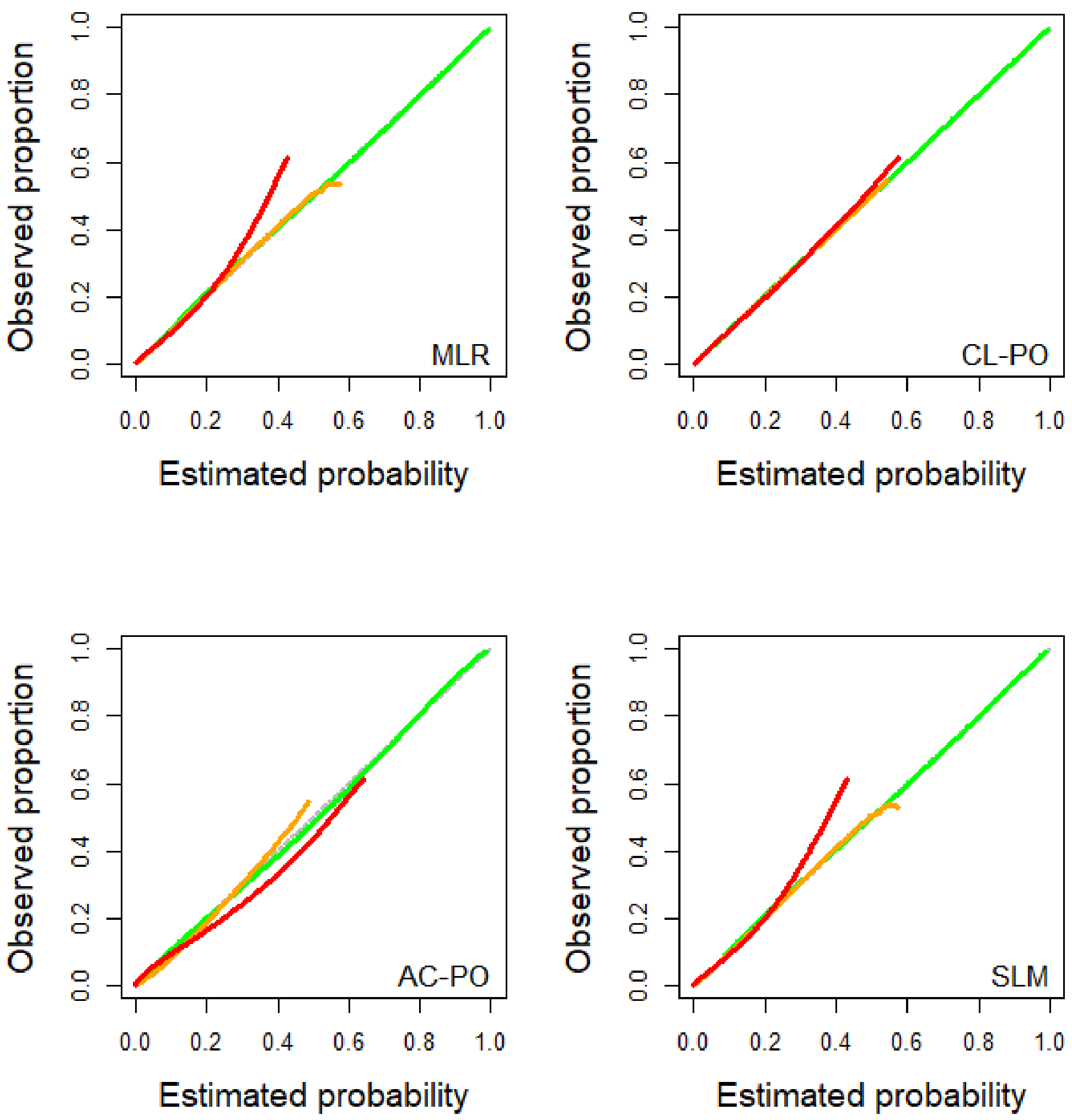

Figure S28. Flexible smoothed calibration curves per outcome category for simulation scenario 4 when the true model has the form of a cumulative logit model with proportional odds (green for category 1, orange for category 2, red for category 3, brown for category 4). These curves are based on the dataset used to develop the model and are therefore apparent (or unvalidated) curves (n=200,000). For some models, lines overlap.

MLR, multinomial logistic regression; CL-PO, cumulative logit model with proportional odds; AC-PO, adjacent category logit model with proportional odds; SLM, stereotype logit model.

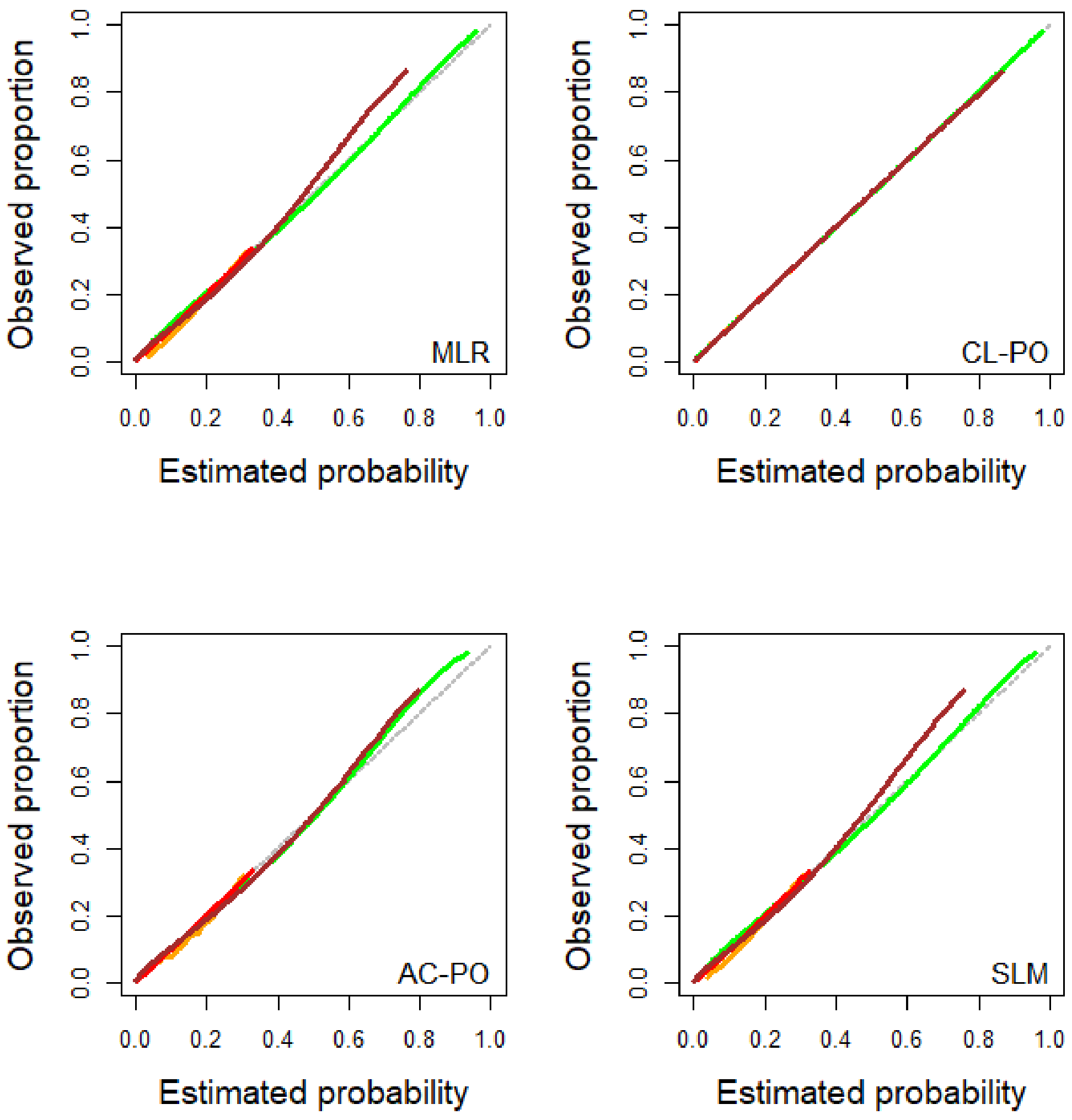

Figure S29. Flexible smoothed calibration curves per outcome category for simulation scenario 5 when the true model has the form of a cumulative logit model with proportional odds (green for category 1, orange for category 2, red for category 3, brown for category 4). These curves are based on the dataset used to develop the model and are therefore apparent (or unvalidated) curves (n=200,000). For some models, lines overlap.

MLR, multinomial logistic regression; CL-PO, cumulative logit model with proportional odds; AC-PO, adjacent category logit model with proportional odds; SLM, stereotype logit model.

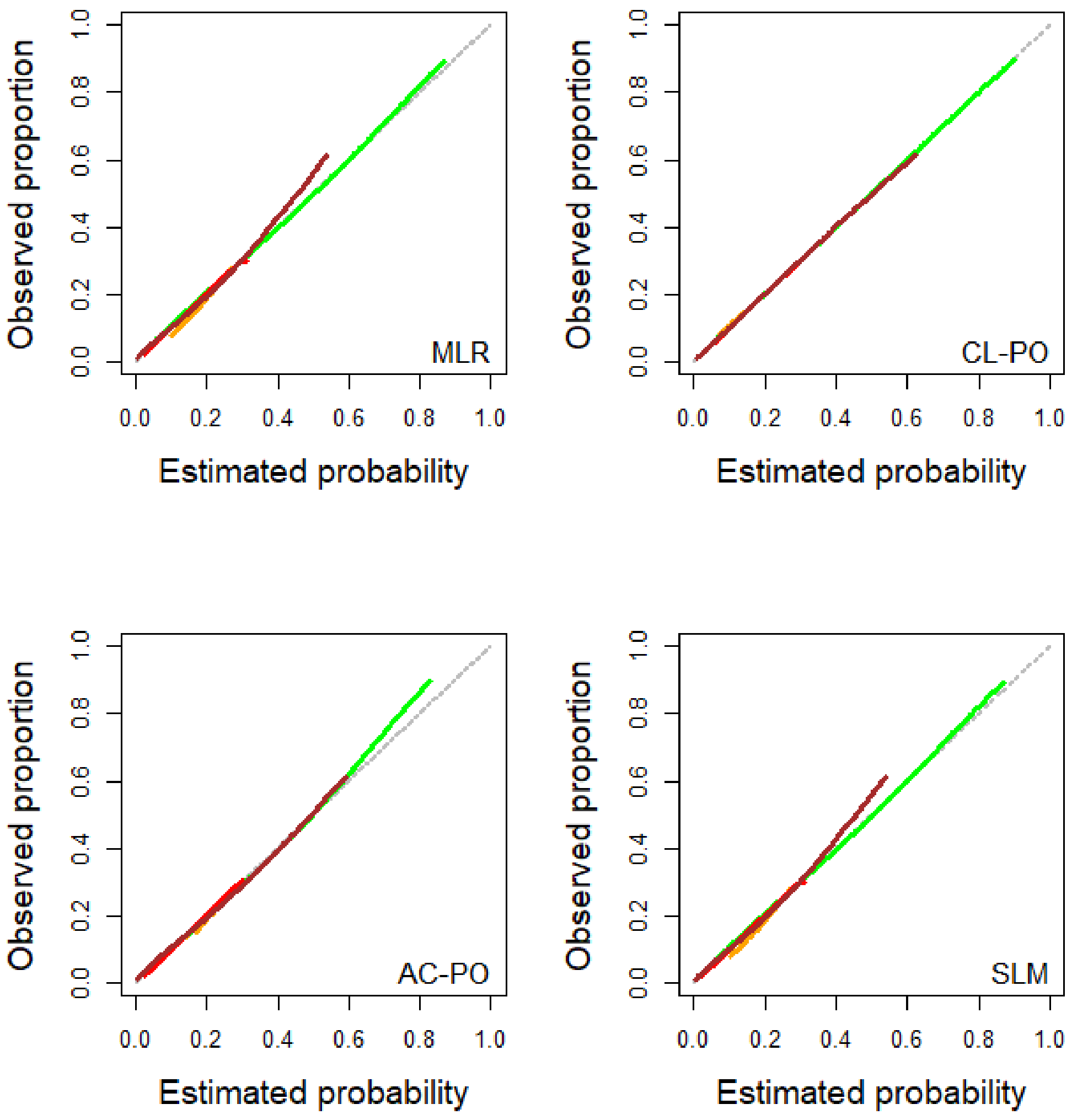

Figure S30. Flexible smoothed calibration curves per outcome category for simulation scenario 6 when the true model has the form of a cumulative logit model with proportional odds (green for category 1, orange for category 2, red for category 3, brown for category 4). These curves are based on the dataset used to develop the model and are therefore apparent (or unvalidated) curves (n=200,000). For some models, lines overlap.

MLR, multinomial logistic regression; CL-PO, cumulative logit model with proportional odds; AC-PO, adjacent category logit model with proportional odds; SLM, stereotype logit model.

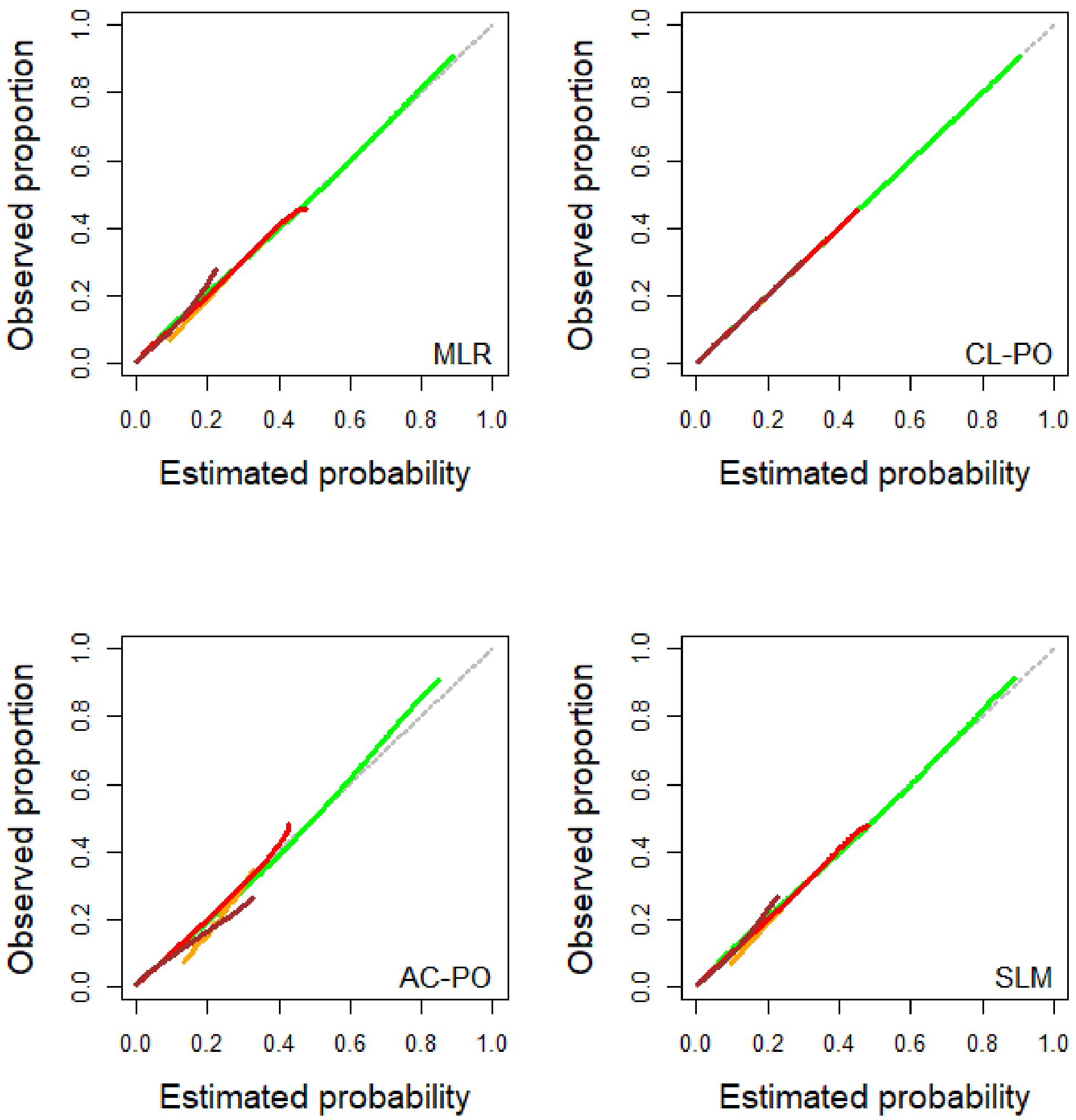

Figure S31. Calibration scatter plots per outcome category for simulation scenario 7 when the true model has the form of a cumulative logit model with proportional odds (green for category 1, orange for category 2, red for category 3). These curves are based on the dataset used to develop the model and are therefore apparent (or unvalidated) curves (n=200,000). Because all predictors are binary, no flexible curves are shown.

MLR, multinomial logistic regression; CL-PO, cumulative logit model with proportional odds; AC-PO, adjacent category logit model with proportional odds; SLM, stereotype logit model.

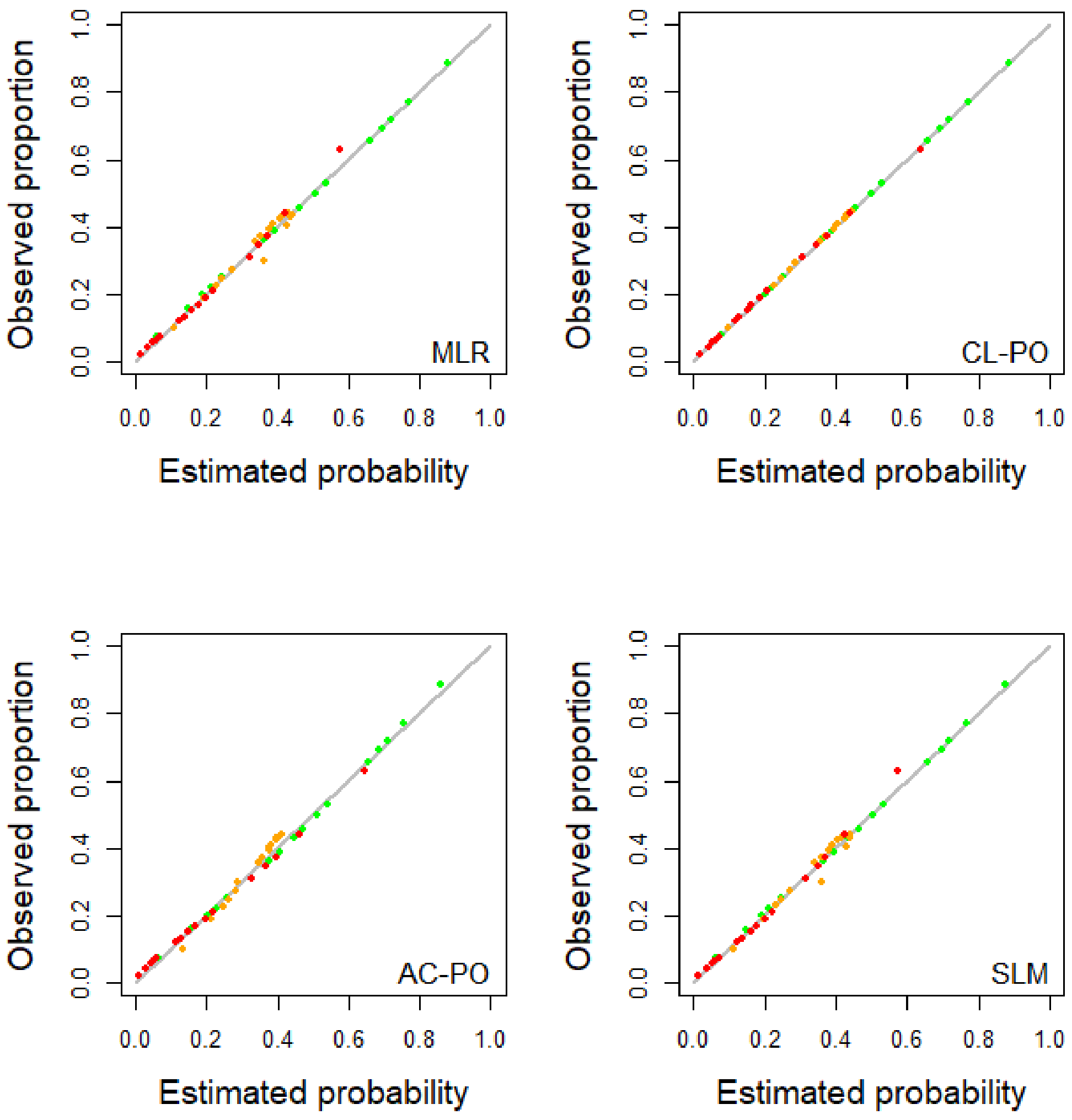

Figure S32. Calibration scatter plots per outcome category for simulation scenario 8 when the true model has the form of a cumulative logit model with proportional odds (green for category 1, orange for category 2, red for category 3, brown for category 4). These curves are based on the dataset used to develop the model and are therefore apparent (or unvalidated) curves (n=200,000). Because all predictors are binary, no flexible curves are shown.

MLR, multinomial logistic regression; CL-PO, cumulative logit model with proportional odds; AC-PO, adjacent category logit model with proportional odds; SLM, stereotype logit model.

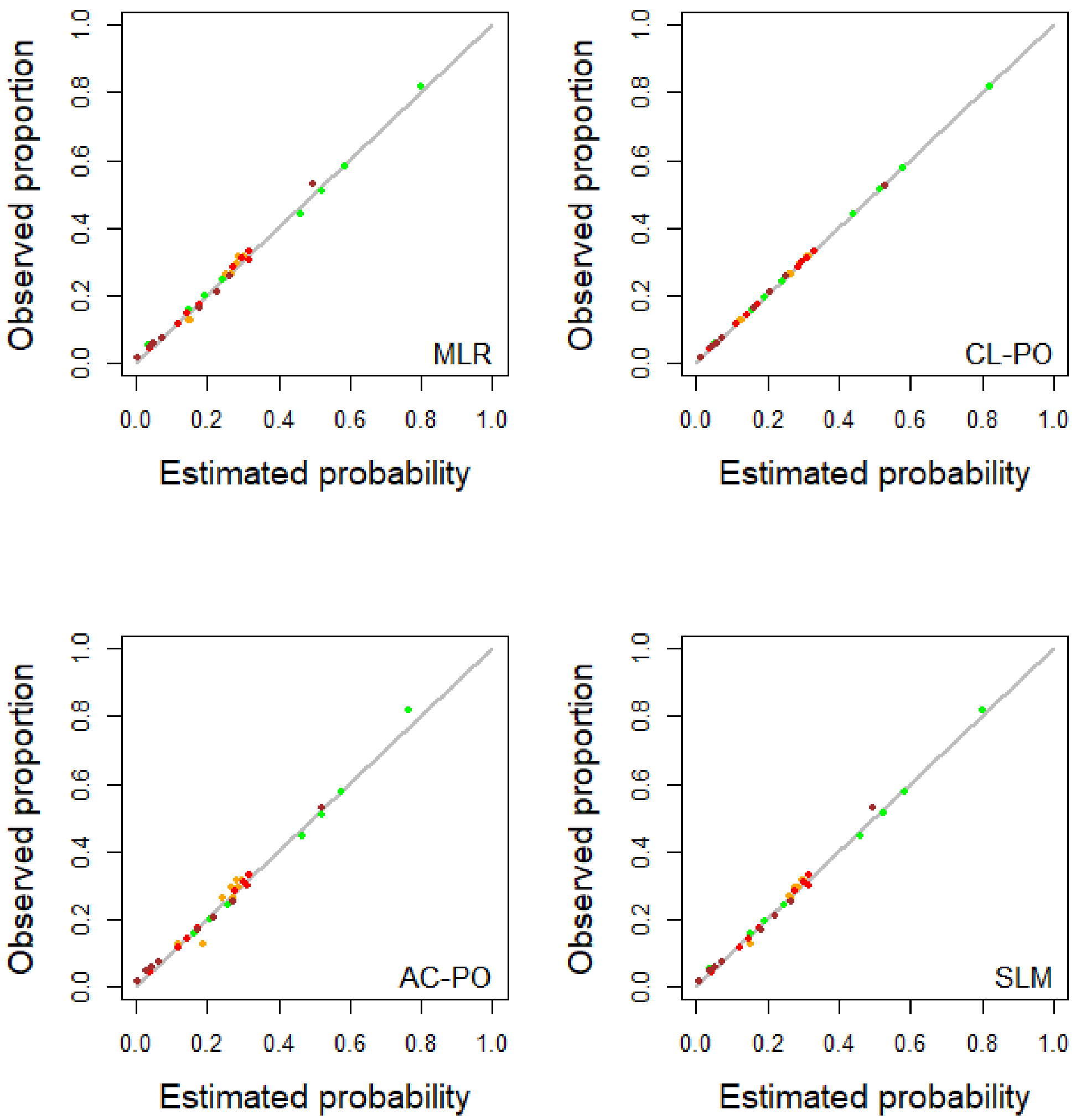

Figure S33. Flexible smoothed calibration curves per outcome category for simulation scenario 9 when the true model has the form of a cumulative logit model with proportional odds (green for category 1, orange for category 2, red for category 3, brown for category 4). These curves are based on the dataset used to develop the model and are therefore apparent (or unvalidated) curves (n=200,000). For some models, lines overlap.

MLR, multinomial logistic regression; CL-PO, cumulative logit model with proportional odds; AC-PO, adjacent category logit model with proportional odds; SLM, stereotype logit model.

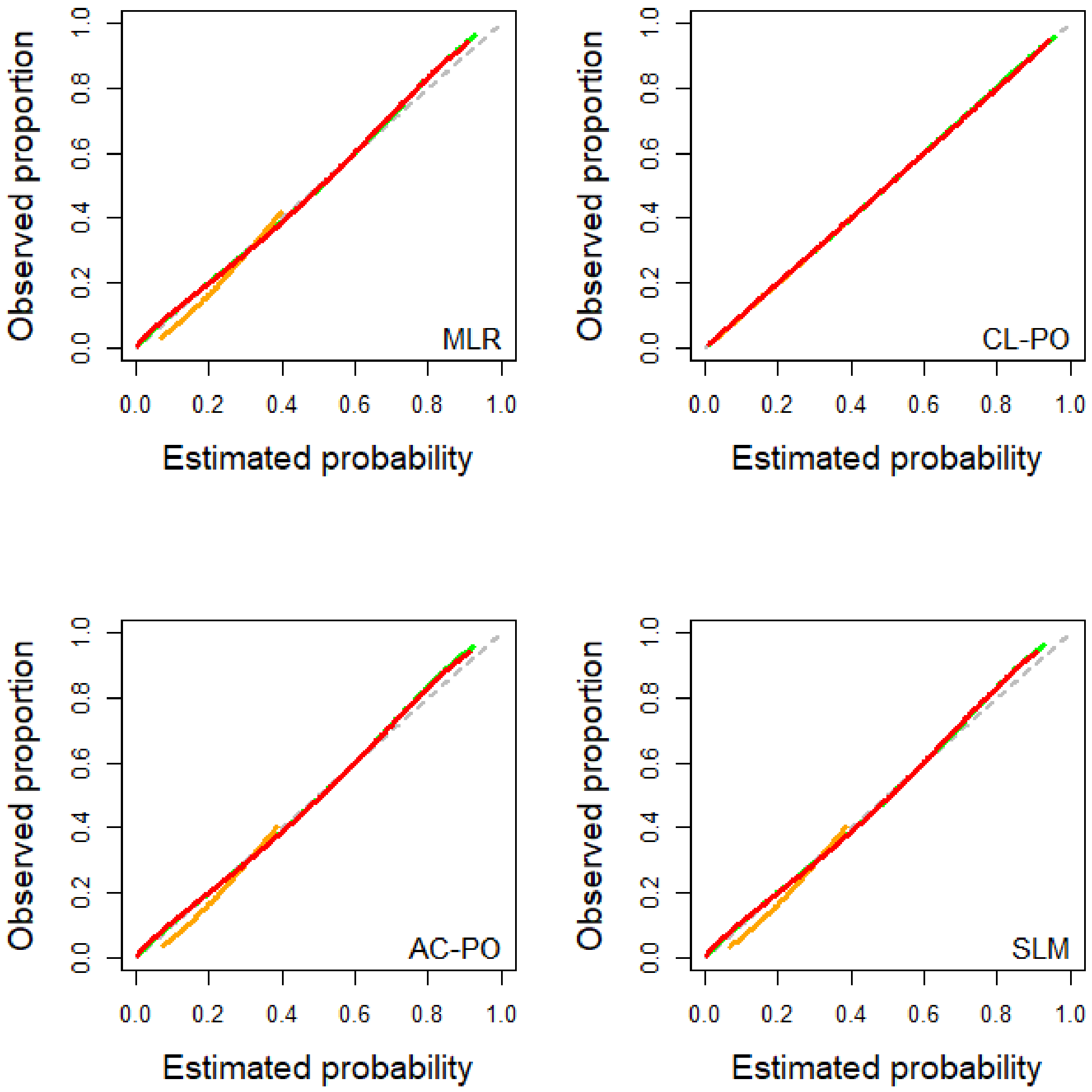

Figure S34. Scatter plot of estimated probabilities for having no coronary artery disease in the case study (n=4,888).

MLR, multinomial logistic regression; CL-PO, cumulative logit model with proportional odds; AC-PO, adjacent category logit model with proportional odds; CR-PO, continuation ratio logit model with proportional odds; CR-NP, continuation ratio logit model without proportional odds; SLM, stereotype logit model

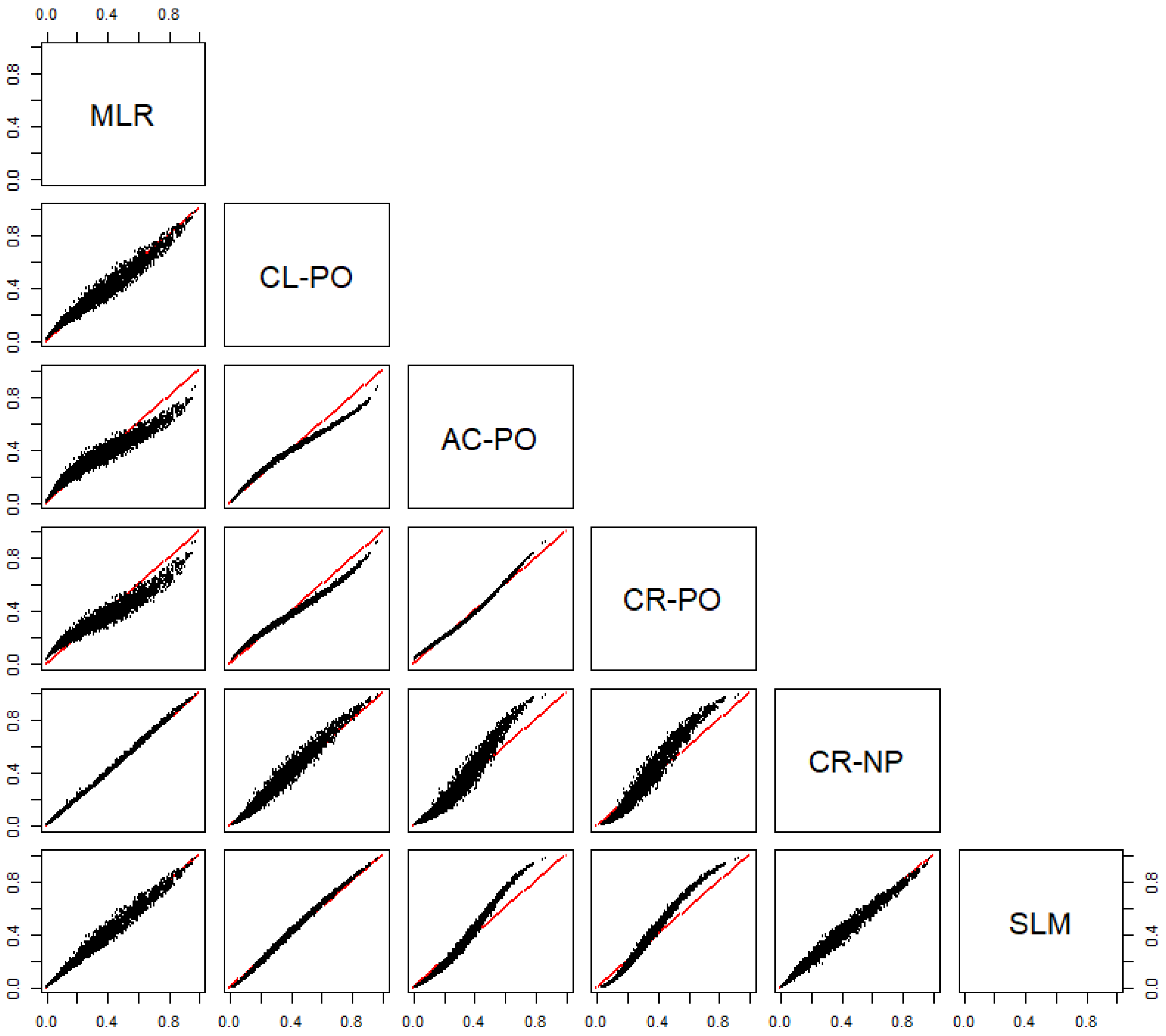

Figure S35. Scatter plot of estimated probabilities for having one-vessel disease in the case study (n=4,888).

MLR, multinomial logistic regression; CL-PO, cumulative logit model with proportional odds; AC-PO, adjacent category logit model with proportional odds; CR-PO, continuation ratio logit model with proportional odds; CR-NP, continuation ratio logit model without proportional odds; SLM, stereotype logit model

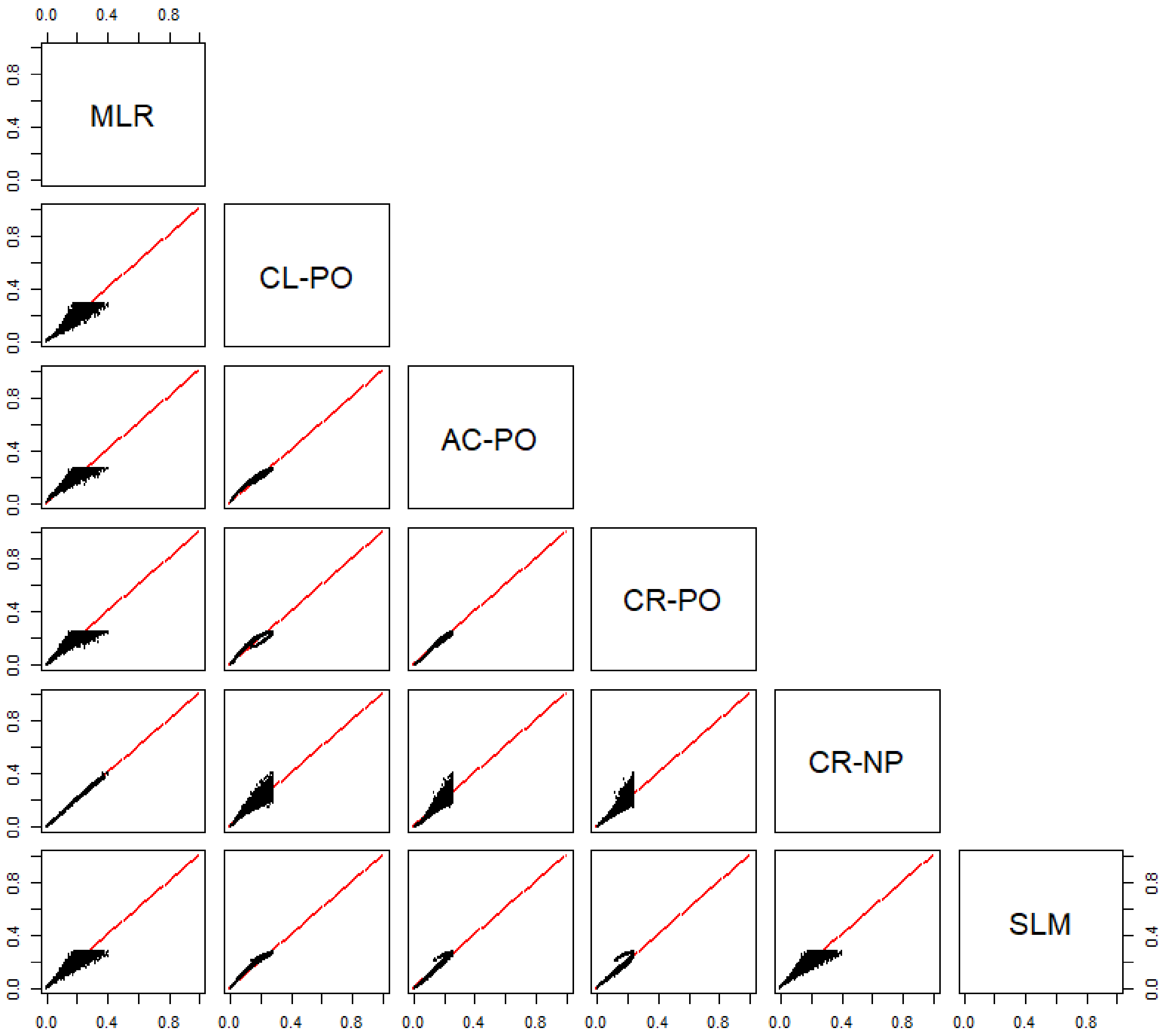

Figure S36. Scatter plot of estimated probabilities for having two-vessel disease in the case study (n=4,888).

MLR, multinomial logistic regression; CL-PO, cumulative logit model with proportional odds; AC-PO, adjacent category logit model with proportional odds; CR-PO, continuation ratio logit model with proportional odds; CR-NP, continuation ratio logit model without proportional odds; SLM, stereotype logit model

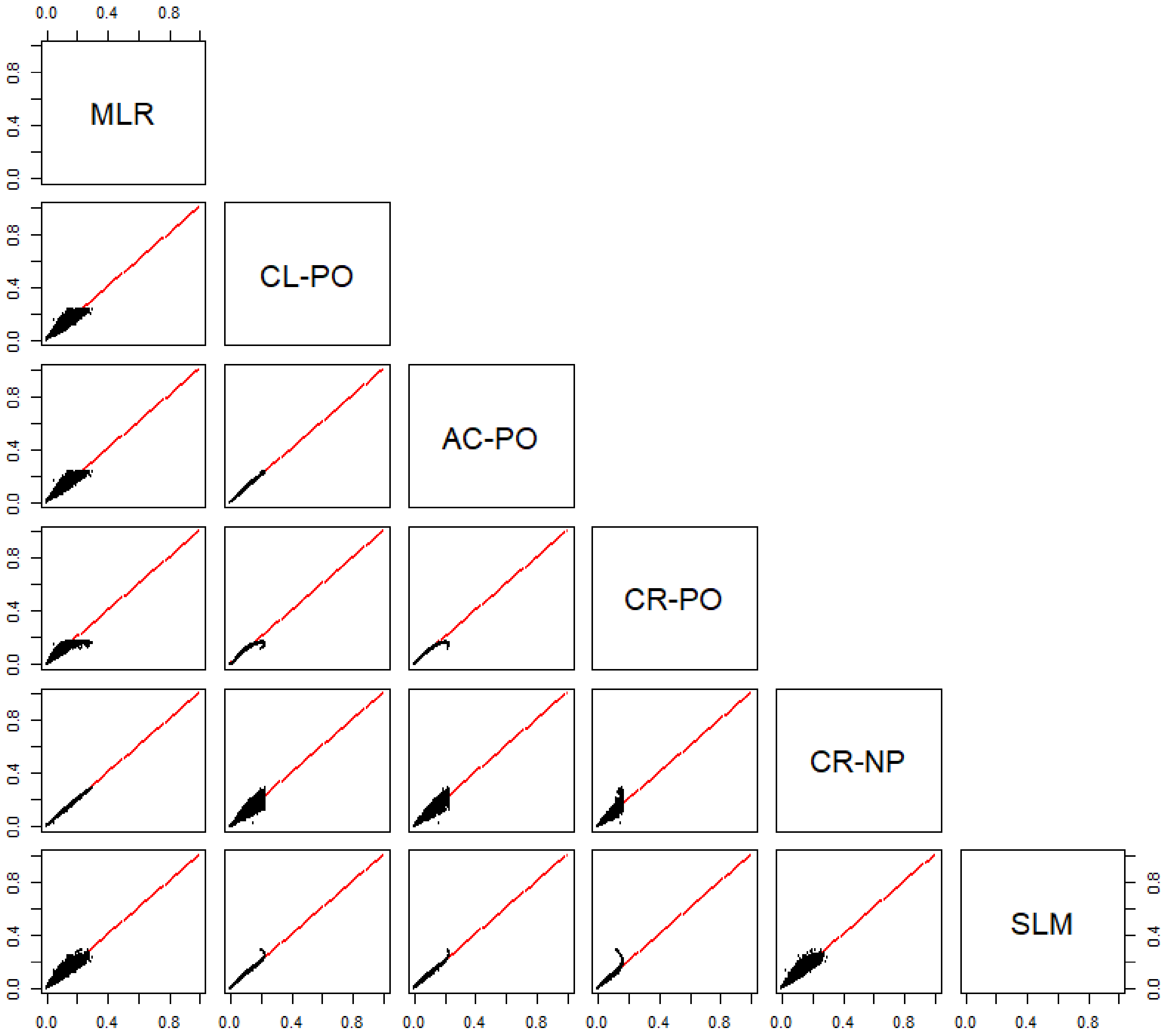

Figure S37. Scatter plot of estimated probabilities for having three-vessel disease in the case study (n=4,888).

MLR, multinomial logistic regression; CL-PO, cumulative logit model with proportional odds; AC-PO, adjacent category logit model with proportional odds; CR-PO, continuation ratio logit model with proportional odds; CR-NP, continuation ratio logit model without proportional odds; SLM, stereotype logit model

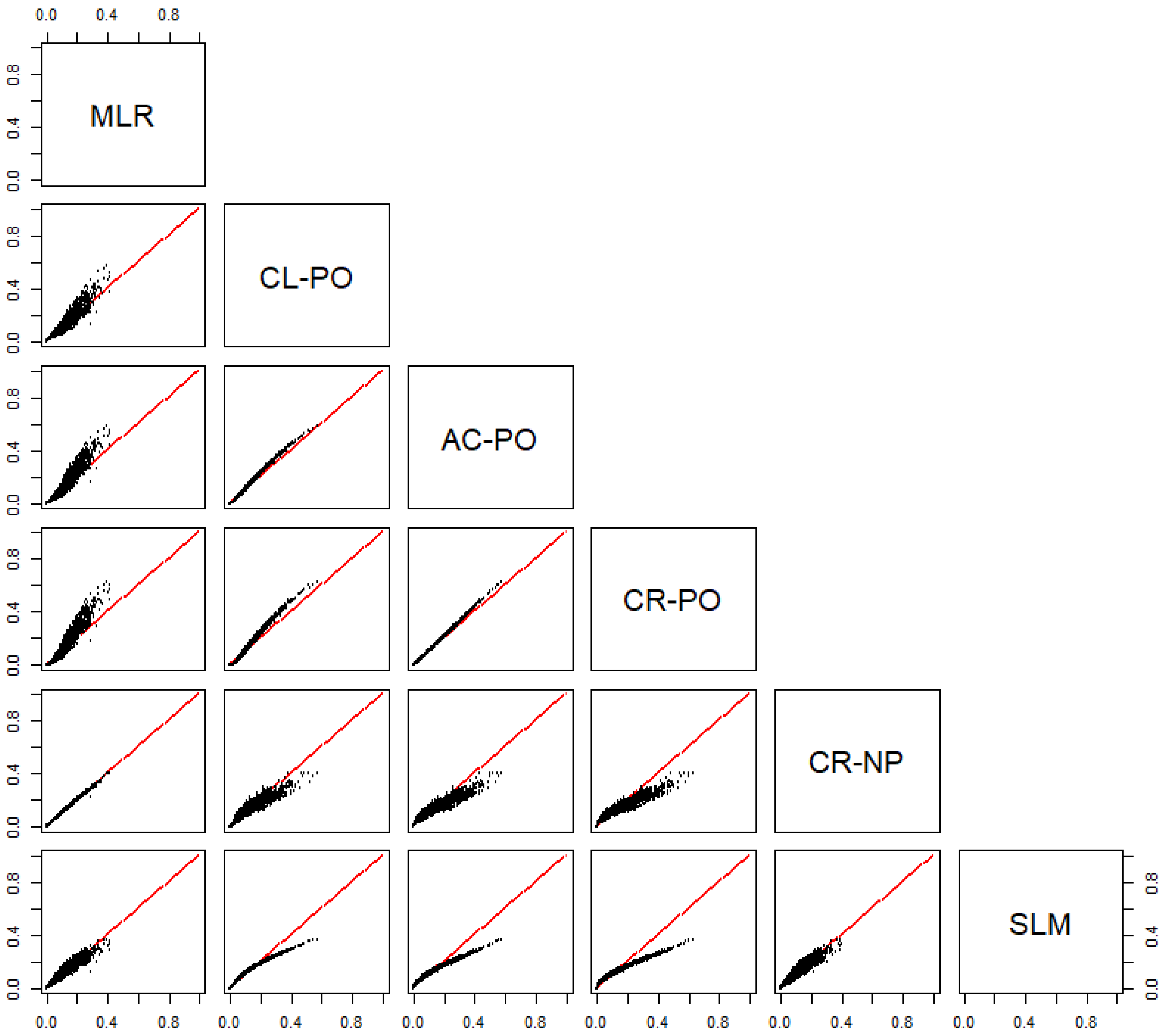

Figure S38. Calibration plots for the MLR model in the case study. These curves are based on the dataset used to develop the model and are therefore apparent (or unvalidated) curves. The top left plot shows the flexible calibration scatter plot per outcome category, the top right plot the flexible calibration curves per outcome category, the bottom left plot the flexible calibration scatter plot per outcome dichotomy, and the bottom right the flexible calibration curves per outcome dichotomy.

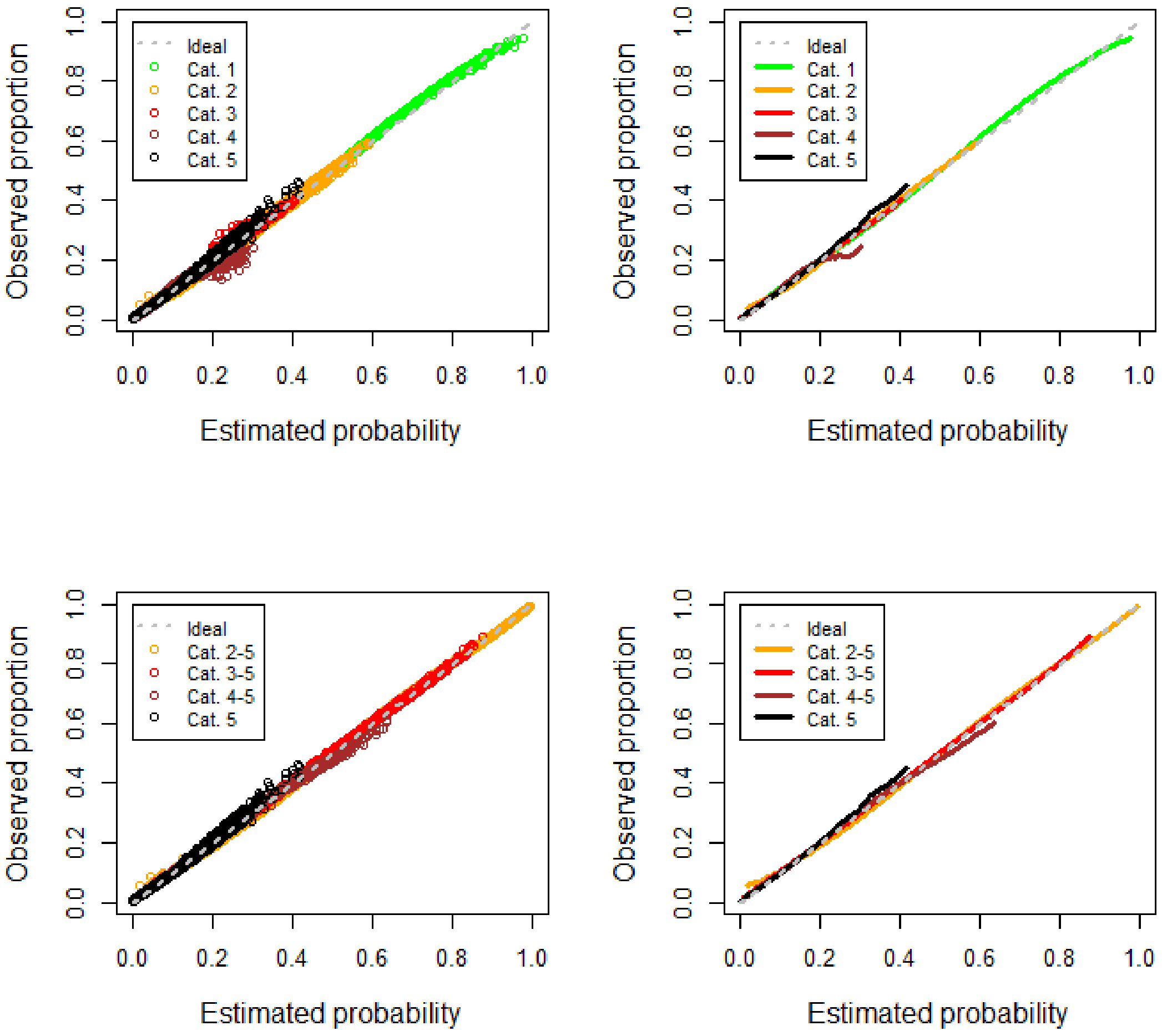

Figure S39. Calibration plots for the CL-PO model in the case study. These curves are based on the dataset used to develop the model and are therefore apparent (or unvalidated) curves. The top left plot shows the flexible calibration scatter plot per outcome category, the top right plot the flexible calibration curves per outcome category, the bottom left plot the flexible calibration scatter plot per outcome dichotomy, and the bottom right the flexible calibration curves per outcome dichotomy.

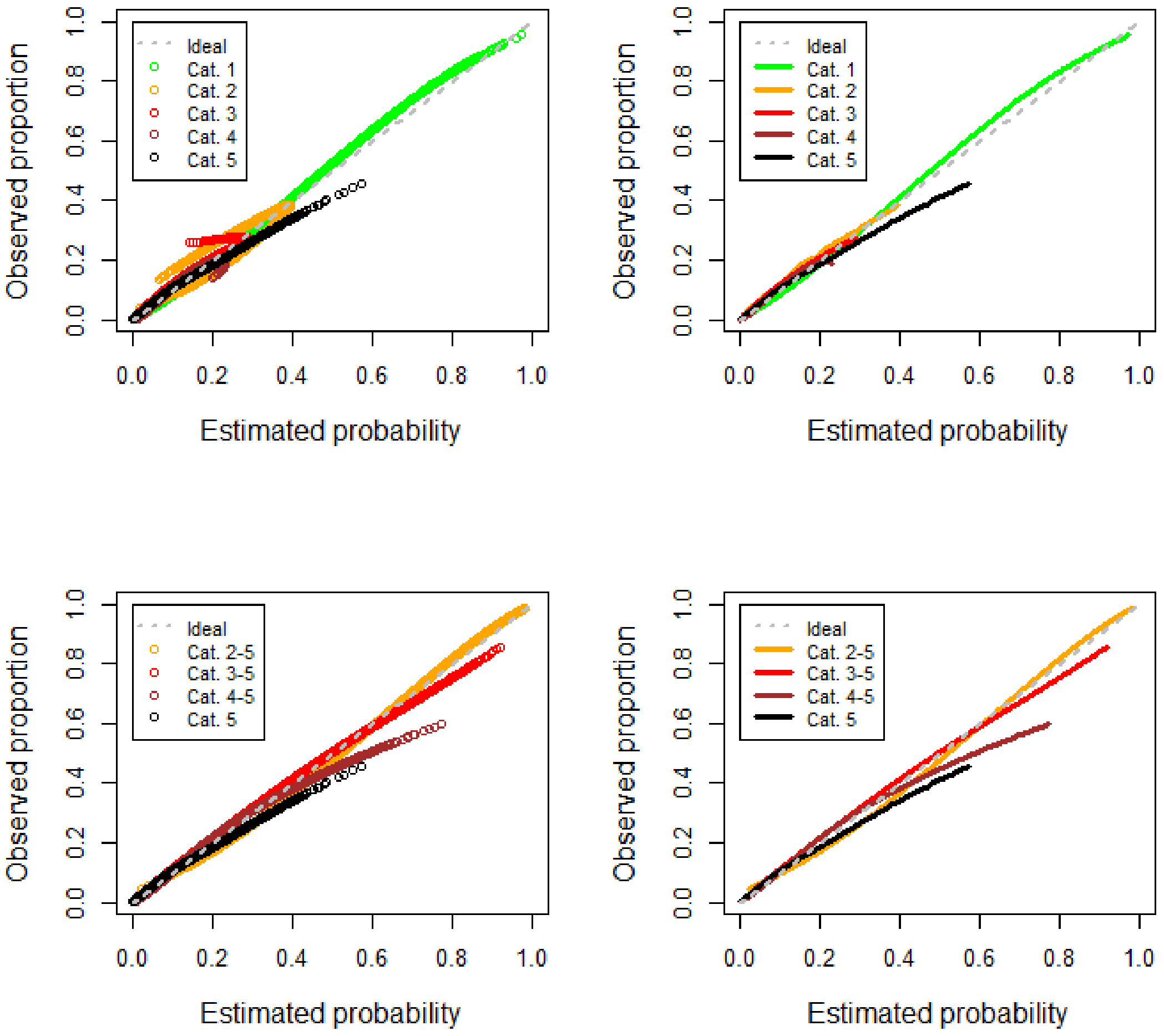

Figure S40. Calibration plots for the AC-PO model in the case study. These curves are based on the dataset used to develop the model and are therefore apparent (or unvalidated) curves. The top left plot shows the flexible calibration scatter plot per outcome category, the top right plot the flexible calibration curves per outcome category, the bottom left plot the flexible calibration scatter plot per outcome dichotomy, and the bottom right the flexible calibration curves per outcome dichotomy.

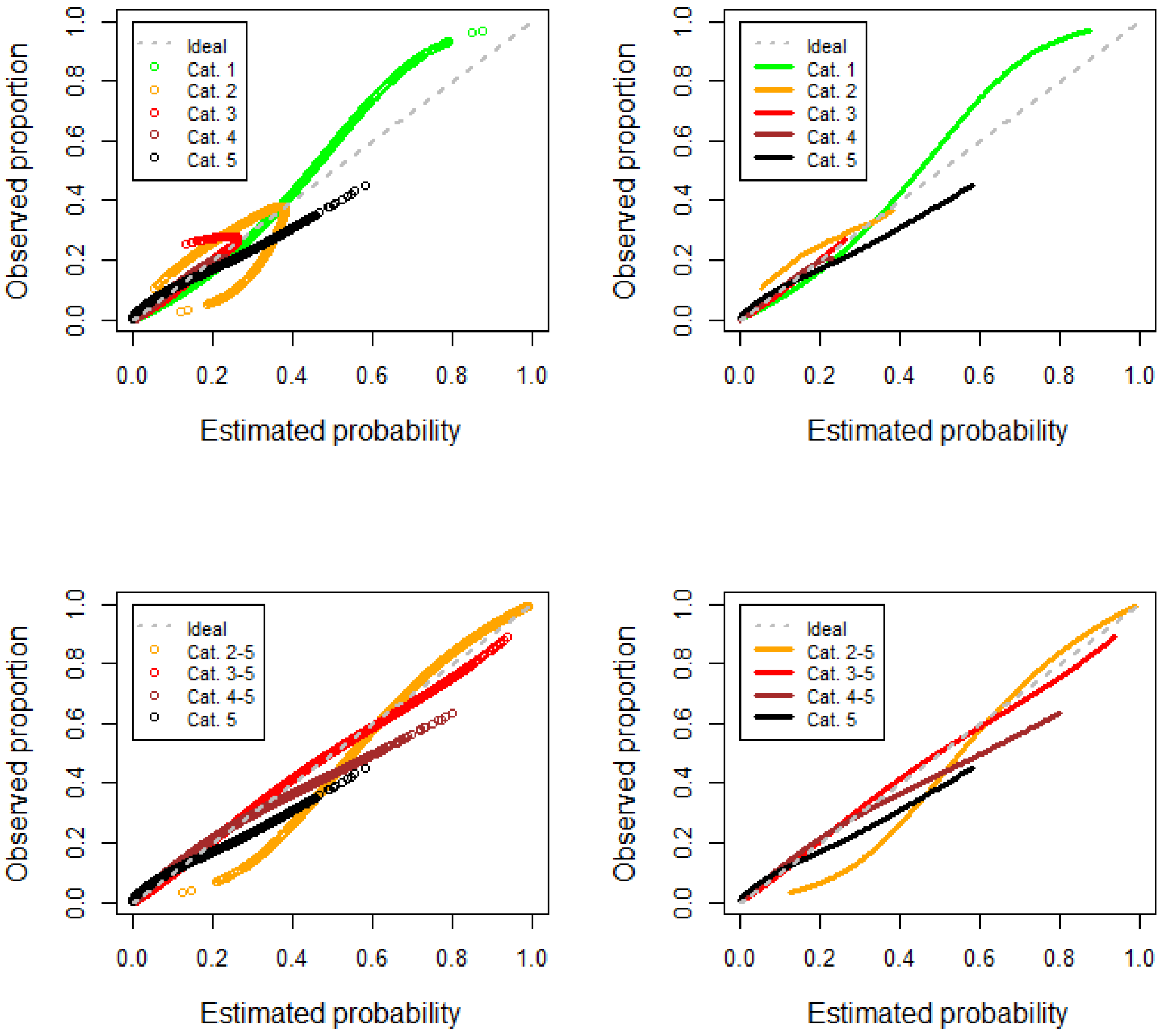

Figure S41. Calibration plots for the CR-PO model in the case study. These curves are based on the dataset used to develop the model and are therefore apparent (or unvalidated) curves. The top left plot shows the flexible calibration scatter plot per outcome category, the top right plot the flexible calibration curves per outcome category, the bottom left plot the flexible calibration scatter plot per outcome dichotomy, and the bottom right the flexible calibration curves per outcome dichotomy.

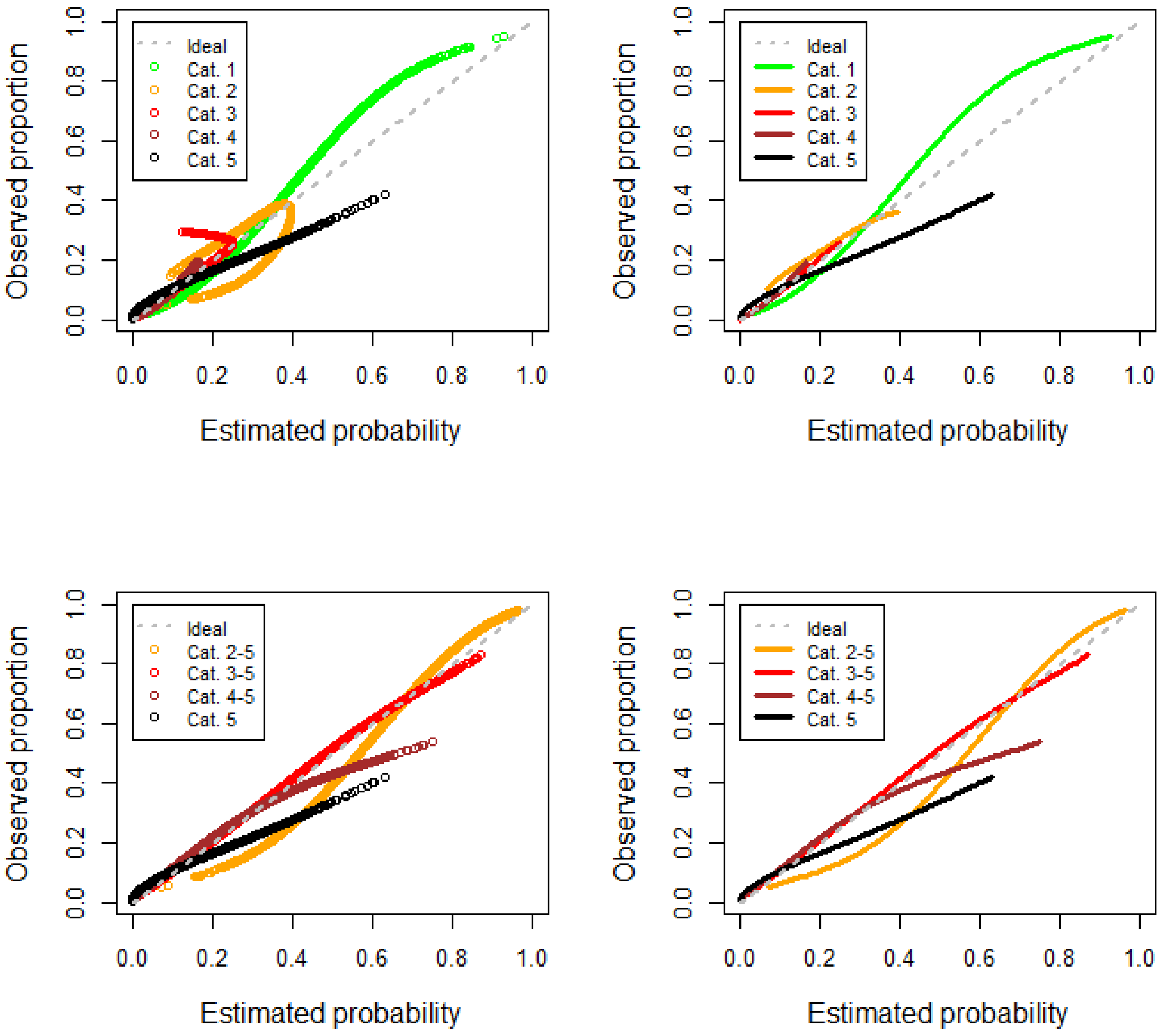

Figure S42. Calibration plots for the CL-NP model in the case study. These curves are based on the dataset used to develop the model and are therefore apparent (or unvalidated) curves. The top left plot shows the flexible calibration scatter plot per outcome category, the top right plot the flexible calibration curves per outcome category, the bottom left plot the flexible calibration scatter plot per outcome dichotomy, and the bottom right the flexible calibration curves per outcome dichotomy.

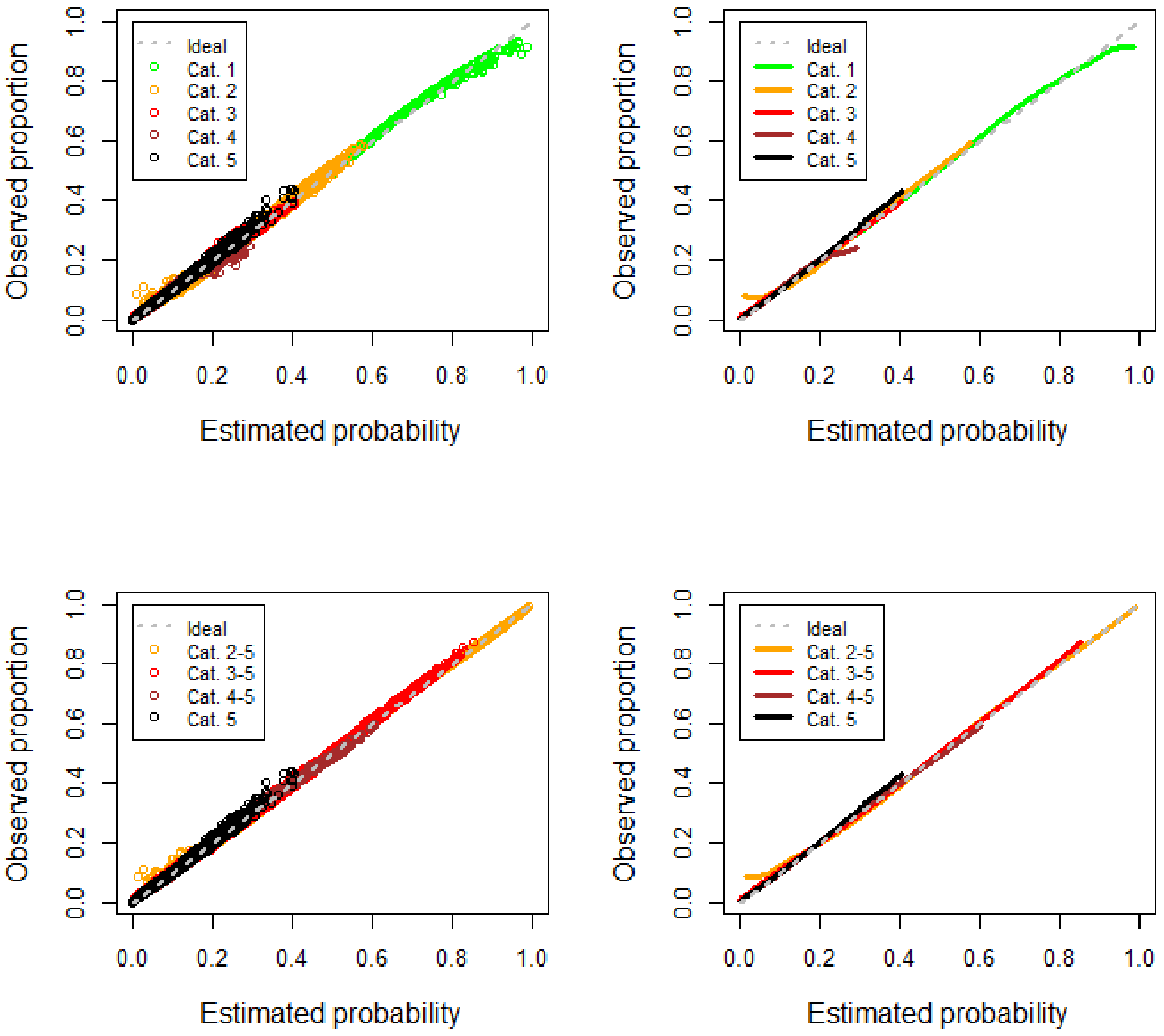

Figure S43. Calibration plots for the SLM model in the case study. These curves are based on the dataset used to develop the model and are therefore apparent (or unvalidated) curves. The top left plot shows the flexible calibration scatter plot per outcome category, the top right plot the flexible calibration curves per outcome category, the bottom left plot the flexible calibration scatter plot per outcome dichotomy, and the bottom right the flexible calibration curves per outcome dichotomy.

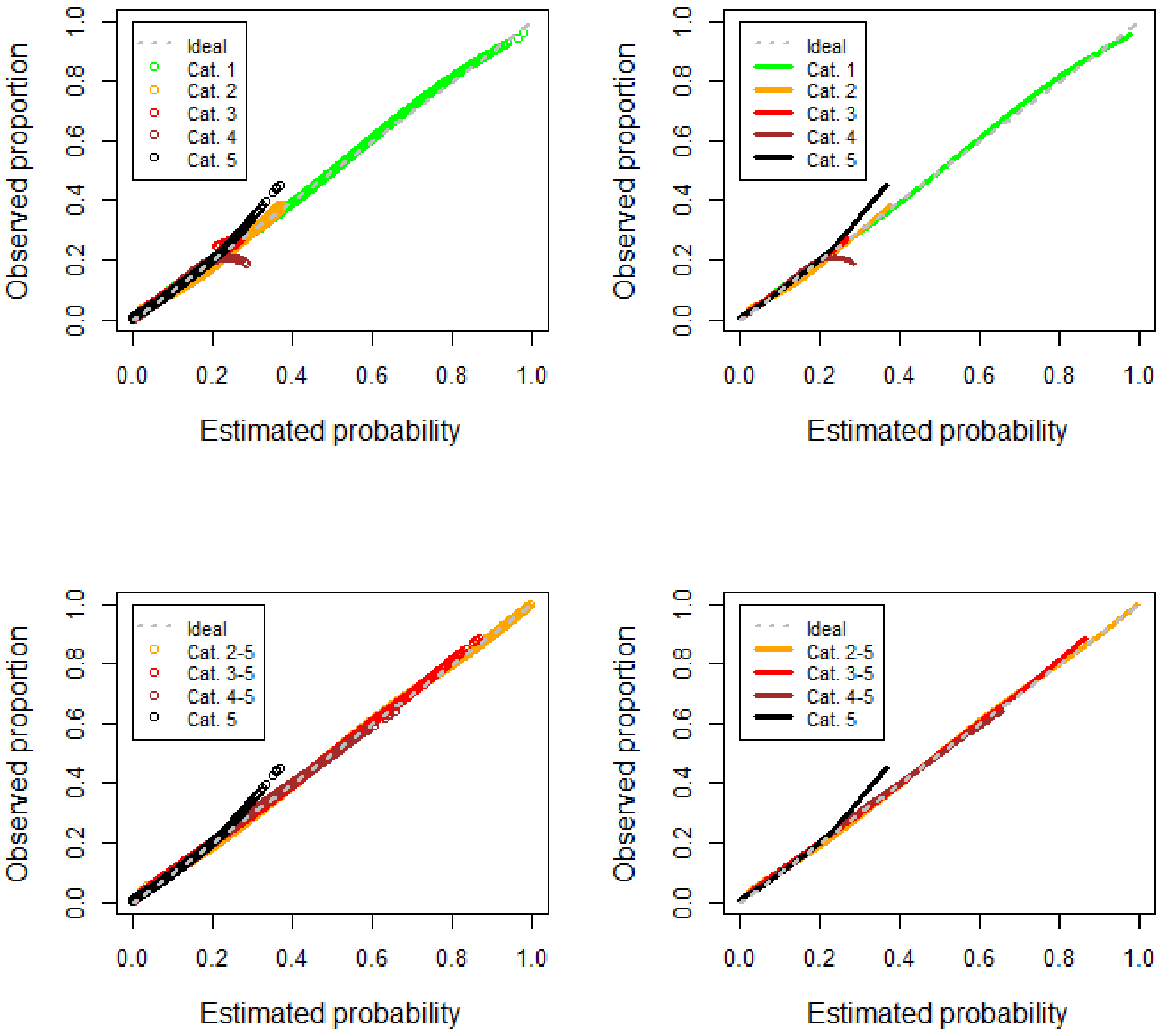